\documentclass[aps,pre,reprint,amsmath,amssymb,showpacs,superscriptaddress]{revtex4-1}

\usepackage{color}

\usepackage{graphicx}
\usepackage{dcolumn}
\usepackage{bm}
\usepackage{braket}%
\usepackage{natbib}
\usepackage{hyperref}
\hypersetup{%
 colorlinks=true, citecolor=blue, filecolor=blue, linkcolor=blue, urlcolor=blue}

\begin{document}

\preprint{Physical Review E}

\title{Lattice Boltzmann modeling and simulation of liquid jet breakup}


\author{Shimpei Saito}
\email[Corresponding author: ]{s1630195@tsukuba.ac.jp}
\affiliation{Graduate School of Systems and Information Engineering, University of Tsukuba, Tsukuba 305-8573, Japan}

\author{Yutaka Abe}
\affiliation{Faculty of Engineering, Information and Systems, University of Tsukuba,  Tsukuba 305-8573, Japan}

\author{Kazuya Koyama}
\affiliation{Reactor Core and Safety Design Department, Mitsubishi FBR Systems, Inc., Shibuya, Tokyo 150-0001, Japan}

\date{\today}

\begin{abstract}
	A three-dimensional color-fluid lattice Boltzmann model for immiscible two-phase flows is developed in the framework of a three-dimensional 27-velocity (D3Q27) lattice. 
	The collision operator comprises the D3Q27 versions of three sub-operators: a multiple-relaxation-time (MRT) collision operator, a generalized Liu--Valocchi--Kang perturbation operator, and a Latva-Kokko--Rothman recoloring operator.
	A D3Q27 version of an enhanced equilibrium distribution function is also incorporated into this model to improve the Galilean invariance.
	Three types of numerical tests, namely, a static droplet, an oscillating droplet, and the Rayleigh--Taylor instability, show a good agreement with analytical solutions and numerical simulations. 
	Following these numerical tests, this model is applied to liquid-jet-breakup simulations.
	The simulation conditions are matched to the conditions of the previous experiments.
	In this case, numerical stability is maintained throughout the simulation, although the kinematic viscosity for the continuous phase is set as low as $1.8\times10^{-4}$, in which case the corresponding Reynolds number is $3.4\times10^{3}$;
	the developed lattice Boltzmann model based on the D3Q27 lattice enables us to perform the simulation with parameters directly matched to the experiments.
	The jet's liquid column transitions from an asymmetrical to an axisymmetrical shape, and entrainment occurs from the side of the jet.
	The measured time history of the jet's leading-edge position shows a good agreement with the experiments.
	Finally, the reproducibility of the regime map for liquid-liquid systems is assessed.
	The present lattice Boltzmann simulations well reproduce the characteristics of predicted regimes, including varicose breakup, sinuous breakup, and atomization.
\end{abstract}

\pacs{47.11.-j, 47.55.df, 47.61.Jd, 47.85.Dh}

\maketitle

\section{\label{sec:level1}Introduction}
	Multiphase and multicomponent flows appear in many natural and industrial processes.
	A liquid jet injected into another fluid is an interesting example of such a flow. 
	Significant efforts have been put into understanding the breakup of a liquid jet for more than a century.
	Since the pioneering works of Plateau~\citep{Plateau1873} and Rayleigh~\citep{Rayleigh1878}, extensive studies on this subject have been performed theoretically, experimentally, and numerically~\citep{McCarthy1974, Lin1998, Villermaux2007,Eggers2008}.

	Drops form directly from the nozzle at low injection velocities, and a liquid jet issues from the nozzle and then breaks into droplets in various patterns at higher injection velocities.
	The occurrence of such a regime is of interest in the study of liquid-jet breakup.
	\citet{Ohnesorge1936} classified his results into four types of breakup regimes: (0) dripping, (I) varicose, (II) sinuous, and (III) atomization~\citep{Kolev2005,McKinley2011}.
	He also provided a regime map of liquid jets in a gas using the Ohnesorge and Reynolds numbers. 
	After Ohnesorge's work, much research on this subject has been performed (e.g., ~\citep{Merrington1947, Tanasawa1954, Grant1966}).
	The majority of investigations have focused upon liquid-gas systems (liquid jet into gaseous atmosphere).
	Breakup of jets in liquid-liquid systems (liquid jets into another liquid) has not been investigated as extensively.
	Our focus in this paper is therefore on the breakup of liquid jets in immiscible liquid-liquid systems.

	The liquid-liquid-jet systems can also be found in several fields, e.g., chemical processing~\citep{Meister1969,Takahashi1971,Das1997} and CO$_2$ storage in oceans~\citep{Riestenberg2004,Tsouris2007}.
	In the field of nuclear engineering, interactions between melt and coolant must be well understood for safety design of nuclear reactors
	and have therefore been extensively investigated in the literature~\citep{Kondo1995,Dinh1999,Abe2006}.
	In experiments, a high-temperature melt is often used to simulate the core melt materials.
	\citet{Abe2006} discussed the relationships between the fragment size and the Rayleigh--Taylor and Kelvin--Helmholtz interfacial instabilities~\citep{Chandrasekhar1961}.

\begin{figure*}[htbp]
	\includegraphics[width=15cm]{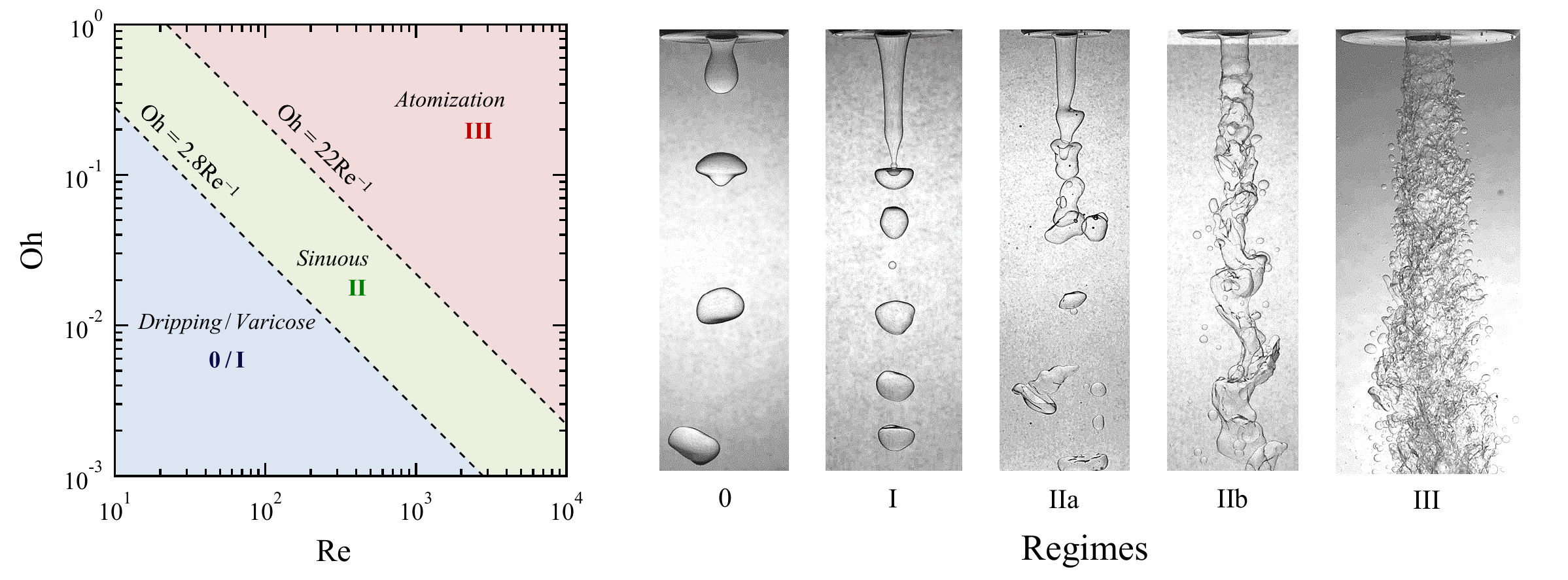}
	\caption{Left: Breakup regime map for jets in immiscible liquid-liquid systems. Right: Corresponding visualization images: Regime 0--dripping, Regime I--varicose breakup, Regime IIa--sinuous breakup {\it without} entrainment, Regime IIb--sinuous breakup {\it with} entrainment, and Regime III--atomization~\citep{Saito2017}. 
	\label{fig:typicalRegimes}}
\end{figure*}

	To better understand the fundamental interactions between melt-jet and coolant interactions, experiments using appropriate test fluids under isothermal conditions are also effective approaches as a separate effect of such interactions~\citep{Dinh1999,Saito2017}.
	\citet{Saito2017} developed a breakup-regime map for a jet in immiscible liquid-liquid systems based on experiments and phenomenological considerations.
	Figure~\ref{fig:typicalRegimes} shows the dimensionless map and the corresponding visual images.
	The Ohnesorge classification~\citep{Ohnesorge1936} was extended to liquid-liquid systems. 
	As can be seen, various flow regimes occur during liquid-jet-breakup processes.
	The breakup transitions from the dropping regime to the atomization regime,
	and the generated droplet size drastically changes depending on the conditions.
	This implies that the breakups of liquid jets are essentially three-dimensional flows and possess multiscale phenomena such as droplet pinch-off and atomization.
%

	Numerical simulations of liquid-liquid jets involve the solution of the Navier--Stokes equations for two fluids with specified boundary and interface conditions.
	Several approaches to solving these types of free-surface problems are available in the literature.
	As a first attempt, \citet{Richards1993} investigated the axisymmetric steady-state laminar jet based on the volume-of-fluid (VOF) method~\citep{Hirt1981}.
	\citet{Thakre2015} also used the VOF method provided by a commercial code, FLUENT, to simulate a melt jet into water.
	They successfully reproduced a variation in the breakup length~\citep{Burger1995};
	later, a similar variation was confirmed by experiments~\citep{Saito2017,Saito2015}.
	\citet{Homma2006} numerically investigated liquid-liquid-jet breakup using a front-tracking method~\citep{Unverdi1992,Tryggvason2001}. 
	They mapped different breakup modes on a plot of Weber number vs. viscosity ratio.
	The drawback of their front-tracking simulations was that they neglected the coalescence of generated drops.
	The aforementioned numerical simulations~\citep{Richards1993,Thakre2015,Homma2006} were limited to two-dimensional cases.

\begin{figure}[b]
	\includegraphics[width=8.5cm]{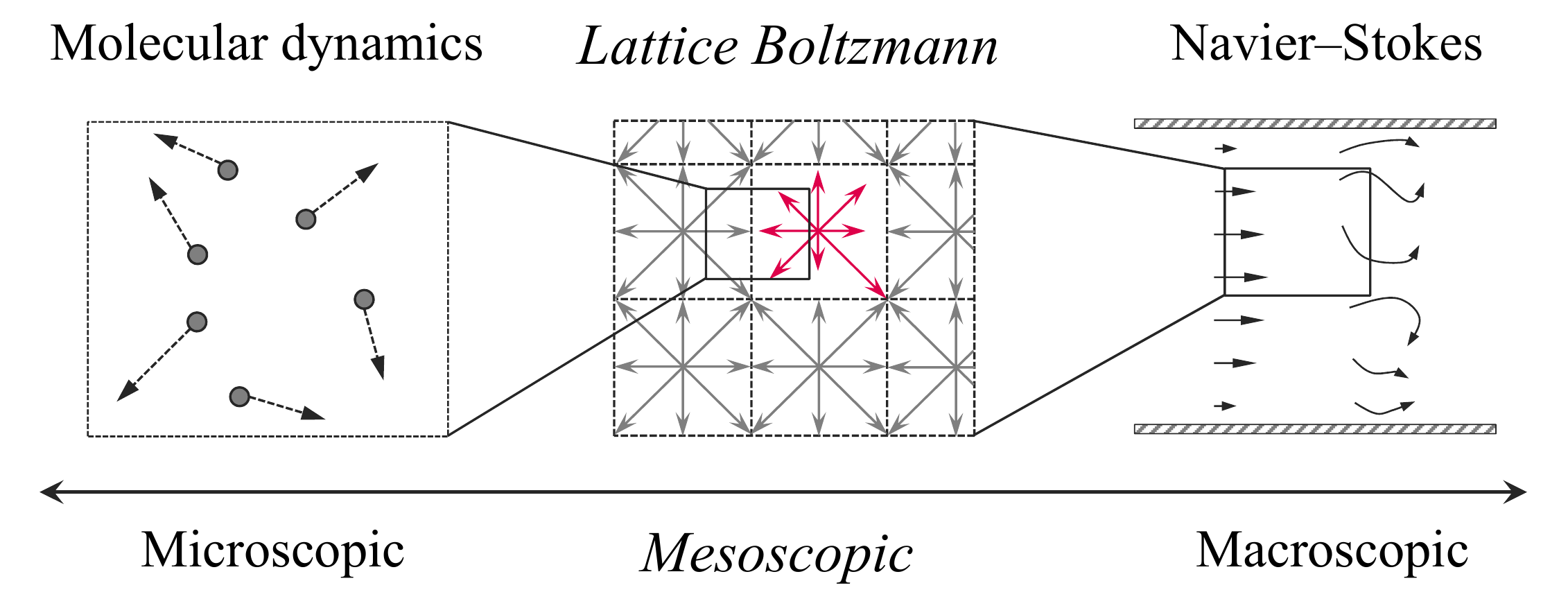}
	\caption{Relationship between multiscale properties of a fluid flow and a simulation method.
	The lattice Boltzmann method is the so-called mesoscopic simulation method between the microscopic particle-based (e.g., molecular dynamics) and macroscopic Navier--Stokes-based methods.
	\label{fig:scale}}
\end{figure}

	A completely different approach is the use of a lattice Boltzmann method.
	Several authors have investigated liquid-liquid-jet flows using multiphase lattice Boltzmann models~\citep{McCracken2005a,Matsuo2015,Saito2016}. 
	In recent years, this method has been recognized as a powerful tool for analysis of complex fluid dynamics, including multicomponent and multiphase flows~\citep{Aidun2010}.
	Compared with other macroscopic CFD methods based on the Navier--Stokes equations, the lattice Boltzmann method, which is constructed using mesoscopic kinetic equations, has several advantages.
	For instance, it is easy to incorporate mesoscale physics such as interfacial breakup or coalescence.
	Moreover, the computational cost for simulating realistic fluid flows is reasonable compared with particle-based methods (e.g., molecular dynamics).
	The relations of the scale properties in fluid flows are schematically illustrated in Fig.~\ref{fig:scale}.
	
	Two-phase or multiphase lattice Boltzmann models can be classified into four categories: 
	\begin{itemize}
		\item Color-fluid model~\citep{Gunstensen1991,Grunau1993},
		\item Pseudo-potential model~\citep{Shan1993,Shan1994},
		\item Free-energy model~\citep{Swift1995,Swift1996},
		\item Mean-field model~\citep{He1999}.
	\end{itemize}
	This classification may not be exhaustive, for instance, the latter two models are sometimes identified as phase-field models~\citep{Li2016} since the Cahn--Hilliard or similar interface tracking equations can be derived from them.
	For details about the multiphase lattice Boltzmann models, interested readers can refer to the comprehensive review papers~\citep{Chen1998, Nourgaliev2003, Aidun2010, Chen2014, Liu2015, Li2016} and references therein.
	In this paper, we focus upon the color-fluid model. 
	This model possesses many strengths in simulations of multiphase/multicomponent flows, including strict mass conservation for each fluid and flexibility in adjusting the interfacial tension~\citep{Ba2016}. 
	A static drop test is no longer needed to determine the interfacial tension; it can be directly obtained without any analysis or assumptions. 
	Moreover, the color-fluid model shows a very small dissolution property for tiny droplets or bubbles \citep{Liu2015}.

	Color-fluid models, which are often referred to as R-K or color-gradient models, were first developed by~\citet{Gunstensen1991}, who extended the two-component lattice gas automata model of~\citet{Rothman1988}. 
	Later, \citet{Grunau1993} enabled the introduction of density and viscosity ratios by modifying the forms of the distribution functions. 
	\citet{Latva-Kokko2005} replaced Gunstensen's maximization-recoloring step with a formulaic segregation algorithm.
	Instead of widening the interface width, Latva-Kokko--Rothman's recoloring algorithm solves some issues with the previous color-fluid-type model, namely, the lattice-pinning problem and the spurious velocities.
	\citet{Reis2007} extended the model to common a two-dimensional nine-velocity (D2Q9) lattice.
	They modified the perturbation operator to recover the Navier--Stokes equations correctly.
	
	\citet{Leclaire2012} demonstrated that integrating Latva-Kokko--Rothman's recoloring operator~\citep{Latva-Kokko2005} into Reis--Phillips' perturbation operator~\citep{Reis2007} greatly improves the numerical stability and accuracy of solutions over a wide range of parameters. 
	Using an isotropic gradient operator also enhanced numerical stability and accuracy~\citep{Leclaire2011}.
	~\citet{Liu2012} derived a generalized perturbation operator using the phase-field (or order parameter) instead of a color-gradient and formulated the color-fluid model in three dimensions.
	Very recently, ~\citet{Leclaire2017} generalized the color-fluid-type lattice Boltzmann model in two and three dimensions.

	Galilean invariance is one of the issues to be improved in the color-fluid family.
	Following ~\citet{Holdych1998}, a source term to improve the Galilean invariance was derived by \citet{Leclaire2013} and incorporated into an equilibrium distribution function.
	The enhanced equilibrium distribution function showed an improvement of the momentum-discontinuity problem	through numerical tests on a layered Couette flow.
	Recently, ~\citet{Ba2016} have modified an equilibrium distribution function based on the third-order Hermite expansion of the Maxwellian distribution.
	They also showed that discontinuous velocities were improved by this modification.

	It is known that the LB method suffers from numerical instability in low-viscosity conditions. 
	Modification of the collision operator is one method for overcoming this issue~\citep{Luo2011}.
	A multiple-relaxation-time (MRT) collision operator or generalized lattice Boltzmann equation~\citep{dHumieres1994, Lallemand2000, dHumieres2002} has been widely used, even for multiphase flows, to enhance numerical stability and accuracy and to reduce spurious current near the interface.
	
	We return to the issue of lattice Boltzmann simulations of liquid-jet breakup.
	~\citet{McCracken2005b} successfully introduced an MRT operator to the multiphase lattice Boltzmann model and performed liquid-jet breakup simulations~\citep{McCracken2005a}.
	They assumed that the flow was axisymmetric in two dimensions. 
	They investigated the influence of interfacial tension, injection velocity, and liquid viscosity under a density ratio of 5.
	However, three-dimensional simulations are required to further understand breakup characteristics, since liquid-jet breakup is an essentially three-dimensional flow, as shown in Fig.~\ref{fig:typicalRegimes}.

	The authors carried out lattice Boltzmann simulations of liquid-jet breakup in three dimensions~\citep{Matsuo2015,Saito2016}. 
	\citet{Matsuo2015} used the three-dimensional two-phase lattice Boltzmann model, which was developed by~\citet{Ebihara2003} based on the model of \citet{He1999}. 
	They compared their simulation results with experiments and investigated the effect of the Froude number upon the jet-breakup length.
	In their simulations, however, the Reynolds number was limited to $O(10^2)$.
	This was an order of magnitude smaller than that in the target experiments.
	In addition, the model of \citet{He1999} suffered from the dissolution of tiny droplets~\citep{Liu2015}; thus, it would not be appropriate for liquid-jet-breakup simulations with tiny-droplet formation.
	
	~\citet{Saito2016} incorporated the MRT operator into the three-dimensional 19-velocity (D3Q19) color-fluid model proposed by~\citet{Liu2012} and applied this model to liquid-jet-breakup simulations.
	Although they could simulate liquid-jet breakup with the Reynolds number up to $O(10^3)$, the kinematic-viscosity ratio was set to unity to avoid numerical instability. 
	This meant that the kinematic viscosity of the surrounding liquid in their simulation was more viscous than that in the target experiment.
	Further improvement is required to compare the numerical results with experiments; this is the motivation of the present study.

	In this paper, we present the three-dimensional two-phase lattice Boltzmann model for immiscible two-phase flows and its application to liquid-jet breakup.
	In Sec.~\ref{sec:methodology}, we formulate the three-dimensional two-phase lattice Boltzmann model for immiscible two-phase flows in the framework of a three-dimensional 27-velocity (D3Q27) lattice. 
	The collision operator consists of D3Q27 versions of three sub-operators: an MRT-collision operator, a generalized-perturbation operator~\citep{Liu2012}, and a formulatic-recoloring operator~\citep{Latva-Kokko2005}.
	A D3Q27 version of the enhanced equilibrium distribution functions~\citep{Leclaire2013} is also incorporated into this model to improve its Galilean invariance.
	In Sec.~\ref{sec:validation}, numerical tests, including those of a static droplet, an oscillating droplet, and the Rayleigh--Taylor instability, are used to validate the developed model. 
	In Sec.~\ref{sec:jet}, this model is applied to liquid-jet-breakup simulations.
	A simulation in which the parameters are exactly matched to the target experiment is performed and compared with experimental data. 
	Finally, we assess the reproducibility of the breakup regimes expected by the dimensionless-regime map~\citep{Saito2017}.
	Sec.~\ref{sec:conclusions} concludes this paper.

\section{\label{sec:methodology}Methodology}

	The present model is formulated on a D3Q27 lattice.
	The key to the formulation is the combination of previous work and their extension to the D3Q27 framework. The main points of this process can be briefly summarized as follows:
\begin{itemize}
 \item Introducing a D3Q27 MRT collision operator~\citep{Geier2015} and relaxation parameters~\citep{Suga2015}
 \item Extending an enhanced equilibrium distribution function~\citep{Leclaire2013} to the D3Q27 lattice
 \item Extending a generalized perturbation operator~\citep{Liu2012} to the D3Q27 lattice
\end{itemize}

\begin{figure}[tb]
	\includegraphics[width=7cm]{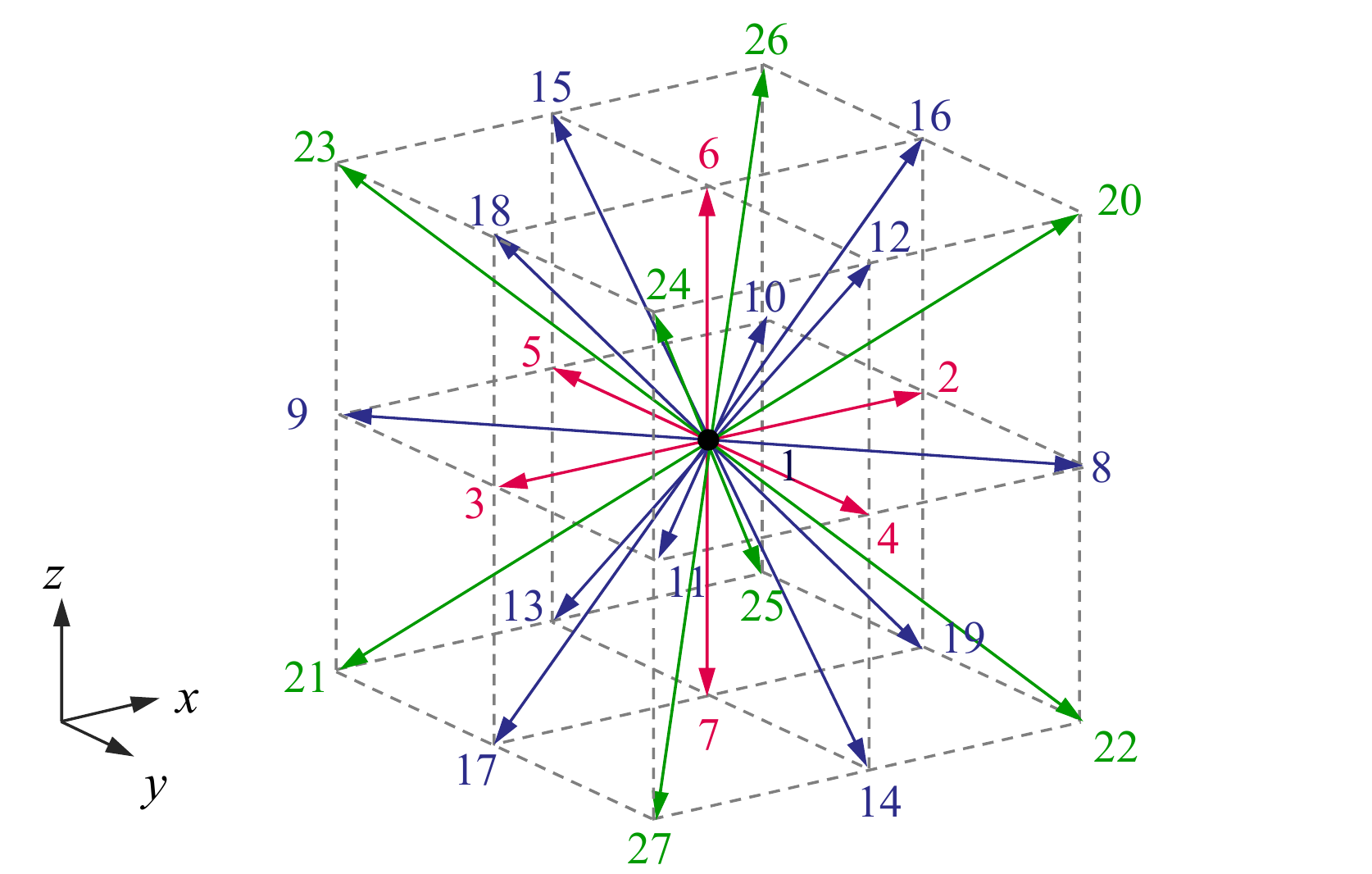}
	\caption{Three-dimensional 27-velocity (D3Q27) lattice. \label{D3Q27}}
\end{figure}

	For the present three-dimensional lattice Boltzmann model, distribution functions move on a D3Q27 lattice with the lattice velocity ${\bf c}_i$ defined as follows:
\begin{widetext}
\begin{equation}
	{\bf c}_i = (c_{ix},c_{iy},c_{iz}) =
	\begin{cases}
		~(0,0,0)c, & i=1 ,\\
		~(\pm1,0,0)c,~(0,\pm1,0)c,~(0,0,\pm1)c, & i = 2,3,\dots,7, \\
		~(\pm1,\pm1,0)c,~(0,\pm1,\pm1)c,~(\pm1,0,\pm1)c, & i = 8,9,\dots,19, \\
		~(\pm1,\pm1,\pm1)c, & i = 20,21,\dots,27, \label{eq:velSet}
	\end{cases}
\end{equation}
\end{widetext}
where $c=\delta_x/\delta_t$, $\delta_x$ is the lattice spacing and $\delta_t$ is the time step. 
	The schematic structure of the D3Q27 lattice is shown in Fig.~\ref{D3Q27}.
	The D3Q27 model is a straightforward extension of the D2Q9 model~\citep{He1997}.

	In this model, two immiscible fluids are represented as pseudo red and blue fluids, respectively. 
	The distribution function, $f_i^k$, is introduced to represent the fluid $k$, 
where $k=r$ and $b$ denote the colors ``red'' and ``blue,'' respectively,
and $i$ is the lattice-velocity direction.
	The total distribution function is defined as $f_i = f_i^r+f_i^b$.
	The evolution of the distribution function is expressed by the following lattice Boltzmann equation:
\begin{equation}
	f_i^k({\bf x}+{\bf c}_i\delta_t,t+\delta_t) - f_i^k({\bf x},t) = \Omega_i^k({\bf x},t), \label{LBE}
\end{equation}
where ${\bf x}$ and $t$ are the position and time, respectively.
	The collision operator $\Omega_i^k$ is made up of three sub-operators~\cite{Tolke2002}:
\begin{equation}
	\Omega_i^k = \left(\Omega_i^k \right)^{(3)} \left[\left(\Omega_i^k \right)^{(1)} + \left(\Omega_i^k \right)^{(2)} \right],
\end{equation}
where $(\Omega_i^k)^{(1)}$ is the single-phase collision operator,
$(\Omega_i^k)^{(2)}$ is the perturbation operator,
and $(\Omega_i^k)^{(3)}$ is the recoloring operator.
Using the MRT operator, the single-phase collision operator can be written as
\begin{equation}
	\left(\Ket{\Omega^k}\right)^{(1)} = -{\bf M}^{-1}{\bf K}{\bf M}
	\left(\Ket{f^k}-\Ket{f^{k(e)}} \right) + \Ket{F}. \label{scoll}
\end{equation}
The density of the fluid $k$ is given by
\begin{equation}
	\rho_k = \sum_i f_i^k.
\end{equation}
The total fluid density is given by $\rho = \sum_k \rho_k$, and the total momentum is defined as
\begin{equation}
	\rho{\bf u} = \sum_i \sum_k f_i^k {\bf c}_i + \frac{1}{2} {\bf F}\delta_t, \label{Momentum}
\end{equation}
where ${\bf F}$ is the body force.
	Note that Eq.~(\ref{Momentum}) indicates that the local velocity is modified to incorporate the spatially varying body force~\citep{Guo2002}.
	In Eq.~(\ref{LBE}), ${\bf M}$ and ${\bf K}$ are, respectively, the $27\times27$ transformation and relaxation matrices.
	In this paper, Dirac's bra-ket notation is employed.
	Here, the ``bra'' operator $\bra{f}$ denotes a row vector along each lattice-velocity direction, i.e., $(f_1,f_2,\dots,f_{27})$,
	and the ``ket'' operator $\ket{f}$ denotes a column vector, i.e., $(f_1,f_2,\dots,f_{27})^{\rm T}$, where the superscript ``${\rm T}$'' is the transpose operator. 

	The MRT collision operator in three dimensions is usually implemented with the D3Q15 or D3Q19 lattices.
	As an early attempt for the D3Q27 MRT collision operator, \citet{Dubois2011} and \citet{Premnath2011} presented the formulation independently in 2011.
	\citet{Dubois2011} arranged the D3Q27 orthogonal basis vectors for the moments according to their character (scalars, vectors, and tensors, etc.) and then used {\it raw moments} to formulate an MRT lattice Boltzmann model. 
	On the other hand, \citet{Premnath2011} arranged the D3Q27 orthogonal basis vectors based on the increasing order of moments and used {\it central moments} to formulate an MRT lattice Boltzmann model.
	Later, \citet{Geier2015} provided the following orthogonal moment set for the D3Q27 lattice:
\begin{widetext}
\begin{equation}
{\bf M} = 
\left [
	\begin{array}{c}
		\bra{1} \\	
		\bra{c_{ix}} \\	
		\bra{c_{iy}} \\	
		\bra{c_{iz}} \\	
		-2\bra{1} + \bra{|{\bf c}_i|^2} \\	
		2\bra{c_{ix}^2} - \bra{c_{iy}^2+c_{iz}^2}\\ 
		\bra{c_{iy}^2-c_{iz}^2} \\ 
		\bra{c_{ix}c_{iy}} \\ 
		\bra{c_{iy}c_{iz}} \\ 
		\bra{c_{iz}c_{ix}} \\ 
		-4\bra{c_{ix}} + 3\bra{c_{ix}c_{iy}^2+c_{iz}^2c_{ix}} \\ 
		-4\bra{c_{iy}} + 3\bra{c_{iy}c_{iz}^2+c_{ix}^2c_{iy}} \\ 
		-4\bra{c_{iz}} + 3\bra{c_{iz}c_{ix}^2+c_{iy}^2c_{iz}} \\ 
		4\bra{c_{ix}} - 6\bra{c_{ix}c_{iy}^2+c_{iz}^2c_{ix}} + 9\bra{c_{ix}c_{iy}^2c_{iz}^2} \\ 
		4\bra{c_{iy}} - 6\bra{c_{iy}c_{iz}^2+c_{ix}^2c_{iy}} + 9\bra{c_{ix}^2c_{iy}c_{iz}^2} \\ 
		4\bra{c_{iz}} - 6\bra{c_{iz}c_{ix}^2+c_{iy}^2c_{iz}} + 9\bra{c_{ix}^2c_{iy}^2c_{iz}} \\ 
		4\bra{1-|{\bf c}_i|^2} +3\bra{c_{ix}^2c_{iy}^2+c_{iy}^2c_{iz}^2+c_{iz}^2c_{ix}^2} \\ 
		-8\bra{1} + 12\bra{|{\bf c}_i|^2}   - 18\bra{c_{ix}^2c_{iy}^2+c_{iy}^2c_{iz}^2+c_{iz}^2c_{ix}^2} + 27\bra{c_{ix}^2c_{iy}^2c_{iz}^2}\\ 
		2\bra{c_{iz}^2+c_{iy}^2} + 3\bra{c_{ix}^2c_{iy}^2+c_{iz}^2c_{ix}^2} - 4\bra{c_{ix}^2} - 6\bra{c_{iy}^2c_{iz}^2} \\ 
		2\bra{c_{iz}^2-c_{iy}^2} + 3\bra{c_{ix}^2c_{iy}^2-c_{iz}^2c_{ix}^2} \\ 
		-2\bra{c_{ix}c_{iy}} + 3\bra{c_{ix}c_{iy}c_{iz}^2} \\ 
		-2\bra{c_{iy}c_{iz}} + 3\bra{c_{ix}^2c_{iy}c_{iz}} \\ 
		-2\bra{c_{iz}c_{ix}} + 3\bra{c_{ix}c_{iy}^2c_{iz}} \\ 
		\bra{c_{ix}c_{iy}^2-c_{iz}^2c_{ix}} \\ 
		\bra{c_{iy}c_{iz}^2-c_{ix}^2c_{iy}} \\ 
		\bra{c_{iz}c_{ix}^2-c_{iy}^2c_{iz}} \\ 
		\bra{c_{ix}c_{iy}c_{iz}} 
		\label{eq:momentSet}
	\end{array}
	\right] 
	,
\end{equation}
\end{widetext}
where $|{\bf c}_i| = \sqrt{c_{ix}^2+c_{iy}^2+c_{iz}^2}$. 
	The transformation matrix transfers the distribution functions from velocity space to moment space.
	Using the MRT collision operator instead of a traditional single-relaxation-time (or BGK) collision operator contributes to enhancement of numerical stability and accuracy, even with additional computational costs. 
	The practical forms of the transformation matrix and its inverse are given in Appendix~\ref{sec:appA}.
The relaxation matrix ${\bf K}$ is the diagonal matrix given by~\citep{Suga2015}
\begin{widetext}
	\begin{equation}
		\begin{split}
			{\bf K} &= {\rm diag} \left[s_1, s_2, \dots, s_{27} \right] \\
			&= {\rm diag} 
			\left[
			s_1,s_1,s_1,s_1, s_5, s_6,s_6, s_8,s_8,s_8, 
			s_{11},s_{11},s_{11}, s_{14},s_{14},s_{14}, s_{17},s_{18},
			s_{19},s_{19}, s_{21},s_{21},s_{21}, s_{24},s_{24},s_{24}, s_{27}
			\right],
		\end{split}
	\end{equation}
\end{widetext}
where the elements $0<s_i<2$ represent both the hydrodynamic and non-hydrodynamic relaxation parameters.
	The hydrodynamic parameters satisfy the following relations:
\begin{align}
	\nu &= \frac{c^2}{3} \left(\frac{1}{s_6} - \frac{1}{2} \right) \delta_t
			 = \frac{c^2}{3} \left(\frac{1}{s_8} - \frac{1}{2} \right) \delta_t, \label{eq:shear}
	\\
	\zeta &= \frac{5-3c^2}{9} \left(\frac{1}{s_5} - \frac{1}{2} \right) \delta_t, \label{eq:bulk}
\end{align}
where $\nu$ and $\zeta$ are the kinematic viscosity and the bulk viscosity, respectively.
	In this paper, we use the optimized parameters proposed by \citet{Suga2015}: 
$s_1 = 0$, $s_5 = 1.5$, $s_{11} = 1.5$, $s_{14} = 1.83$,  $s_{17} = 1.4$,  $s_{18} = 1.61$,  $s_{19} = s_{21} =  1.98$, and $s_{24} = s_{27} = 1.74$.
	We confirmed that these parameters significantly enhanced the numerical stability even at extremely low kinematic viscosity with the order of $O(10^{-4})$.

	For the single-phase collision operator, an enhanced equilibrium distribution function proposed by \citet{Leclaire2013} is used in this paper:
\begin{align}
	&f_i^{k(e)}(\rho_k,{\bf u},\alpha_k) \nonumber
	\\
	&= \rho_k \left(\varphi_i^k 
	+ w_i \left[\frac{3}{c^2} ({\bf c}_i \cdot {\bf u})
		+ \frac{9}{2c^4}({\bf c}_i \cdot {\bf u})^2
		- \frac{3}{2c^2}{\bf u}^2 \right] \right) + \Phi_i^k. 
		\label{eq:EES}
\end{align}
	In the case of $\Phi_i^k=0$, Eq.~(\ref{eq:EES}) recovers the common form of an equilibrium distribution function.
	Using the form of Eq.~(\ref{eq:EES}), the Galilean invariance is improved for variable density and viscosity ratios under the hypothesis of a small pressure gradient \citep{Leclaire2013,Leclaire2014,Leclaire2015}.
	However, it should be noted that this can only partly restore Galilean invariance, as still the third order diagonal equilibrium moments are not independently supported, and are related to the corresponding first moments. 
	Consequently, there will be cubic velocity errors in Galilean invariance even for the D3Q27 lattice, which could become perceptible for flows under high shear. 
	This issue can be solved by making some additional corrections to the collision operator, as suggested recently by \citet{Dellar2014}.

	The weights, $w_i$, are those of a standard D3Q27 lattice~\cite{He1997}:
\begin{equation}
	w_i = 
	\begin{cases}
		~8/27, & i=1 ,\\
		~2/27, & i = 2,3,\dots,7, \\
		~1/54, & i = 8,9,\dots,19, \\
		~1/216, & i = 20,21,\dots,27.
\end{cases}
\end{equation}
	Moreover, in the D3Q27 lattice, one can derive
\begin{equation}
	\varphi_i^k = 
	\begin{cases}
		~\alpha_k, & i=1 ,\\
		~2(1-\alpha_k)/19, & i = 2,3,\dots,7, \\
		~(1-\alpha_k)/38, & i = 8,9,\dots,19, \\
		~(1-\alpha_k)/152, & i = 20,21,\dots,27,
		\label{eq:varphi_ik}
\end{cases}
\end{equation}
and 
\begin{equation}
	\Phi_i^k = 
	\begin{cases}
		~-3 \bar{\nu}({\bf u}\cdot \nabla \rho_k)/c, & i=1 ,\\
		~+16\bar{\nu}({\bf G}_k : {\bf c}_i \otimes {\bf c}_i)/c^3, & i = 2,3,\dots,7, \\
		~+4 \bar{\nu}({\bf G}_k : {\bf c}_i \otimes {\bf c}_i)/c^3, & i = 8,9,\dots,19, \\
		~+1 \bar{\nu}({\bf G}_k : {\bf c}_i \otimes {\bf c}_i)/c^3, & i = 20,21,\dots,27, \label{eq:Phi}
\end{cases}
\end{equation}
where $\otimes$ is the tensor product and the symbol ``$:$'' indicates tensor contraction; 
$\bar{\nu}$ is the kinematic viscosity interpolated by~\citep{Liu2012,Tolke2002}
\begin{equation}
	\bar{\nu} = \frac{1+\phi}{2}\nu_r + \frac{1-\phi}{2}\nu_b.
\end{equation}
Here, $\phi$ is the order parameter to distinguish the two components in a multicomponent flow, defined as~\citep{Liu2012}
\begin{equation}
	\phi({\bf x},t) = \frac{\rho_r({\bf x},t)-\rho_b({\bf x},t)}{\rho_r({\bf x},t)+\rho_b({\bf x},t)}.
	\label{eq:order}
\end{equation}
	The values of the order parameter $\phi=1,-1$, and $0$ correspond to a purely red fluid, a purely blue fluid, and the interface, respectively~\citep{Tolke2002}.
	In the framework of D3Q27 lattice, the tensor ${\bf G}_k$ in Eq.~(\ref{eq:Phi}) is defined as
\begin{equation}
	{\bf G}_k = \frac{1}{48} \left[{\bf u}\otimes \nabla \rho_k 
	+ ({\bf u}\otimes \nabla \rho_k)^{{\rm T}} \right].
\end{equation}
	As established in~\citep{Grunau1993}, the density ratio between the fluids, $\gamma$, must be taken into account as follows to obtain a stable interface:
\begin{equation}
	\gamma = \frac{\rho_r^0}{\rho_b^0} = \frac{1-\alpha_b}{1-\alpha_r},
\end{equation}
where the superscript ``0'' indicates the initial value of the density at the beginning of the simulation~\citep{Leclaire2013}.
In each homogeneous phase region, the pressure of the fluid $k$ is given by
\begin{equation}
	p_k = \rho_k \left(c_s^k \right)^2 = \rho_k \frac{9(1-\alpha_k)}{19}c^2,
	\label{eq:press}
\end{equation}
for the D3Q27 lattice. 
	This corresponds to an isothermal equation of state. 
	In this paper, we choose $\alpha_b = 8/27$, in which $c_s^b = 1/\sqrt{3}$~\citep{Leclaire2014, Saito2016}.

	The term $\Ket{F}$ in Eq.~(\ref{scoll}) represents the discrete forcing term accounting for the body force ${\bf F}$. 
	In the MRT framework, the forcing term reads as~\citep{Yu2010}
\begin{equation}
	\Ket{F} = {\bf M}^{-1} \left({\bf I} - \frac{1}{2} {\bf K} \right) {\bf M} \Ket{F'},
	\label{eq:ketF}
\end{equation}
where ${\bf I}$ is a unit matrix, $\ket{F} = (F_1,F_2,\dots,F_{27})^{{\rm T}}$, and $\ket{F'} = (F'_1,F'_2,\dots,F'_{27})^{{\rm T}}$ is given by
\begin{equation}
	\Ket{F'} = w_i \left[3\frac{{\bf c}_i - {\bf u}}{c^2} 
	+ 9\frac{({\bf c}_i \cdot {\bf u}) {\bf c}_i}{c^4} \right] \cdot {\bf F}\delta_t.
	\label{eq:ketFd}
\end{equation}
	Eqs.~(\ref{eq:ketF}) and (\ref{eq:ketFd}) reduce to Guo {\it et al}.'s original forcing scheme~\citep{Guo2002} when using a single-relaxation time~\cite{Yu2010}

	To model the interfacial tension, \citet{Liu2012} derived a generalized perturbation operator based on the CSF~\citep{Brackbill1992} concept, and the work of~\citet{Reis2007} is employed to obtain the interfacial tension:
\begin{equation}
	\left(\Omega_i^k \right)^{(2)} = \frac{A_k}{2} |\nabla \phi| 
	\left[w_i \frac{({\bf c}_i \cdot \nabla \phi)}{|\nabla \phi|^2} - B_i \right].
	\label{eq:perturb}
\end{equation}
	Eq.~(\ref{eq:perturb}) satisfies the correct form of the interfacial-tension force in the Navier--Stokes equations when the lattice-specific variables $B_i$ are chosen correctly.
	We derived the values of $B_i$ in the framework of the D3Q27 lattice as follows:
\begin{equation}
	B_i = 
	\begin{cases}
		~-(10/27)c^2, & i=1 ,\\
		~+(2/27)c^2, & i = 2,3,\dots,7, \\
		~+(1/54)c^2, & i = 8,9,\dots,19, \\
		~+(1/216)c^2, & i = 20,21,\dots,27. \label{eq:B_i}
\end{cases}
\end{equation} 
	In this model, the interfacial tension can be directly given by
\begin{equation}
	\sigma = \frac{4}{9} A \tau c^4 \delta_t, \label{eq:sigma}
\end{equation}
where we assumed that $A=A_r=A_b$, $\tau$ is the relaxation time.
Parameter $A$ controls the strength of interfacial tension, $\sigma$.

	Although the perturbation operator, $(\Omega_i^k)^{(2)}$, generates interfacial tension, it does not guarantee the immiscibility of both fluids. To promote phase segregation and maintain the interface, the following recoloring operator is applied~\cite{Latva-Kokko2005, Halliday2007, Leclaire2012}
\begin{eqnarray}
	(\Omega_i^r)^{(3)} = \frac{\rho_r}{\rho}f_i 
	+ \beta \frac{\rho_r \rho_b}{\rho^2} \cos(\theta_i)
	f_i^{(e)}(\rho,\bf{0},\bar{\alpha}),
\\
	(\Omega_i^b)^{(3)} = \frac{\rho_b}{\rho}f_i 
	- \beta \frac{\rho_r \rho_b}{\rho^2} \cos(\theta_i)
	f_i^{(e)}(\rho,\bf{0},\bar{\alpha}),
\end{eqnarray}
where
\begin{equation}
	\cos(\theta_i) = \frac{{\bf c}_i \cdot \nabla \phi}{|{\bf c}_i|  |\nabla \phi|},
\end{equation}
and $f_i^{(e)}$ is evaluated using Eq.~(\ref{eq:EES}), a zero velocity, and
\begin{equation}
	\bar{\alpha} = \frac{1}{2} (1 + \phi) \alpha_r + \frac{1}{2} (1 - \phi) \alpha_b.
\end{equation}

In the present model, the following continuity and Navier--Stokes equations can be derived via Chapman--Enskog analysis~\citep{Liu2012,Guo2002,Kruger2017}
\begin{align}
	\frac{\partial \rho}{\partial t} &+ \nabla \cdot (\rho {\bf u}) = 0, \label{Continuity}
	\\
	\frac{\partial (\rho {\bf u})}{\partial t} + \nabla \cdot (\rho {\bf u} {\bf u})
	&= -\nabla p + \nabla \cdot {\bf \Pi} + \nabla \cdot {\bf S} + {\bf F}, \label{N-S}
\end{align}
where 
\begin{equation}
	{\bf \Pi} = \rho \nu [\nabla {\bf u} + (\nabla {\bf u})^{{\rm T}}] + \rho (\zeta - \nu)(\nabla \cdot {\bf u}){\bf I},
\end{equation}
is the viscous stress tensor 
with the shear viscosity $\nu$ given by Eq.~(\ref{eq:shear}) and the bulk viscosity $\zeta$ given by Eq.~(\ref{eq:bulk});
$p=p_r+p_b$ is the pressure.
	In Eq.~(\ref{N-S}), the term $\nabla \cdot {\bf S}$ arises from the perturbation operator given by Eq.~(\ref{eq:perturb}) and is equivalent to the interfacial force based on the CSF concept~\citep{Liu2012};
the capillary stress tensor, ${\bf S}$, is given by
\begin{equation}
	{\bf S} = -\tau \delta_t \sum_i \sum_k \left(\Omega_i^k \right)^{(2)} {\bf c}_i {\bf c}_i.
\end{equation}

	For the computation of the gradient operator for an arbitrary function $\chi$, 
the following second-order isotropic central scheme~\citep{Liu2012,Guo2011,Liang2014,Lou2012}
\begin{equation}
	\nabla \chi({\bf x},t) = \frac{3}{c^2} \sum_i \frac{w_i \chi({\bf x}+{\bf c}_i \delta_t,t) {\bf c}_i}{\delta_t},
\end{equation}
is adopted.

	All the simulations in this paper are carried out in lattice units.
	In this paper, $\delta_x$ and $\delta_t$ are set to 1 as in the usual lattice Boltzmann simulations.
	The aforementioned formulation focuses on the {\it two}-component systems.
	It should be straightforward to implement the present model in three or more component systems according to the work of \citet{Leclaire2013b}.

	One can consider the key to the developed lattice Boltzmann model in this paper to be based on a combination of previous works and their extension to the framework of the D3Q27 lattice.
	A brief procedure to determine the related lattice-specific coefficients [Eqs.(\ref{eq:varphi_ik}), (\ref{eq:Phi}), and (\ref{eq:B_i})] is described in Appendix~\ref{sec:appB}.

\section{\label{sec:validation}Numerical tests}

\subsection{\label{sec:staticDrop}Static droplet}

	Static-droplet tests are performed to test the validity of interfacial tension predicted by Eq.~(\ref{eq:sigma}).
	The computational domain is discretized into an $85\times85\times85$ lattice.
	A steady red droplet with radius $R$ is immersed in a blue fluid.
	The density field of each phase is initialized as follows:
\begin{align}
	\rho_r(x,y,z) = \frac{\rho_r^0}{2} \left[1 -  \tanh 
	\left( \frac{2(r-R)}{W} \right) \right],
\\
	\rho_b(x,y,z) = \frac{\rho_b^0}{2} \left[1 +  \tanh 
	\left( \frac{2(r-R)}{W} \right) \right],
\end{align}
where $W=4$ and $r=\sqrt{(x-x_c)^2+(y-y_c)^2+(z-z_c)^2}$ with the central position of the computational domain $(x_c,y_c,z_c)$.
	We set the density and kinematic-viscosity ratios as 1.5 and 1, respectively.
	The kinematic viscosity for each phase is set to be $1.0\times10^{-3}$, and gravity is neglected throughout the simulations.
	The parameters are $A$ in Eq.~(\ref{eq:sigma}) and initial droplet radius $R$ (see Table \ref{tab:laplace}). 
	Periodic boundary conditions are imposed on all sides of the computational domain.

\begin{figure}[htbp]
	\includegraphics[width=8.5cm]{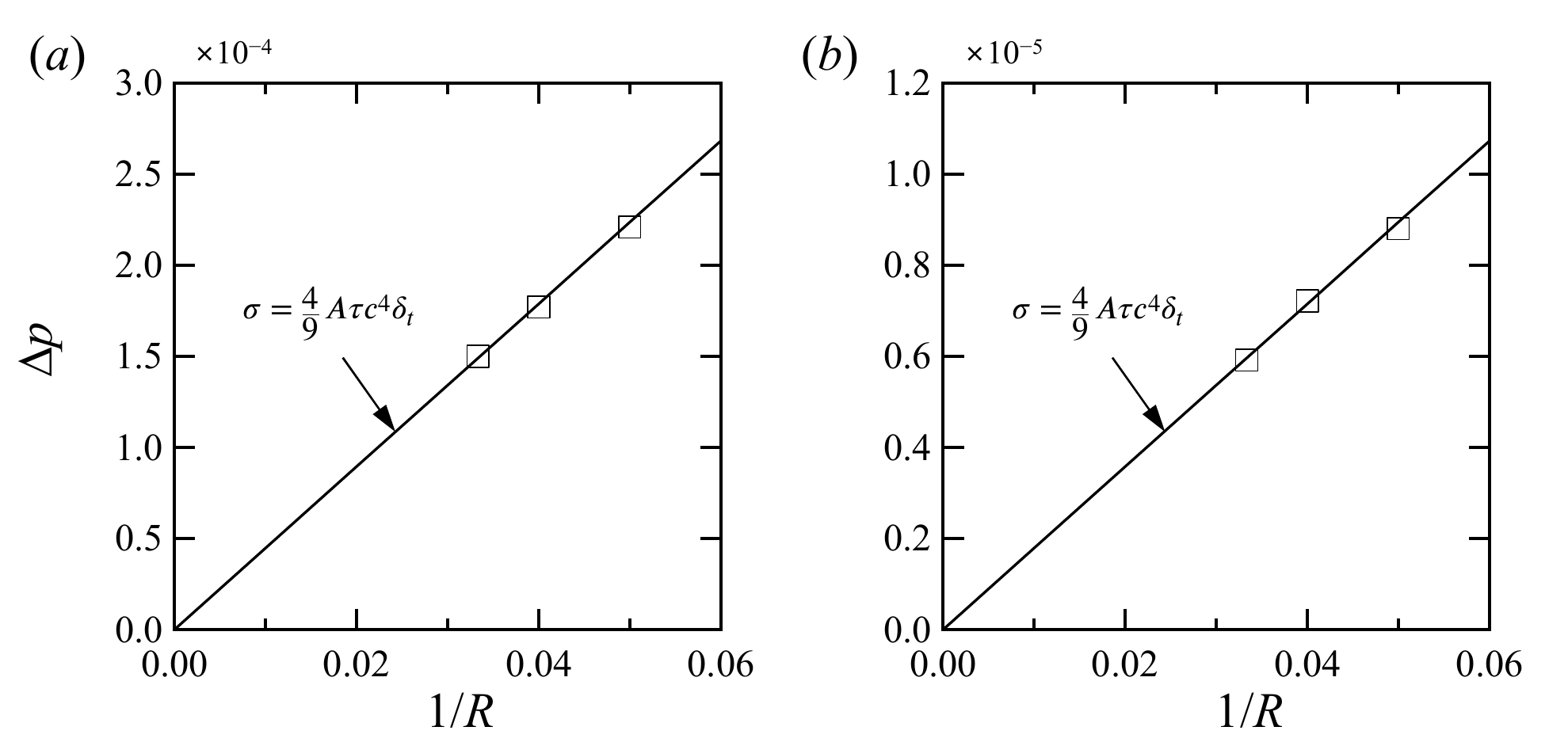}
	\caption{Laplace's law for a static droplet: (a) $A=1.0\times10^{-2}$ and (b) $A=4.0\times10^{-4}$. The solid line is the theoretical prediction given by Eq.~(\ref{eq:sigma}). 
	The present simulations show a good agreement with the theoretical prediction.
	\label{fig:Laplace}}
\end{figure}

The Laplace equation in three dimensions is given by 
\begin{equation}
	\Delta p = \frac{2\sigma}{R},
	\label{eq:lap}
\end{equation}
where $\Delta p$ is the pressure difference across the droplet interface.
The pressure for each phase is evaluated by Eq.~(\ref{eq:press}) and is measured after 80,000 iterations using the same procedure as~\citet{Leclaire2012}.
	Figure~\ref{fig:Laplace} shows the measured pressure differences.
	For both the higher [Fig.~\ref{fig:Laplace}(a)] and lower [Fig.~\ref{fig:Laplace}(b)] interfacial tensions, the results are in proportion to the droplet curvature $1/R$,
with which the Laplace law was satisfied.
	The theoretical prediction by Eq.~(\ref{eq:sigma}), shown in Fig.~\ref{fig:Laplace} as a solid line, also agrees well with the measured pressure differences.
	Table~\ref{tab:laplace} summarizes the simulation parameters and the evaluated error.
	The error $E$ is calculated as~\citep{Liu2012}
\begin{equation}
	E = \frac{|\sigma_{th} - \sigma_{Lap}|}{\sigma_{th}},
\end{equation}
	where $\sigma_{th}$ and $\sigma_{Lap}$ are the interfaceial tension predicted by Eq.~(\ref{eq:sigma}) and that measured by the Laplace equation Eq.~(\ref{eq:lap}), respectively.
	We confirm that the lattice Boltzmann model developed in Sec.~\ref{sec:methodology} can predict the interfacial tension for a static case within a maximum error of 1.6\%.
	
\begin{figure*}[htbp]
	\includegraphics[width=16cm]{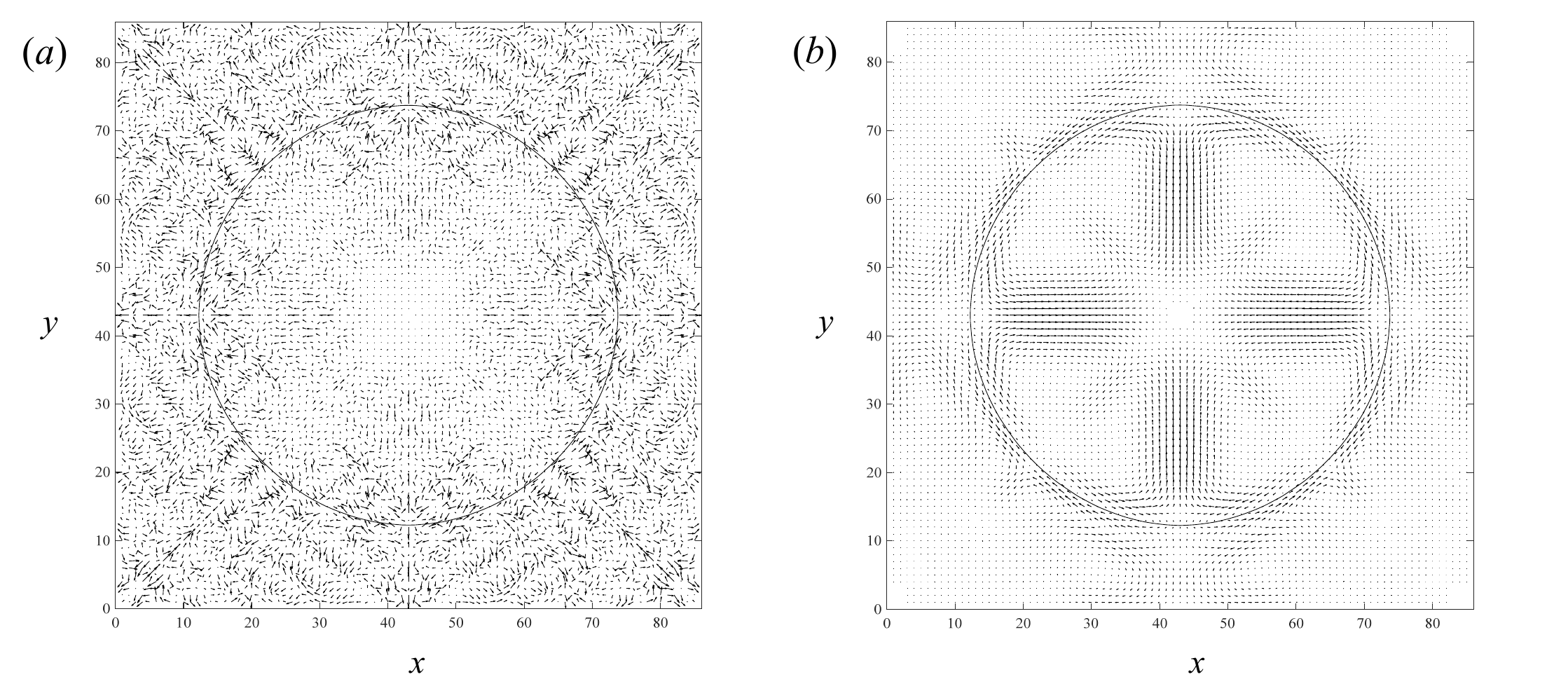}
	\caption{Influence of lattice isotropy on the spurious velocities: (a) D3Q19~\citep{Saito2016} and (b) D3Q27.
	The parameters are $R=30$, $A=1.0\times 10^{-2}$, $\rho_r/\rho_b = 1.5$, and $\nu_r/\nu_b=1.0$ ($\nu_r = \nu_b = 1.0\times10^{-3}$).
	The velocity vectors are magnified 200 times.
	The maximum spurious velocities $|{\bf u}|_{max}$ are $1.2\times 10^{-2}$ for D3Q19 and $5.8\times 10^{-3}$ for D3Q27.
	Enhancing the lattice isotropy from D3Q19 to D3Q27 contributes to reducing the spurious velocity.
	\label{fig:spurious}}
\end{figure*}

	We should mention the influence of lattice isotropy on the so-called spurious velocity.
	Fig.~\ref{fig:spurious} compares the droplet shape and velocity field at the equilibrium state for D3Q19 [Fig.~\ref{fig:spurious}(a)] and D3Q27 lattices [Fig.~\ref{fig:spurious}(b)].
	The numerical test using D3Q19 is based on \citep{Saito2016}.
	The maximum spurious velocities $|{\bf u}|_{max}$ are $1.2\times 10^{-2}$ for D3Q19 and $5.8\times 10^{-3}$ for D3Q27, respectively.
	Although the conditions are same except for the employed lattice geometry, the simulation using the D3Q27 lattice shows better result.
	Enhancing the lattice isotropy (from D3Q19 to D3Q27) contributes to reducing the spurious velocity.

\begin{table}[tb]
\caption{\label{tab:laplace} Parameters and evaluated errors of static droplet tests.}
\begin{ruledtabular}
\begin{tabular}{ccc}
\textrm{$R$}& \textrm{$A$}& \textrm{$E$\%}\\
\colrule
$30$ & $1.0\times10^{-2}$ & 0.77\\
$25$ & $1.0\times10^{-2}$ & 0.92\\
$20$ & $1.0\times10^{-2}$ & 1.6\\
$30$ & $4.0\times10^{-4}$ & 0.63\\
$25$ & $4.0\times10^{-4}$ & 0.77\\
$20$ & $4.0\times10^{-4}$ & 1.1
\end{tabular}
\end{ruledtabular}
\end{table}

\subsection{\label{sec:dynamicDrop}Oscillating droplet}

	Oscillatory-droplet tests are performed to test the model validity in an unsteady case. 
	The simulation setup and parameters for the present numerical tests follow \citet{Premnath2007}, except for the interfacial-tension coefficient.
	A droplet with a prolate-spheroid shape is placed at the center of a computational domain discretized by a $41\times41\times41$ lattice.
	The droplet's minimum and maximum radii are 11 and 15, respectively. 
	We set a density ratio of 4.
	The kinematic viscosity ratio is set to be unity ($\nu = \nu_r = \nu_b$), and  gravity is neglected throughout the simulations.
	The parameters are $\nu$ and $A$ (see Table \ref{tab:table0}).
	Parameter $A$ is the same as the ones provided in Sec.~\ref{sec:staticDrop}.

	The analytical solution of \citet{Miller1968} is used for comparison with the computed time periods.
	The frequency of the $n$-th mode is given by
\begin{equation}
	\omega_n = \omega_n^* - \frac{1}{2}\chi {\omega_n^*}^{\frac{1}{2}} + \frac{1}{4}\chi^2, \label{eq:resoFreq}
\end{equation}
where $\omega_n$ is Lamb's natural resonance frequency~\citep{Lamb1945}
\begin{equation}
	(\omega_n^*)^2 = \frac{n(n+1)(n-1)(n+2)}{R_e^3 [n\rho_2+(n+1)\rho_1]} \sigma. \label{eq:Lamb}
\end{equation}
	Note that $R_e$ in Eq.~(\ref{eq:Lamb}) is the equivalent radius of a droplet. 
	Subscripts 1 and 2 refer to the dispersed and continuous phases, respectively, so they can be replaced by $r$ and $b$ in this paper.
	Parameter $\chi$ is given by
\begin{equation}
	\chi = \frac{(2n+1)^2 (\mu_1 \mu_2 \rho_1 \rho_2)^{\frac{1}{2}}}
	{2^{\frac{1}{2}} R_e [n \rho_2 + (n+1)\rho_1][(\mu_1 \rho_1)^{1/2}+(\mu_2 \rho_2)^{1/2}]}.
\end{equation}
	We are only interested in the second mode $(n=2)$ here.
	The analytical expression for the time period is obtained by $T_{th} = 2 \pi / \omega_2$.

\begin{figure}[htbp]
	\includegraphics[width=8.5cm]{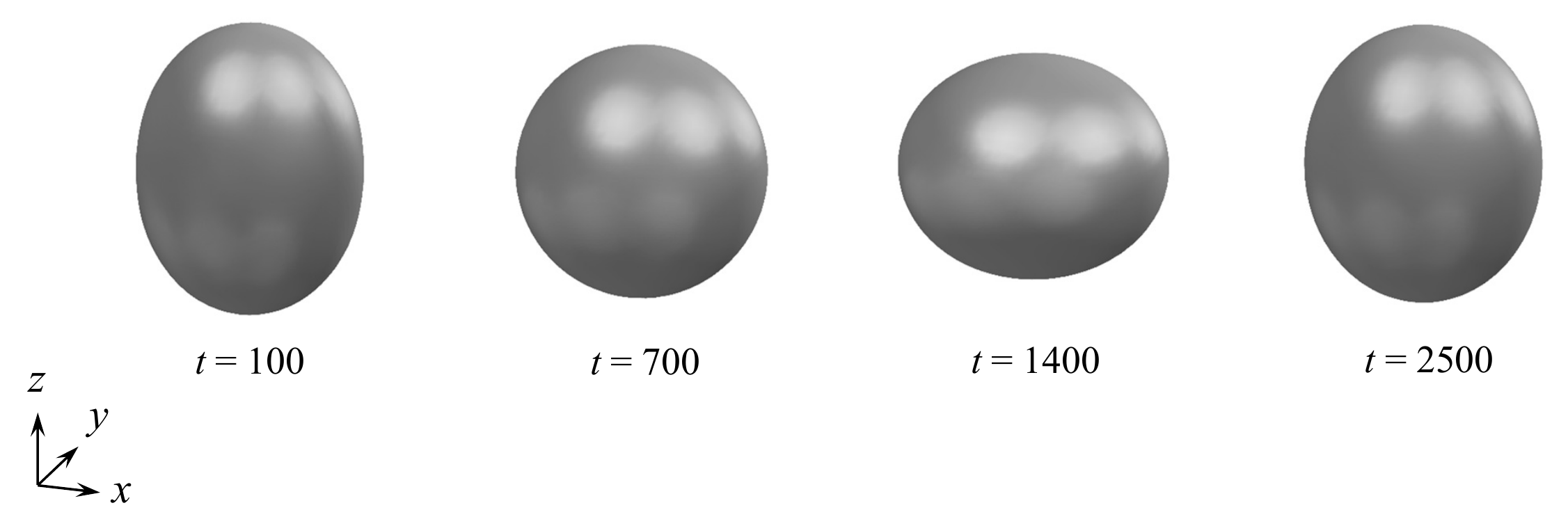}
	\caption{Typical snapshots of shape change of an initially elongated drop. 
	After a spherical shape ($t=700$), the droplet becomes an oblate spheroid ($t=1,400$).
	The configuration turns into a prolate spheroid ($t=2,500$).
	\label{fig:dropOsci}}
\end{figure}

	Figure~\ref{fig:dropOsci} shows the transient change of the oscillating droplet shapes with $\nu=3.333\times10^{-3}$ and $A = 1.0\times10^{-2}$.
	After assuming a spherical shape at $t=700$, the droplet becomes an oblate spheroid at $t=1,400$.
	The configuration turns into a prolate spheroid at $t=2,500$.
	Such a series of oscillations can be also seen in Ref.~\citep{Premnath2007}

\begin{figure}[htbp]
	\includegraphics[width=8.5cm]{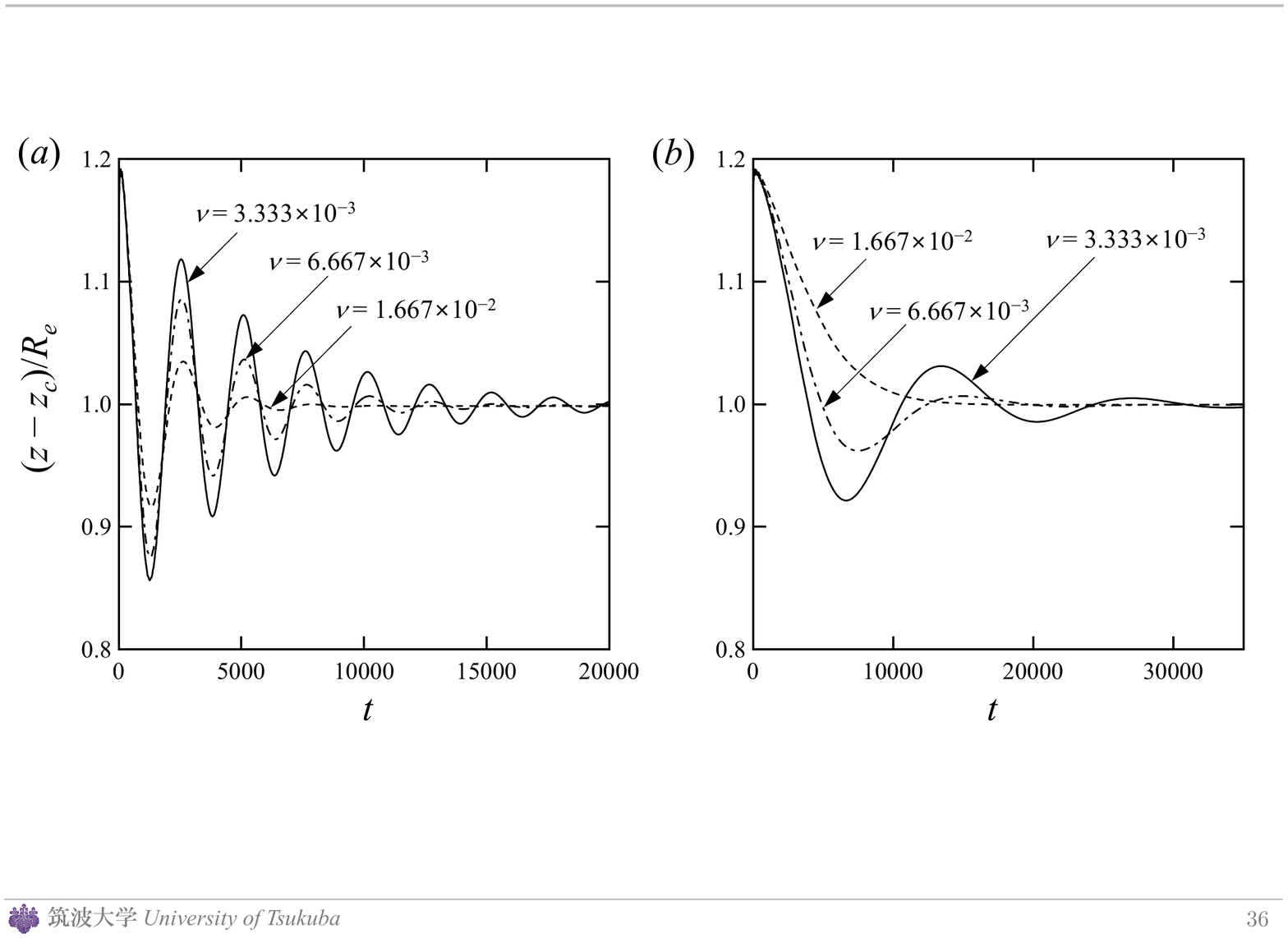}
	\caption{Time history of the interfacial location of an oscillating droplet: (a) $A=1.0\times10^{-2}$ and (b) $A=4.0\times10^{-4}$.  
	(a) Under higher-interfacial tension, all cases attenuate with oscillation regardless of the kinematic viscosity.
	(b) For the lower-interfacial-tension case, the timescale of oscillation becomes longer;
	this tendency agrees with the theory.
	\label{fig:dropOsci2}}
\end{figure}

	The interfacial location is measured, and the results are shown in Fig.~\ref{fig:dropOsci2} as a function of time.
	The interfacial locations are recorded per 10 time steps.
	Under higher-interfacial tension [Fig.~\ref{fig:dropOsci2}(a)], all cases attenuate with oscillation regardless of the kinematic viscosity. 
	The higher the kinematic viscosity, the earlier attenuation occurs.
	For the lower-interfacial-tension case [Fig.~\ref{fig:dropOsci2}(b)], the timescale of oscillation becomes qualitatively longer;
	this tendency agrees with the theoretical expectation by Eqs.~(\ref{eq:resoFreq}) and (\ref{eq:Lamb}).
	Only in the case of $\nu = 1.667\times10^{-2}$ does the interfacial location reach an equilibrium spherical shape without a series of oscillations, as shown in Fig.~\ref{fig:dropOsci}.
	The effect of viscous damping distinguishes rather than the effect of interfacial tension in this case.

	The simulation parameters and evaluated errors in the oscillating period are summarized in Table~\ref{tab:table0}.
The error $E$ is calculated as
\begin{equation}
	E = \frac{|T_{th}-T_{sim}|}{T_{th}},
\end{equation}
where $T_{sim}$ is the measured oscillation period.
	Note that, in the case of $\nu = 1.667\times10^{-2}$ and $A=4.0\times10^{-4}$, we cannot measure the oscillation period since no oscillation occurred.
	The maximum errors are 2.4\% for $A=1.0\times10^{-2}$ and 10.8\% for $A=4.0\times10^{-4}$, respectively.
	Throughout, the accuracy is better when the kinematic viscosity is lower.
	It is found that the accuracy of the low-interfacial-tension case is difficult to assess  via droplet-oscillation tests; however, the tests show that the present lattice Boltzmann model can be applied to unsteady two-phase-flow simulations.

\begin{table}[tb]
\caption{\label{tab:table0} Parameters and evaluated errors of oscillating droplet  tests.}
\begin{ruledtabular}
\begin{tabular}{cccc}
\textrm{$\nu$}& \textrm{$A$}& \textrm{$\sigma$}& \textrm{$E$\%}\\
\colrule
$3.333\times10^{-3}$ & $1.0\times10^{-2}$ & $2.27\times10^{-3}$ & 0.60\\
$6.667\times10^{-3}$ & $1.0\times10^{-2}$ & $2.31\times10^{-3}$ & 0.83\\
$1.667\times10^{-2}$ & $1.0\times10^{-2}$ & $2.44\times10^{-3}$ & 2.4\\
$3.333\times10^{-3}$ & $4.0\times10^{-4}$ & $9.07\times10^{-5}$ & 2.7\\
$6.667\times10^{-3}$ & $4.0\times10^{-4}$ & $9.24\times10^{-5}$ & 10.8\\
$1.667\times10^{-2}$ & $4.0\times10^{-4}$ & $9.78\times10^{-5}$ & --
\end{tabular}
\end{ruledtabular}
\end{table}

\subsection{Rayleigh--Taylor instability}

	To assess the validity of the body force implementation [Eqs.~(\ref{eq:ketF}) and (\ref{eq:ketFd})], the Rayleigh--Taylor instability is selected as the next numerical test.
	The Rayleigh--Taylor instability is a fundamental interfacial instability that is induced when a heavy fluid is placed over a light one subjected to a slightly disturbed interface in a gravitational field~\citep{Chandrasekhar1961}. 
	The Rayleigh--Taylor instability has received considerable attention owing to its extensive applications, e.g., in the fundamental process of melt jet breakup~\citep{Abe2006,Saito2017}.
	To our knowledge, this is the first time a color-fluid model has been applied to the three-dimensional Rayleigh--Taylor instability.

\begin{figure}[tb]
	\includegraphics[width=8cm]{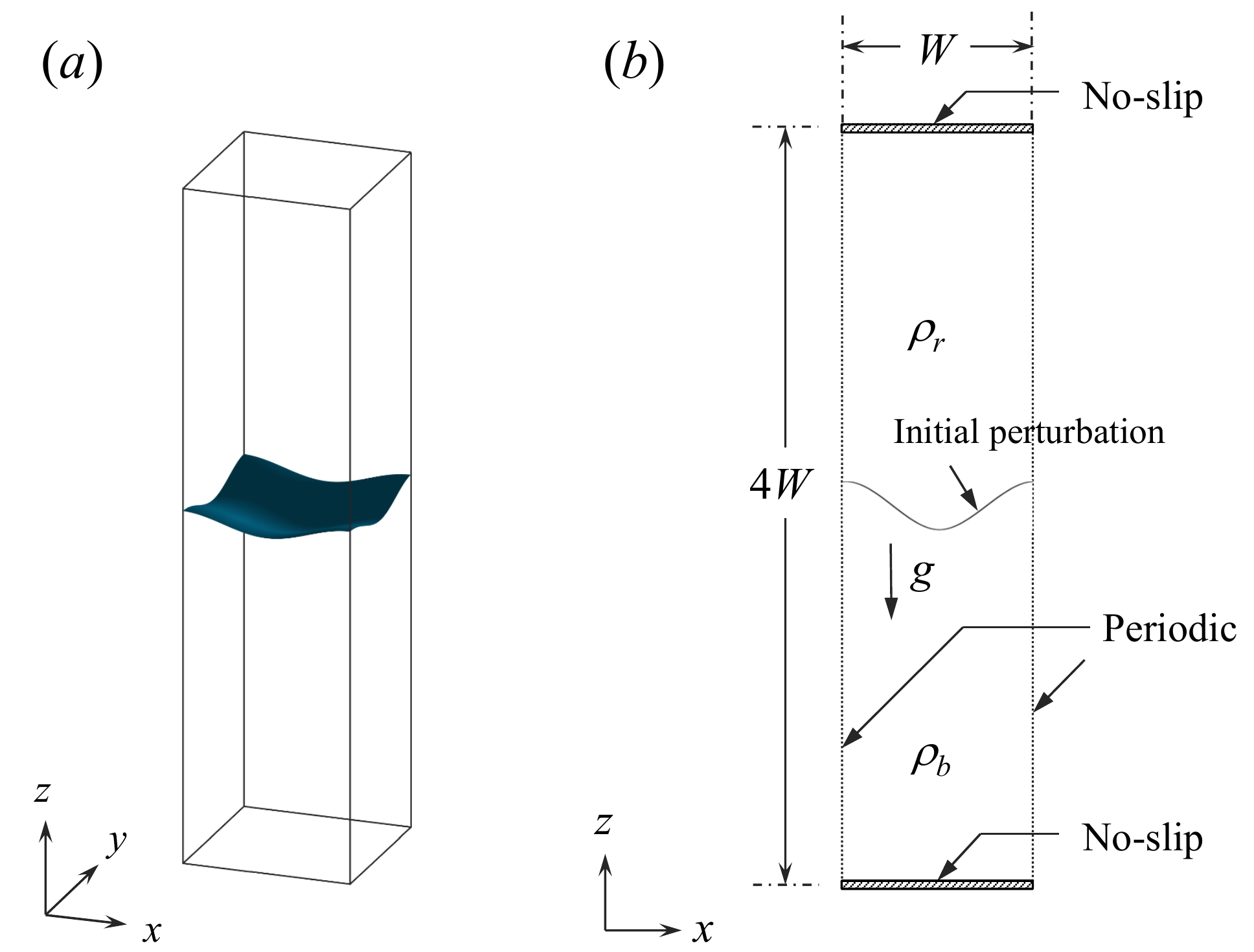}
	\caption{Boundary conditions for the Rayleigh--Taylor instability. (a) The computational domain is discretized into an $128\times128\times512$ lattice and the single-mode initial perturbation is initially imposed. (b) The top and bottom are no-slip boundaries and the lateral boundaries are periodic. \label{fig:setupRT}}
\end{figure}

	We refer to the computational setup adopted in \citet{He1999b}.
A schematic diagram of the computational setup is illustrated in Fig.~\ref{fig:setupRT}.
	The top and bottom boundaries are no-slip walls;
	the lateral boundaries are periodic.
	As in the manner of~\citep{He1999b}, the single-mode initial perturbation is initially imposed as
\begin{equation}
	h(x,y) = 0.05W\left[\cos \left(\frac{2\pi x}{W}\right) + \left(\frac{2\pi y}{W}\right) \right],
\end{equation}
at the mid-plane, where $W$ is the width of the computational domain.
	The body force in Eq.~(\ref{eq:ketFd}) for this problem is incorporated as
\begin{equation}
	{\bf F}({\bf x},t) = -\left(\rho({\bf x},t)-\frac{\rho_r^0+\rho_b^0}{2} \right){\bf g},
\end{equation}
with ${\bf g}=(0,0,-g)$.
	The gravitational acceleration $g$ is chosen to satisfy $(Wg)^{1/2}=0.04$~\citep{He1999}. 
	The computational domain is set to be $W\times W \times 4W$ with $W=128$.
	The domain's size resulted in a $128\times128\times512$ lattice.

	The Atwood number
\begin{equation}
	{\rm At} = \frac{\rho_r^0 - \rho_b^0}{\rho_r^0 + \rho_b^0}, \label{eq:atw}
\end{equation}
is fixed at 0.5 throughout the simulations.
	The interface tension is neglected; thus, the perturbed interface is expected to always be unstable in the inviscid case.
	The kinematic-viscosity ratio is set to unity.
Another dimensionless group is the Reynolds number, defined as
\begin{equation}
	{\rm Re} = \frac{\sqrt{Wg}W}{\nu}.
\end{equation}
We use three patterns of Reynolds numbers: ${\rm Re} = 512$, $1,024$, and $5,120$.

\begin{figure}[htbp]
	\includegraphics[width=6.5cm]{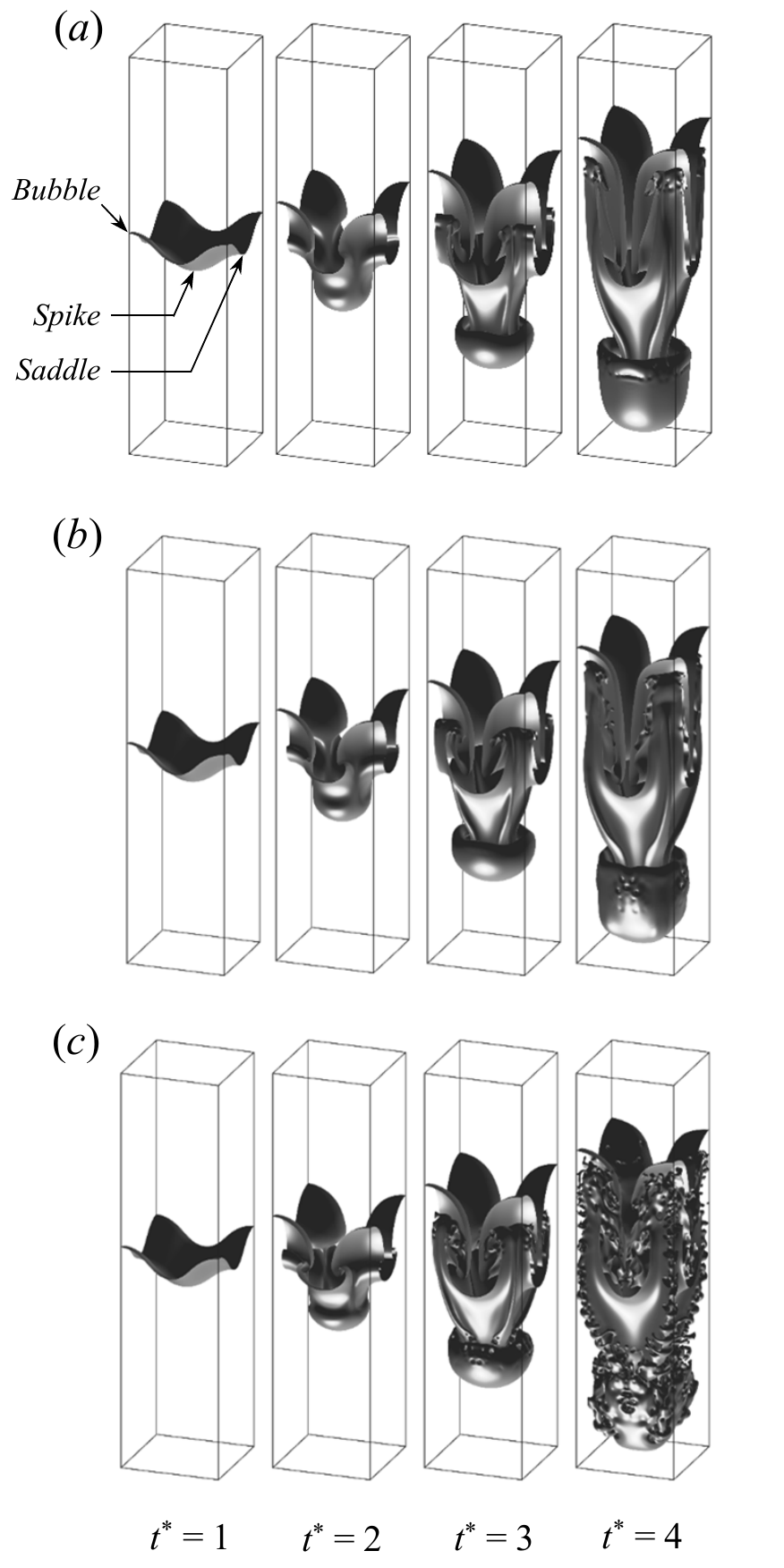}
	\caption{Interface evolution of the single-mode Rayleigh--Taylor instability with (a) ${\rm Re}=512$, (b) ${\rm Re}=1,024$, and (c) ${\rm Re}=5,120$. The Atwood number [Eq.~(\ref{eq:atw})] is fixed at 0.5 in all simulations.
	At later stages, the higher the Reynolds number is, the more unstable the interface becomes. \label{fig:interfaceRT}}
\end{figure}

	Figure~\ref{fig:interfaceRT} shows the interfacial evolution of the Rayleigh--Taylor instability.
	The dimensionless time is given by 
\begin{equation}
	t^* = \frac{t}{\sqrt{W/g}}.
\end{equation}
	In the initial stages by $t^*=2$, the Reynolds number dependence on the interfacial configuration is small.
We can see the edge of the spike rolled up at $t^*=3$.
	At later stages ($t^*=3,4$), the higher the Reynolds number is, the more unstable the interface becomes. 
	For the ${\rm Re}=5,120$ case [Fig.~\ref{fig:interfaceRT}(c)], the interface becomes especially complicated. 
	The Kelvin--Helmholtz instability appears conspicuously as the Reynolds number increases.

\begin{figure*}[htbp]
	\includegraphics[width=12cm]{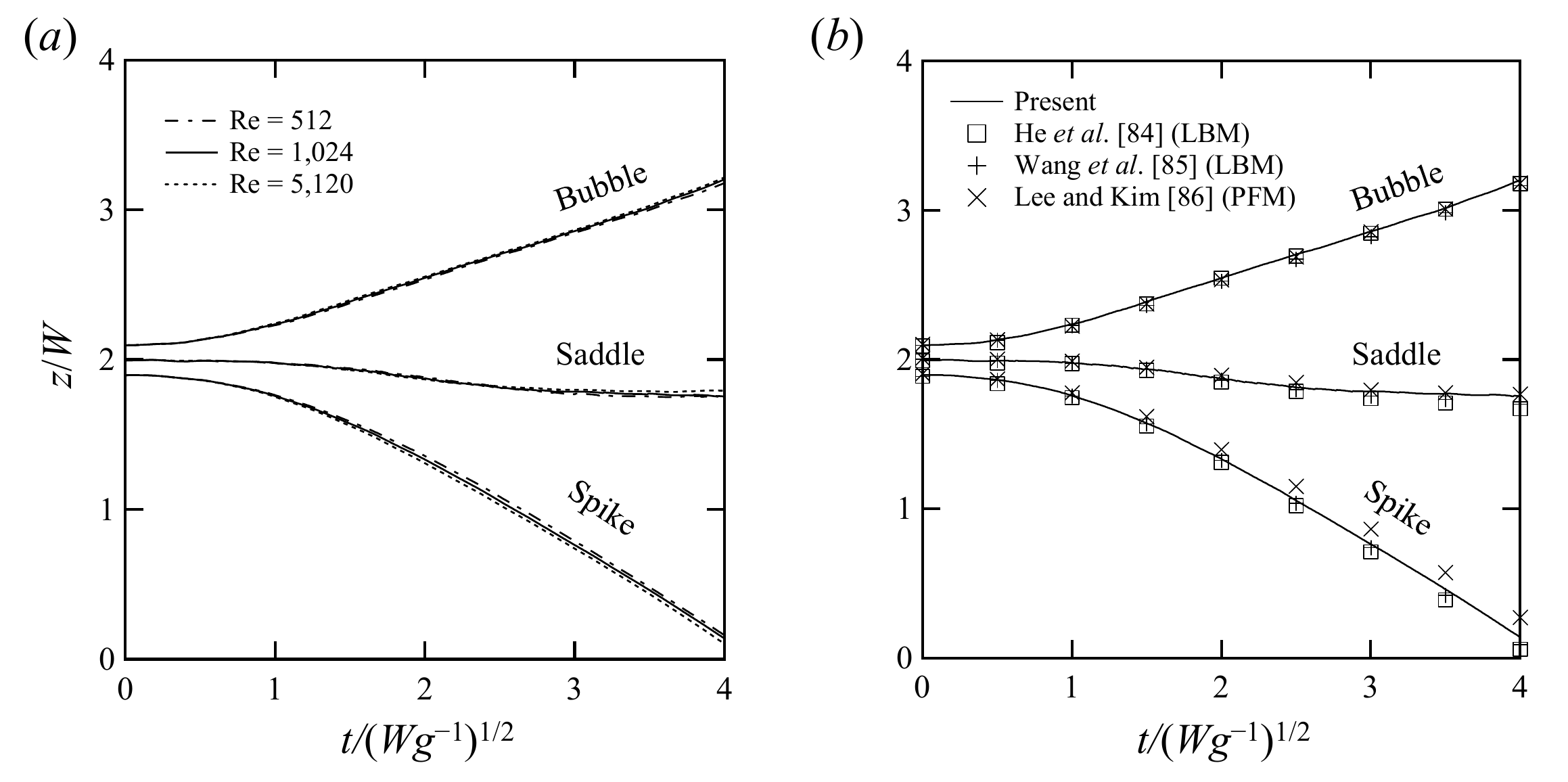}
	\caption{Time history of the positions of the bubble front, spike tip, and saddle point. (a) Reynolds number dependence of the positions with ${\rm Re}=512$ (dash-dotted line), ${\rm Re}=1,024$ (solid line), and ${\rm Re}=5,120$ (dotted line). 
	As pointed out by \citet{He1999b}, when ${\rm Re}>512$, the Reynolds number dependence is negligible.
	(b) Comparison of the present study (solid line) and previous works with the lattice Boltzmann~\citep{He1999b,Wang2016} and phase-field methods~\citep{Lee2013}, where all simulations are performed with ${\rm Re}=512$.
	For the spike-tip evolution, no significant difference is observed among the lattice Boltzmann simulations. 
	At later stages, these simulations are found to penetrate more deeply than the phase-field simulation.
	 \label{fig:positionRT}}
\end{figure*}

	The time history of the positions of the bubble front, spike tip, and saddle point are calculated and plotted in Fig.~\ref{fig:positionRT}(a) (see Fig.~\ref{fig:interfaceRT} for the locations of the bubble front, spike tip, and saddle point). 
	As can be seen in Fig.~\ref{fig:positionRT}(a), the differences in the positions are very small in our simulations, unlike the interface structure shown in Fig.~\ref{fig:interfaceRT}.
	~\citet{He1999b} pointed out that, when ${\rm Re}>512$, the Reynolds number dependence is negligible.
	The present simulation, using the forcing scheme in Eqs.~(\ref{eq:ketF}) and (\ref{eq:ketFd}), shows a similar trend, supporting their suggestion.

	Using the same computational setup and the parameters of \citet{He1999b}, three-dimensional Rayleigh--Taylor simulations were carried out using the lattice Boltzmann~\citep{Wang2016} and phase-field methods~\citep{Lee2013}.
	\citet{Wang2016} used the phase-field-based MRT lattice Boltzmann model; 
	\citet{Lee2013} directly solved the Navier--Stokes--Cahn--Hilliard equations.
	Here, we compare our results with previous works~\citep{He1999b, Lee2013, Wang2016}.
	Figure~\ref{fig:positionRT}(b) shows the time histories of the positions at ${\rm Re}=1,024$ and ${\rm At}=0.5$.
	Comparing the results, the time changes of the bubble front and the saddle point are almost the same, irrespective of the method used.
	For the spike-tip evolution, no significant difference is observed among the lattice Boltzmann simulations. 
	At later stages, these simulations are found to penetrate more deeply than the phase-field simulation.
	The difference between the lattice Boltzmann and phase-field methods is considered to arise from the different wall-boundary conditions implemented.

\section{\label{sec:jet} Liquid jet breakup}

\subsection{\label{sec:setup}Setup}
\begin{figure}[htbp]
	\includegraphics[width=8.5cm]{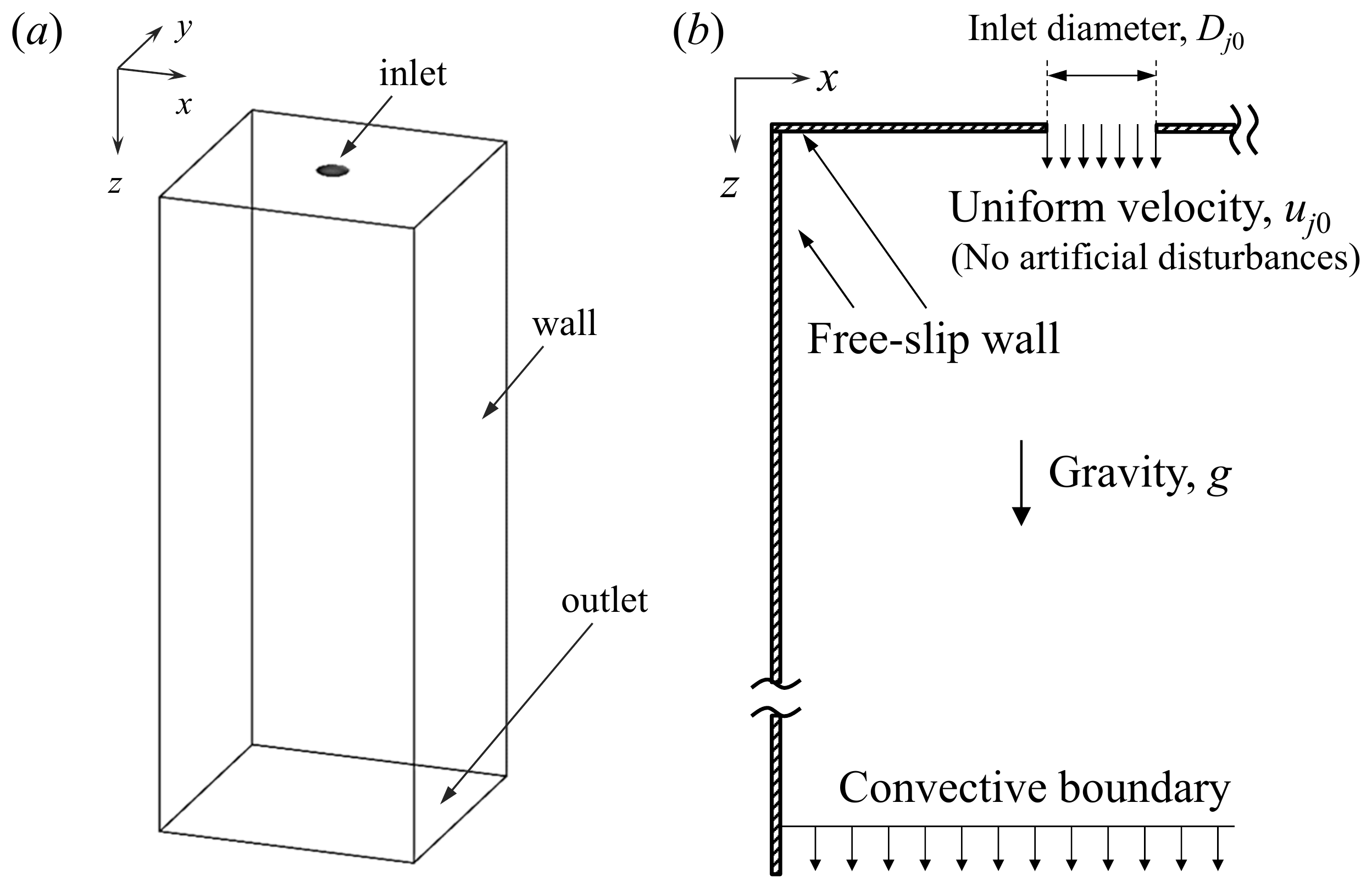}
	\caption{Boundary conditions for a liquid-jet simulation. (a) The computational domain is discretized into a $8D_{j0}\times8D_{j0}\times20D_{j0}$ lattice. 
	The boundaries consist of an inflow boundary, a wall boundary, and an outflow boundary. 
	(b) A circular inflow boundary with a uniform velocity $u_{j0}$ is implemented at the top within $D_{j0}$.
	Free-slip~\citep{Succi2001} and convective boundary conditions~\citep{Lou2013} are implemented as the wall and outflow boundaries, respectively.
	\label{fig:boundaryJet}}
\end{figure}

	Figure~\ref{fig:boundaryJet} illustrates a schematic diagram of the boundary conditions for liquid-jet-breakup simulations.
	This computational setup is the same as that of \citet{Saito2016}, except for the outflow boundary.
	In the initial state, the computational domain is filled with blue-particle-distribution functions, $f_i^b$, with zero velocity.
	The boundaries consist of an inflow boundary, a wall boundary, and an outflow boundary.
	A circular inflow boundary is implemented at the top within $(x-x_c)^2 + (y-y_c)^2 < (D_{j0}/2)^2$, where $x_c$ and $y_c$ are central locations on the $x$-$y$ plane.
	The uniform velocity $u_{j0}$ is implemented, with the corresponding equilibrium functions given at this site.
	No artificial disturbances are considered at the inflow boundary.
	A wall boundary is implemented on the top (except for an inflow-boundary site) and on the lateral sites.
	A free-slip condition~\citep{Succi2001} is implemented as a wall boundary condition.
	At the outflow, a convective boundary condition~\citep{Lou2013} is used,
in which the following convective equation for the distribution functions
\begin{equation}
	\frac{\partial f_i^k}{\partial t} + U_c \frac{\partial f_i^k}{\partial z} = 0, ~ {\rm at} ~z=N,
\end{equation} 
is solved, where $N$ is the node on the outflow boundary.
	Following \citet{Lou2013}, we set two ghost-nodes $N+1$ and $N+2$.
	The discretized form can be given by the first-order implicit scheme,
\begin{align}
	&f_i^k(x,y,N,t+\delta_t) \nonumber
	\\
	&= \frac{f_i^k(x,y,N,t)+\lambda f_i^k(x,y,N-1,t+\delta_t)}{1+\lambda},
\end{align}
where $\lambda=U_c(t+\delta_t)\delta_t/\delta_x$.
	For the convective velocity normal to the outflow boundary $U_c$, several choices can be considered, e.g., the local velocity, the average velocity, and maximum velocity~\citep{Orlanski1976}.
	Through some numerical tests, we determine that the local velocity is suitable to the present system, that is,
\begin{equation}
	U_c(x,y,N,t) = u_z (x,y,N-1,t),
\end{equation}
where $u_z({\bf x},t) = u_z (x,y,z,t)$ is the $z$-direction component of fluid velocity ${\bf u}$.

	The body force in Eq.~(\ref{eq:ketFd}) for the liquid-jet simulations is set as 
\begin{equation}
	{\bf F}({\bf x},t) = \left(\rho({\bf x},t)-\rho_b^0 \right){\bf g}, 
	\label{eq:bodyForceJet}
\end{equation}
with ${\bf g} = (0,0,g)$.
	Eq.~(\ref{eq:bodyForceJet}) means that the gravitational force acts only in the dispersed phase~\citep{Chen2014}.

%

\subsection{\label{sec:comExp}Comparison with experiments}
	
	Using the lattice Boltzmann model developed in Sec.~\ref{sec:methodology} and the boundary conditions provided in Sec.~\ref{sec:setup}, we perform here numerical simulations of liquid-jet breakup.		
	The parameters for this simulation are set to be exactly the same as in the experiments~\citep{Saito2016}, and the results are compared.
	In the target experiments, a glycerin-water-mixture jet was injected into a silicon-oil pool.
	These test fluids were immiscible with each other.
	In \citep{Saito2016}, the lattice Boltzmann simulation was also carried out using the MRT color-fluid model based on the D3Q19 lattice.
	
	In the framework of linear theory, one can choose the following dimensionless groups to describe the problem~\cite{Lin1998}
\begin{align}
	\gamma &= \frac{\rho_j}{\rho_c}, \label{eq:densityRatio}
	\\
	\eta &= \frac{\nu_j}{\nu_c}, \label{eq:viscosityRatio}
	\\
	{\rm Re} =& \frac{\rho_j  u_{j0} D_{j0}}{\mu_j}, \label{eq:jetReynolds}
	\\
	{\rm We} =& \frac{\rho_j u_{j0}^2 D_{j0}}{\sigma}, \label{eq:jetWeber}
	\\
	{\rm Fr} &= \frac{u_{j0}^2}{gD_{j0}}. \label{eq:Froude}
\end{align}
	The conditions of the experiments and simulations are summarized in Table~\ref{tab:table1} and in the dimensionless groups in Eqs.~(\ref{eq:densityRatio})--(\ref{eq:Froude}).
	The parameters for the target experiments and the present simulation match exactly.
	In Table~\ref{tab:table1}, only the kinematic-viscosity ratio $\eta$ in the previous simulation differs from the others.
	This was because the previous D3Q19 MRT color-fluid model could not maintain the numerical stability when $\eta=4.2$.
	Note that the present lattice Boltzmann model remains stable under the condition in Table~\ref{tab:table1}.
	
	According to the grid refinement study in~\citep{Saito2016}, we determine the inlet diameter $D_{j0}=25$ and the computational domain  $8D_{j0}\times8D_{j0}\times20D_{j0}$.
	In this paper, the density of the dispersed phase $\rho_j (=\rho_r^0)$ and the inlet velocity $u_{j0}$ are set to be 1.0 and 0.1, respectively.
	Other parameters can be determined using the relations of dimensionless groups [Eqs.~(\ref{eq:densityRatio})--(\ref{eq:Froude})] in lattice units:
	$\rho_c = \rho_b^0 = 0.70$, $\sigma = 1.1\times10^{-3}$, $\nu_j = \nu_r = 7.4\times10^{-4}$, $\nu_c = \nu_b = 1.8\times10^{-4}$, and $g = 4.7\times10^{-5}$.

\begin{table}[b]
\caption{\label{tab:table1} Conditions of the experiments and simulations.}
\begin{ruledtabular}
\begin{tabular}{ccc}
\textrm{}&
\textrm{Experiment and present simulation}&
\textrm{Previous simulation\footnote{The D3Q19 MRT color-fluid model was used. In the simulation, the kinematic-viscosity ratio $\eta$ was set to be unity owing to the limitation of numerical stability.}}\\
\colrule
$\gamma$ & 1.4 & 1.4\\
$\eta$ & 4.2 & 1.0 \\
${\rm Re}$ & $3.4\times10^{3}$ & $3.4\times10^{3}$\\
${\rm We}$ & $2.2\times10^{2}$ & $2.2\times10^{2}$\\
${\rm Fr}$ & 8.5 & 8.5\\
\end{tabular}
\end{ruledtabular}
\end{table}

\begin{figure*}[tb]
	\includegraphics[width=12cm]{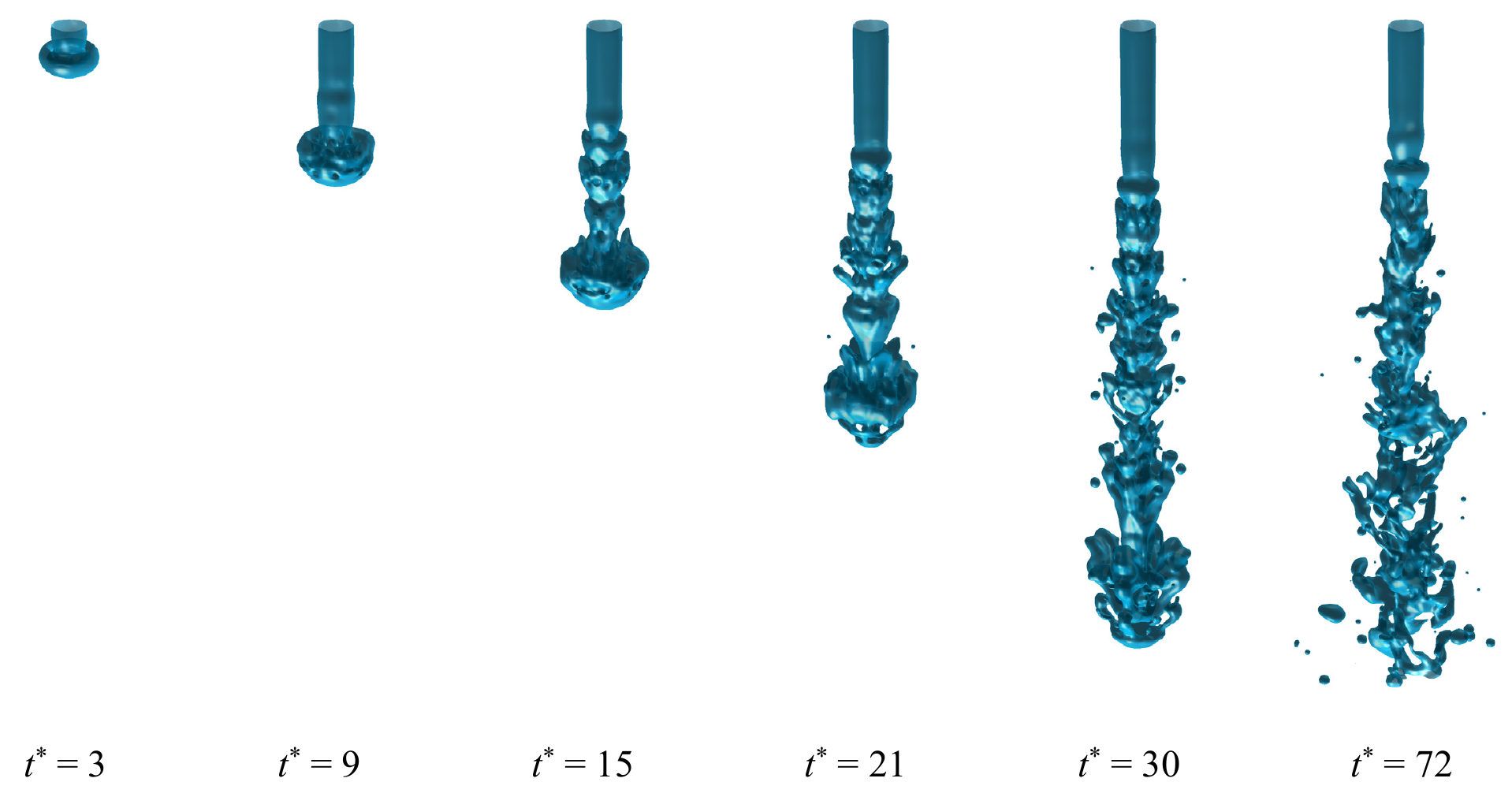}
	\caption{Interface evolution of liquid-jet breakup under the same conditions as the target experiment: $\gamma=1.4$, $\eta=4.2$, ${\rm Re}=3.4\times10^3$, ${\rm We}=2.2\times10^2$, and ${\rm Fr}=8.5$.
	The computational domain is set to be $200\times 200 \times 500$.
	The interfacial instability is triggered by the return flow of a mushroom head generated in an initial stage of jet injection.
	The liquid core finally becomes asymmetric with entrainment behavior.
	Numerical stability is maintained throughout the simulation, although the kinematic viscosity for the continuous phase is set as low as $1.8\times10^{-4}$.
	\label{fig:sinuous}}
\end{figure*}

	Figure~\ref{fig:sinuous} shows the interface evolution of the present simulation.
	Time is non-dimensionalized by the inlet diameter $D_{j0}$ and the inlet velocity $u_{j0}$ as
\begin{equation}
	t^* = \frac{t}{D_{j0}/u_{j0}},
\end{equation}
	in the liquid-jet-breakup simulations.
	At $t^*=3$, a mushroom-like head shape is created. 
	At $t^*=9$ to 15, the interface of the jet becomes unstable. 
	Observation shows that this interfacial instability is triggered by the return flow of a mushroom head generated in an initial stage of jet injection.
	Such a characteristic flow pattern leads to the onset of interfacial instability, although no artificial spatial or temporal perturbation has been assumed in the initial or boundary conditions.
	At later stages ($t^*=30$ and 72), the jet's leading-edge collapses, and
	the entrainment behavior, namely, the droplet formation from the side of the jet, is also observed.
	Finally, the liquid core becomes asymmetric, as can be seen at $t^*=72$.
	Such a series of processes is similar to an experimental observation~\citep{Saito2016}.
	
\begin{figure}[tb]
	\includegraphics[width=7.5cm]{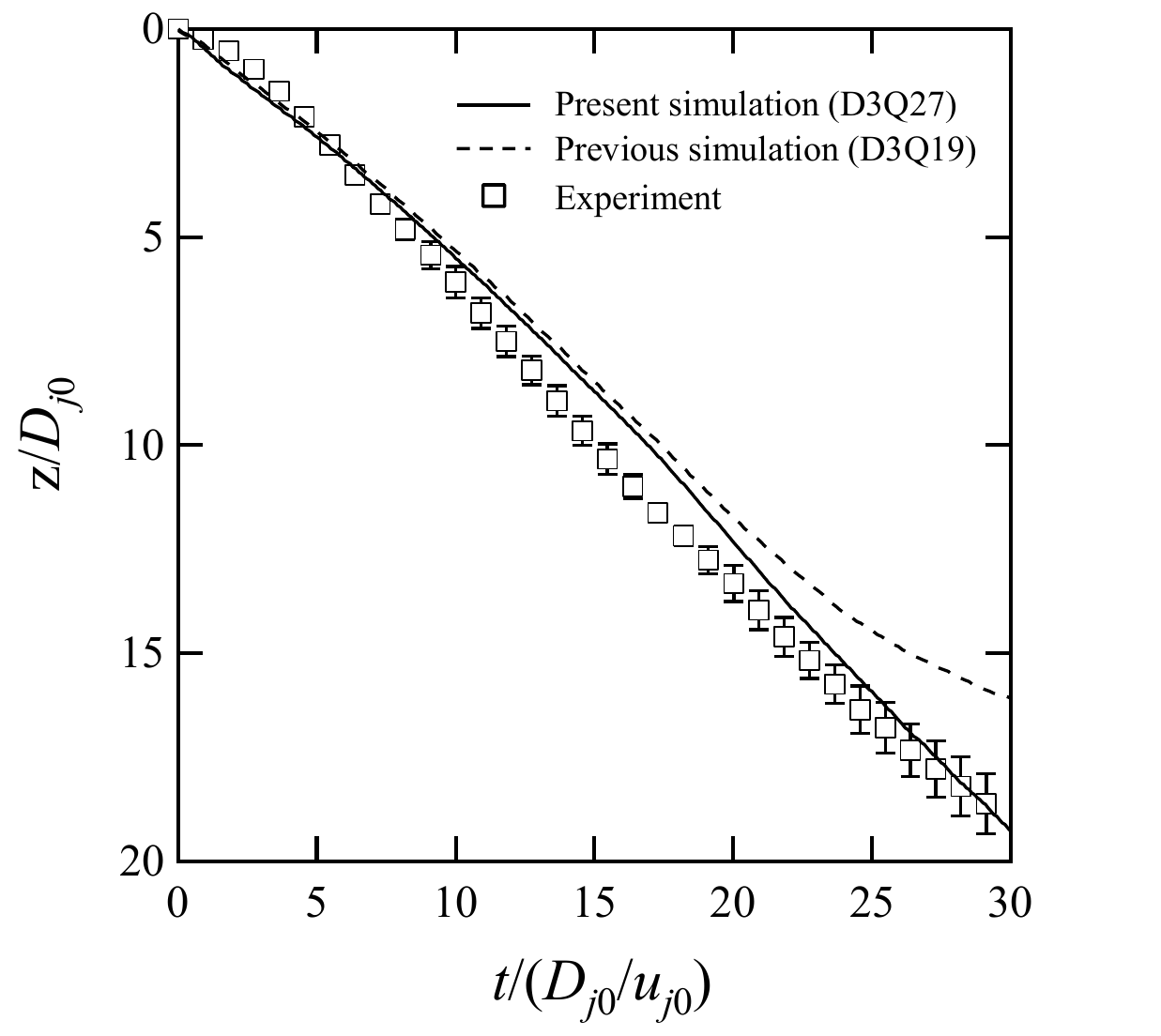}
	\caption{Time history of jet penetration: the present simulation (solid line), previous simulation (dashed line)~\citep{Saito2016}, and experimental result (square symbol)~\citep{Saito2016}.
	The present simulation shows a better agreement with the experimental data than that using the previous D3Q19 model.
	 \label{fig:penetration}}
\end{figure}

	The history of jet penetration can be evaluated by both experiments and simulations.
A comparison between experiments and simulations in terms of the position of the jet's leading-edge is provided in Fig.~\ref{fig:penetration},
	where the position is normalized by the inlet diameter $D_{j0}$.
	Focusing on the results of lattice Boltzmann simulations, the time histories are almost the same until $t^*=15$.
	Afterward, the differences in positions gradually increase and the present simulation results tend to be close to the experimental data.
	This is thought to be due to the influence of ambient viscosity.
	In the present simulation, the lower viscosity of the surrounding fluid facilitates the penetration of the jet.
	The simulation using the developed lattice Boltzmann model shows a better agreement with the experimental data than that using the previous D3Q19 model~\citep{Saito2016}.
	The enhanced numerical stability of the model enables us to simulate more realistic conditions.

\subsection{Transition of breakup regimes}
	Recently, \citet{Saito2017} have classified jet breakup regimes in liquid-liquid systems based on observations.
	This was an extension of Ohnesorge's classification for liquid-gas systems~\cite{Ohnesorge1936,Kolev2005,McKinley2011}.
	The classified breakup regimes are as follows: 0--dripping, I--varicose breakup, II--sinuous breakup, and III--atomization.
	Regime II was further divided into two sub-regimes: IIa--sinuous {\it without} entrainment, IIb--sinuous {\it with} entrainment.
	On the basis of the observations and phenomenological considerations, the flow-transition criteria were derived in~\cite{Saito2017}:
\begin{equation}
	{\rm Oh} = 2.8 {\rm Re}^{-1}, 
\end{equation}
for Regimes I and II, and
\begin{equation}
	{\rm Oh} = 22 {\rm Re}^{-1}, 
\end{equation}
for Regimes II and III, where the Reynolds number ${\rm Re}$ is defined as in Eq.~(\ref{eq:jetReynolds}) and the Ohnesorge number ${\rm Oh}$ can be given by Eqs.~(\ref{eq:jetReynolds}) and (\ref{eq:jetWeber}) as ${\rm Oh} = {\rm We}^{1/2}/{\rm Re}$.
	By using the above transition criteria, one can predict the breakup regimes of an immiscible liquid-liquid jet from initial parameters.
	With the present lattice Boltzmann model, we assess here the potential for the reproducibility of the breakup regimes.

\begin{figure}[tb]
	\includegraphics[width=7.2cm]{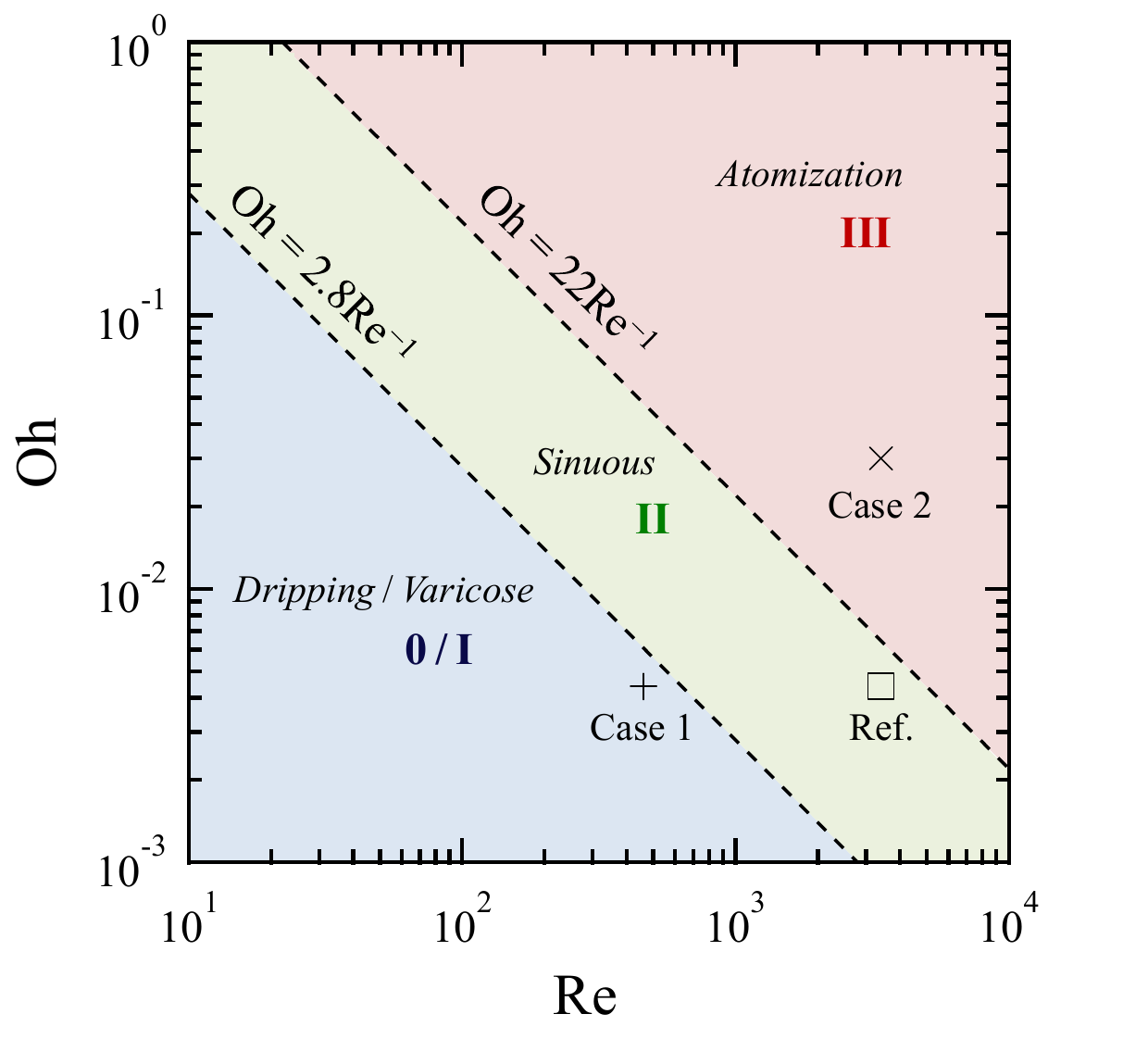}
	\caption{Location of simulation parameters on the regime map proposed in~\citep{Saito2017}.
	The description ``Ref.'' represents the same condition as mentioned in Sec.~\ref{sec:comExp}. 
	In Ref. case, sinuous-type regime is expected and will appear;
	in Case 1 and Case 2, it is expected that dripping/varicose and atomization, respectively, will appear.
	 \label{fig:conditionsAndSketch}}
\end{figure}

	The simulation conditions are summarized in Table~\ref{tab:table2}.
	The corresponding values on the dimensionless diagram are shown in Fig.~\ref{fig:conditionsAndSketch}. 
	The description ``Ref. case'' represents the same condition as mentioned in Sec.~\ref{sec:comExp}. 
	As can be seen, the simulation results provided in the previous section are in Regime II, where sinuous breakup with or without entrainment is expected.
	The other parameters, namely, density ratio $\gamma$, viscosity ratio $\eta$, and Froude number ${\rm Fr}$, are set to be the same as in Table~\ref{tab:table1}. 
	In Case 1, ${\rm Re}$ is decreased by increasing the jet viscosity $\nu_j$ and decreasing jet velocity $u_{j0}$ while ${\rm Oh}$ is fixed at the same value as in the Ref. case. 
	As can be seen in Fig.~\ref{fig:conditionsAndSketch}, Case 1 is in Regime 0 or I. 
	It is expected that the dripping- or varicose-breakup regimes will appear. 
	In Case 2, ${\rm Re}$ is fixed at the same value as in the Ref. case while ${\rm Oh}$ is increased by decreasing the interfacial tension $\sigma$. 
	Since Case 2 is in Regime III in Fig.~\ref{fig:conditionsAndSketch}, it is expected that atomization will appear.
	Note that it is impossible to change only {\it one} parameter in the experimental procedures; 
	only the effect of a focused parameter can be taken into account in the numerical simulations. 	
\begin{table}[b]
\caption{\label{tab:table2} Simulation conditions for dimensionless numbers to be investigated. The density ratio, viscosity ratio, and Froude number are set to be the same as in Table~\ref{tab:table1}}
\begin{ruledtabular}
\begin{tabular}{cccc}
\textrm{}&
\textrm{${\rm Re}$}&
\textrm{${\rm Oh}$}&
\textrm{${\rm We}=({\rm Re} \cdot {\rm Oh})^2$}\\
\colrule
Ref. case & $3.4\times10^{3}$ & $4.4\times10^{-3}$ & $2.2\times10^{2}$\\
Case 1    & $4.6\times10^{2}$ & $4.4\times10^{-3}$ & 4.1\\
Case 2    & $3.4\times10^{3}$ & $3.0\times10^{-3}$ & $1.0\times10^{4}$\\
\end{tabular}
\end{ruledtabular}
\end{table}

\begin{figure*}[tb]
	\includegraphics[width=11cm]{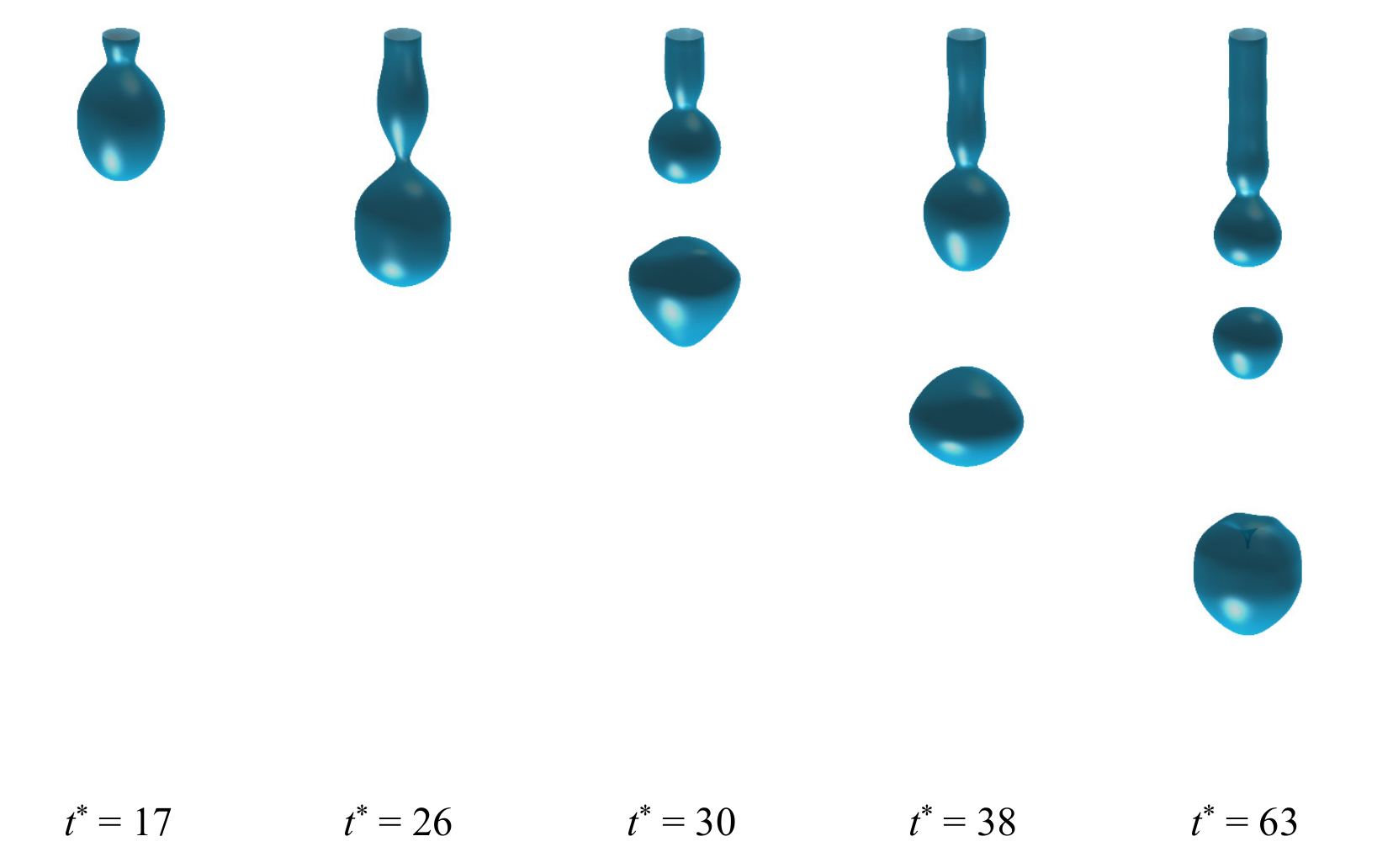}
	\caption{Simulation results of Case 1 in Table~\ref{tab:table2}: $\gamma=1.4$, $\eta=4.2$, ${\rm Re}=4.6\times10^2$, ${\rm We}=4.1$, and ${\rm Fr}=8.5$.
	The computational domain is set to be $120\times 120 \times 300$.
	A droplet forms mainly at the tip of the jet; the characteristics of varicose breakup (Regime I) appear. \label{fig:varicose}}
\end{figure*}

	Figure~\ref{fig:varicose} shows the simulation results for Case 1 in Table~\ref{tab:table2}. 
	We set the inlet diameter and velocity to $D_{j0}=15$ and $u_{j0}=0.05$, respectively.
	The other parameters are as follows:
	$\sigma = 9.1\times10^{-3}$, $\nu_j = \nu_r = 1.6\times10^{-3}$, $\nu_c = \nu_b = 3.9\times10^{-4}$, and $g = 2.0\times10^{-5}$.
	At $t^*=17$, the swollen part is generated around the inlet.
	The mushroom-like head does not appear. 
	The swollen part moves downward with the growth of the neck part at $t^*=26$,
and the corresponding part breaks up into a single droplet at $t^*=30$.
	At this time, the following swollen part is generated on a liquid column.
	The formation of the swollen part, the growth of the neck part, and breakup into a single droplet are observed through simulation.
	In this case, the so-called satellite-droplet formation just after the primary-droplet formation is not observed.
	According to the experimental data in liquid-liquid systems~\citep{Saito2017}, the average size of satellite droplet is about 0.3 times smaller than the nozzle diameter $D_{j0}$.
	To reproduce the satellite-droplet formation, the higher resolution would be required. 
	The order of the droplet sizes has the same extent as the inlet diameter or is larger than it.
	This series of processes is a characteristic of the so-called pinch-off behavior, which is similar to the scenario considered in Rayleigh's breakup for a liquid column~\citep{Rayleigh1878}.
	This corresponds to the varicose breakup or Regime I in Fig.~(\ref{fig:conditionsAndSketch}).
	The characteristics of the varicose breakup regime are reproduced by the present lattice Boltzmann simulation.

\begin{figure*}[tb]
	\includegraphics[width=11cm]{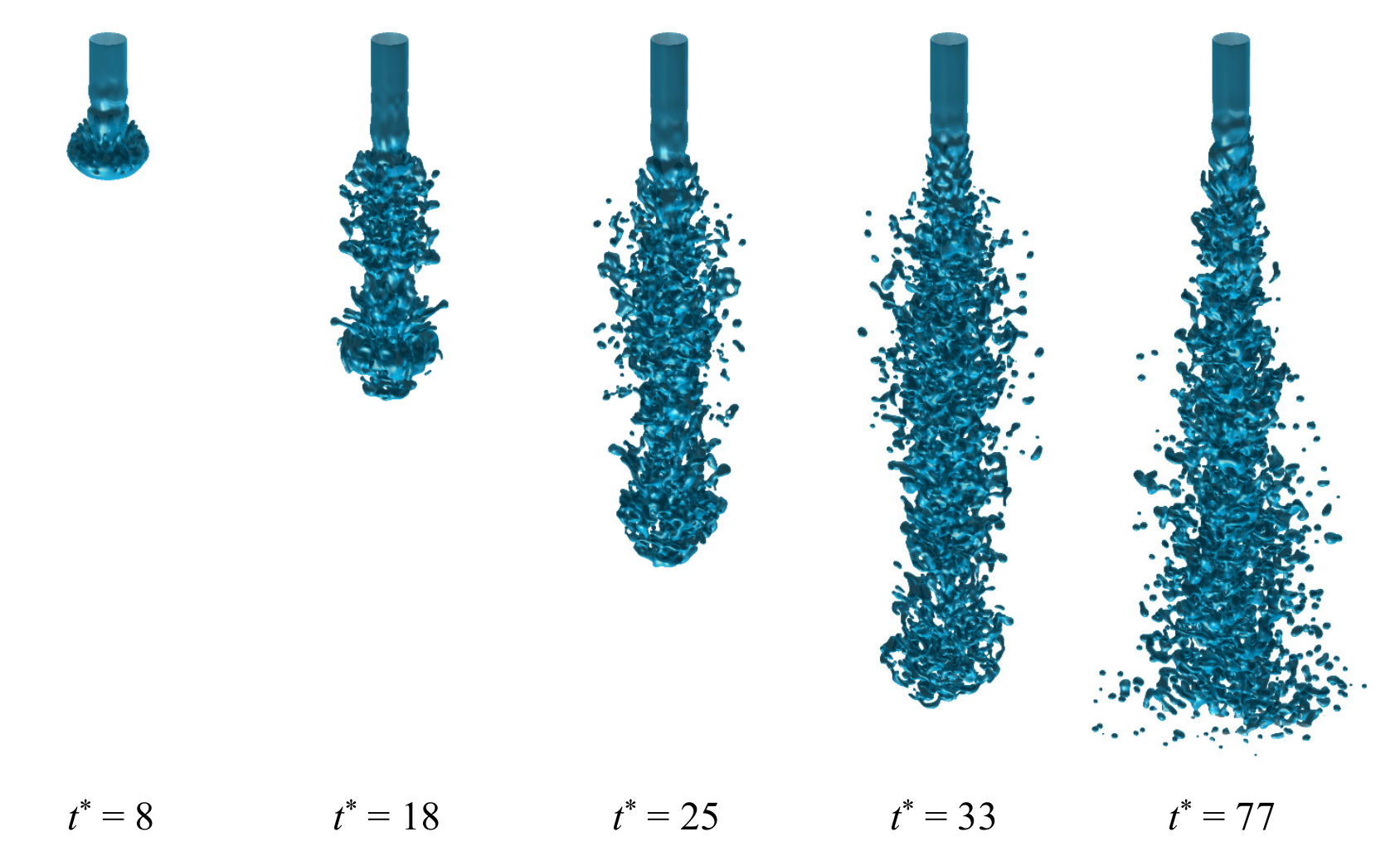}
	\caption{Simulation results of Case 2 in Table~\ref{tab:table2}: $\gamma=1.4$, $\eta=4.2$, ${\rm Re}=3.4\times10^3$, ${\rm We}=1.0\times10^4$, and ${\rm Fr}=8.5$.
	The computational domain is set to be $240\times 240 \times 600$.
	A large number of droplets are entrained from the jet surface; the characteristics of atomization (Regime III) appear. \label{fig:atomization}}
\end{figure*}

	Figure~\ref{fig:atomization} shows the simulation results for Case 2 in Table~\ref{tab:table2}. 
	We set the inlet diameter and velocity to $D_{j0}=30$ and $u_{j0}=0.1$, respectively.
	The other parameters are as follows:
	$\sigma = 3.0\times10^{-5}$, $\nu_j = \nu_r = 8.8\times10^{-4}$, $\nu_c = \nu_b = 2.1\times10^{-4}$, and $g = 3.9\times10^{-5}$.
	As in the Ref. case, a mushroom-like head appears at $t^*=8$. 
	Through $t^*=18$--$33$, the jet continues penetration with active entrainment. 
	The sizes of the generated droplets are much smaller than the inlet diameter, and the number of droplets is much greater than in the Ref. case and Case 1. 
	Owing to the entrainment from a jet interface, the liquid column is fully covered, and it is difficult to identify this column in the snapshots shown in Fig.~\ref{fig:atomization}. 
	The series of processes corresponds to the atomization or Regime III in Fig.~(\ref{fig:conditionsAndSketch}).
	The characteristics of the atomization regime are also reproduced by the present lattice Boltzmann simulations.
	
	Before summarizing the simulation results, let us mention the jet-breakup regime shown in Fig.~\ref{fig:sinuous} again.
	During the penetration, the jet maintains an axisymmetric configuration.
	This is reasonable because the computational setup described in Sec.~\ref{sec:setup} is exactly symmetric on the $x$-$y$ plane.
	However, the jet column finally winds and becomes an asymmetric shape (see $t^*=72$ in Fig.~\ref{fig:sinuous}).
	In addition, the entrainment also can be seen in this case.	
	This type of breakup regime corresponds to Regime IIb in Fig.~(\ref{fig:conditionsAndSketch}).
	This implies that the physical balance among hydrodynamic forces, including inertia, viscous force, and interfacial-tension force can be naturally reproduced via the lattice Boltzmann simulation.
	This fact numerically supports the validity of the dimensionless diagram proposed in Ref.~\citep{Saito2017}.
	
\begin{figure*}[htbp]
	\includegraphics[width=10cm]{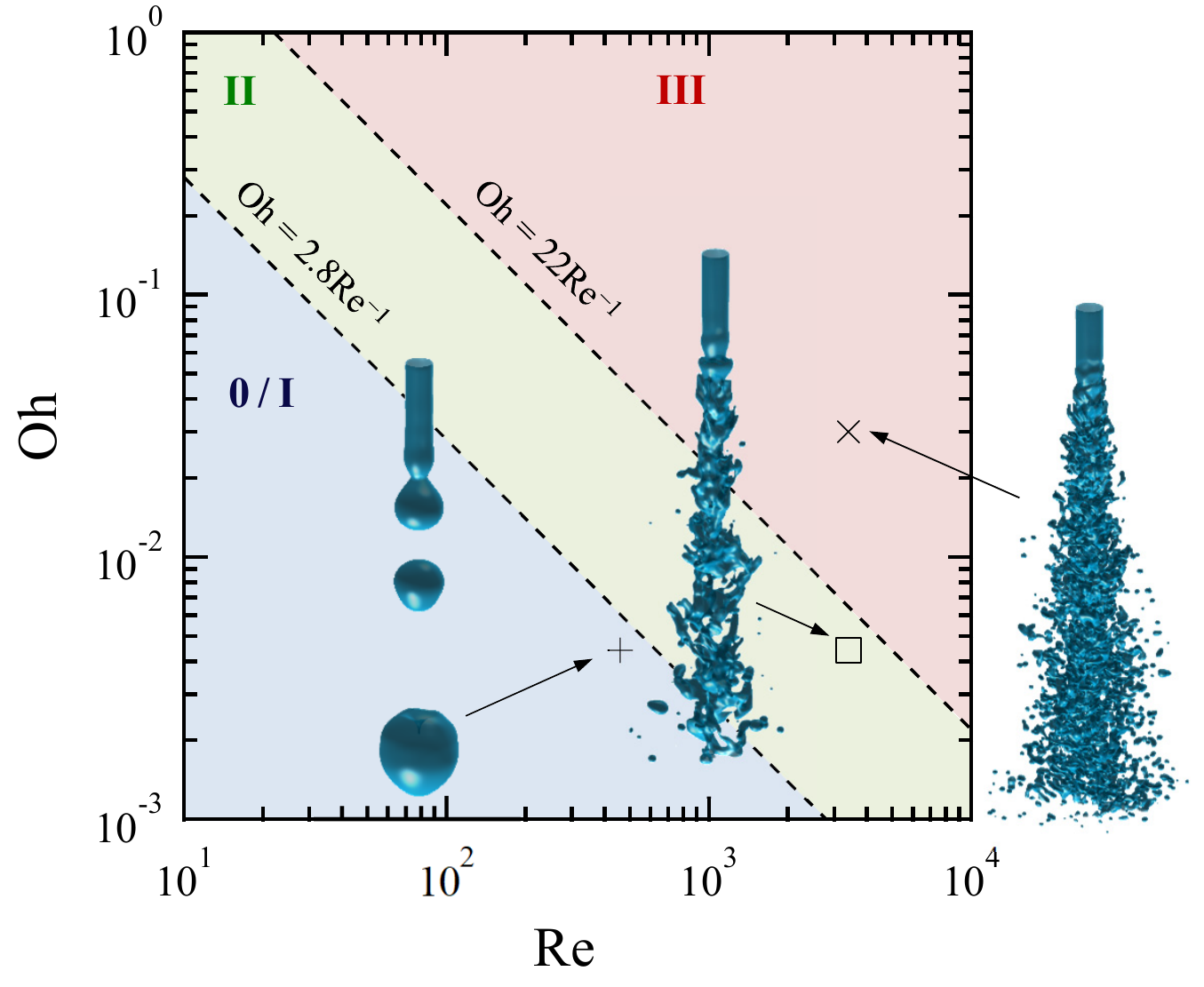}
	\caption{Summary of the present results of liquid-jet-breakup simulations in the dimensionless diagram for liquid-liquid systems~\citep{Saito2017}. 
	The lattice Boltzmann model developed in this paper well reproduce the breakup characteristics expected by the regime map.
	\label{fig:summary}}
\end{figure*}

	Finally, we summarize the simulation results in the dimensionless diagram of Fig.~\ref{fig:conditionsAndSketch}.
	Typical snapshots of liquid-jet breakup from Figs.~\ref{fig:sinuous}, \ref{fig:varicose}, and \ref{fig:atomization} are summarized in Fig~\ref{fig:summary}.
	The characteristics of varicose breakup (Regime I), sinuous breakup (Regime II), and atomization (Regime III) are successfully simulated. 
	On the breakup regimes, we can conclude that the lattice Boltzmann model developed in Sec.~\ref{sec:methodology} under the boundary conditions described in Sec.~\ref{sec:setup} well reproduce the breakup characteristics expected by the regime map proposed in Ref.~\citep{Saito2017}.

\section{\label{sec:conclusions}Conclusions}
	In this paper, a three-dimensional lattice Boltzmann model for immiscible two-phase flows was developed in the framework of a D3Q27 lattice.
	An MRT collision operator for the D3Q27 lattice~\citep{Dubois2011, Geier2015} was introduced.
	The choice of relaxation parameters optimized by \citet{Suga2015} greatly improved the numerical stability of the model.
	A generalized perturbation operator~\citep{Liu2012} and an enhanced equilibrium distribution function~\citep{Leclaire2013} were also successfully incorporated into the present D3Q27 model. 
	
	Using the formulated lattice Boltzmann model, three types of numerical tests were carried out: a static droplet, an oscillating droplet, and the Rayleigh--Taylor instability.
	The static-droplet test shows that the measured pressure difference satisfied the Laplace law,
	and the interfacial tension agreed well with that predicted by Eq.~(\ref{eq:sigma}) within a maximum error of 1.6\%.
	The oscillating-droplet test was performed to compare the oscillation period with the analytical solution.
	For the high-interfacial-tension case, the error with the analytical solution was within 2.4\%.
	Under low interfacial tension, a droplet reached equilibrium without oscillations in the most viscous case; the maximum error for the available data was 10.8\%.
	The Rayleigh--Taylor simulations were performed with a computational setup and parameters that were strictly similar to those of \citet{He1999b}. 
	The positions of the bubble front, spike tip, and saddle point were measured for comparison with previous works using the lattice Boltzmann~\citep{He1999b, Wang2016} and phase-field methods~\citep{Lee2013}.
	The time changes of the bubble front and the saddle point were almost the same, irrespective of the method used.
	At later stages, a little difference between the lattice Boltzmann and phase-field simulations was observed for the spike-tip evolution owing to the difference in implementation of the wall-boundary conditions.
	
	The developed model was applied to liquid-jet-breakup simulations.
	First, we chose the parameters to match the experimental conditions~\citep{Saito2016}. 
	The five dimensionless groups were employed to determine the physical properties in lattice units.
	The developed D3Q27 model maintained numerical stability throughout the simulations; the previous work using a D3Q19 color-fluid lattice Boltzmann model~\citep{Saito2016} failed to simulate stably under the same conditions.
	The present results showed that the characteristic interfacial evolution captured the experimental results. 
	A mushroom-like head was formed at the early stages;
	later, the configuration of the liquid core transitioned from asymmetric to axisymmetric, and interface entrainment also naturally occurred.
	The time history of the jet's leading-edge was compared with that obtained by experiment. Quantitative comparisons agreed well with the experimental data.

	By choosing the parameters based on the regime map for jet breakup in liquid-liquid systems~\citep{Saito2017}, we performed simulations to evaluate the reproducibility of the regime map.
	In the varicose regime, pinch-off-type breakup, i.e., the droplet formation in the tip of the liquid column, occurred.
	Satellite-drop formation was not confirmed in the present simulation.
	In the atomization regime, entrainment from the liquid column was distinguished.
	The liquid column was fully covered with entrainment droplets, which were much smaller than the inlet diameter.
	In conclusion, the breakup regimes appearing in the simulations successfully reproduced the predicted regimes, including the varicose, sinuous, and atomization regimes.

	The authors also tested the stable ranges of the developed color-fluid model through simple test cases as in Sec.~\ref{sec:staticDrop}.
	For such a static case, at least, it is confirmed that the available maximum density ratio $\gamma ~(=\rho_r/\rho_b)$ was up to 1,000 at the unit kinematic-viscosity ratio ($\nu_r = \nu_b = 1.0\times 10^{-3}$);
	the maximum kinematic-viscosity ratio $\eta~(=\nu_r/\nu_b)$ was up to 1,000 at $\gamma = 1.5$.
	From such a point of view, the authors believe that the three-dimensional color-fluid lattice Boltzmann model developed in this paper is generally suitable for various other applications within the aforementioned stability ranges.


%



\begin{acknowledgments}
	This work was supported by Mitsubishi Heavy Industries, Ltd. 
	The authors are grateful to H. Sakaba and H. Sato for their invaluable support. 
	S.S. is grateful to A. Kaneko and Y. Iwasawa for the helpful discussions. 
	The support of JSPS KAKENHI Grant Number 16J02077 is also acknowledged.
\end{acknowledgments}

\appendix

\section{\label{sec:appA} Transformation matrix}
A transformation matrix in practical form is obtained by substituting the velocity set [Eq.~(\ref{eq:velSet})] into the moment set [Eq.~(\ref{eq:momentSet})]:
\begin{widetext}
\begin{equation}
\scalebox{0.66}{$\displaystyle
{\bf M} = 
\left [
	\begin{array}{ccccccccccccccccccccccccccc}
		1 & 1 & 1 & 1 & 1 & 1 & 1 & 1 & 1 & 1 & 1 & 1 & 1 & 1 & 1 & 1 & 1 & 1 & 1 & 1 & 1 & 1 & 1 & 1 & 1 & 1 & 1 \\	
		0 & 1 & -1	& 0	 & 0	& 0 	& 0 	& 1	 & -1	& 1	& -1	& 0 	& 0	&0	& 0	& 1	& -1	& -1	& 1	& 1	& -1	& 1	& -1	& -1&	1	&1	&-1 \\	
		0	&0	&0	&1	&-1	&0	&0	&1	&-1	&-1	&1	&1	&-1	&1	&-1	&0	&0	&0	&0	&1	&-1	&1	&-1	&1	&-1	&-1	&1 \\	
		0	&0	&0	&0	&0	&1	&-1	&0	&0	&0	&0	&1	&-1	&-1	&1	&1	&-1	&1	&-1	&1	&-1	&-1	&1	&1	&-1	&1	&-1 \\	
		-2	&-1	&-1	&-1	&-1	&-1	&-1	&0	&0	&0	&0	&0	&0	&0	&0	&0	&0	&0	&0	&1	&1	&1	&1	&1	&1	&1	&1 \\	
		0 &2	&2	&-1	&-1	&-1	&-1	&1	&1	&1	&1	&-2	&-2	&-2	&-2	&1	&1	&1	&1	&0	&0	&0	&0	&0	&0	&0	&0 \\	
		0	&0	&0	&1	&1	&-1	&-1	&1	&1	&1	&1	&0	&0	&0	&0	&-1	&-1	&-1	&-1	&0	&0	&0	&0	&0	&0	&0	&0 \\	
		0	&0	&0	&0	&0	&0	&0	&1	&1	&-1	&-1	&0	&0	&0	&0	&0	&0	&0	&0	&1	&1	&1	&1	&-1	&-1	&-1	&-1 \\	
		0	&0	&0	&0	&0	&0	&0	&0	&0	&0	&0	&1	&1	&-1	&-1	&0	&0	&0	&0	&1	&1	&-1	&-1	&1	&1	&-1	&-1 \\	
		0	&0	&0	&0	&0	&0	&0	&0	&0	&0	&0	&0	&0	&0	&0	&1	&1	&-1	&-1	&1	&1	&-1	&-1	&-1	&-1	&1	&1 \\	
		0	&-4	&4	&0	&0	&0	&0	&-1	&1	&-1	&1	&0	&0	&0	&0	&-1	&1	&1	&-1	&2	&-2	&2	&-2	&-2	&2	&2	&-2 \\	
		0	&0	&0	&-4	&4	&0	&0	&-1	&1	&1	&-1	&-1	&1	&-1	&1	&0	&0	&0	&0	&2	&-2	&2	&-2	&2	&-2	&-2	&2 \\	
		0	&0	&0	&0	&0	&-4	&4	&0	&0	&0	&0	&-1	&1	&1	&-1	&-1	&1	&-1	&1	&2	&-2	&-2	&2	&2	&-2	&2	&-2 \\	
		0	&4	&-4	&0	&0	&0	&0	&-2	&2	&-2	&2	&0	&0	&0	&0	&-2	&2	&2	&-2	&1	&-1	&1	&-1	&-1	&1	&1	&-1 \\	
		0	&0	&0	&4	&-4	&0	&0	&-2	&2	&2	&-2	&-2	&2	&-2	&2	&0	&0	&0	&0	&1	&-1	&1	&-1	&1	&-1	&-1	&1 \\	
		0	&0	&0	&0	&0	&4	&-4&	0&	0&	0&	0&	-2&	2&	2&	-2&	-2&	2	&-2&	2&	1&	-1&	-1&	1&	1&	-1&	1&	-1 \\	
		4	&0	&0	&0	&0	&0	&0	&-1	&-1	&-1	&-1	&-1	&-1	&-1	&-1	&-1	&-1	&-1	&-1	&1	&1	&1	&1	&1	&1	&1	&1 \\	
		-8	&4	&4	&4	&4	&4	&4	&-2	&-2	&-2	&-2	&-2	&-2	&-2	&-2	&-2	&-2	&-2	&-2	&1	&1	&1	&1	&1	&1	&1	&1 \\	
		0	&-4	&-4	&2	&2	&2	&2	&1	&1	&1	&1	&-2	&-2	&-2	&-2	&1	&1	&1	&1	&0	&0	&0	&0	&0	&0	&0	&0 \\	
		0	&0	&0	&-2	&-2	&2	&2	&1	&1	&1	&1	&0	&0	&0	&0	&-1	&-1	&-1	&-1	&0	&0	&0	&0	&0	&0	&0	&0 \\	
		0	&0	&0	&0	&0	&0	&0	&-2	&-2	&2	&2	&0	&0	&0	&0	&0	&0	&0	&0	&1	&1	&1	&1	&-1	&-1	&-1	&-1 \\	
		0	&0	&0	&0	&0	&0	&0	&0	&0	&0	&0	&-2	&-2	&2	&2	&0	&0	&0	&0	&1	&1	&-1	&-1	&1	&1	&-1	&-1 \\	
		0	&0	&0	&0	&0	&0	&0	&0	&0	&0	&0	&0	&0	&0	&0	&-2	&-2	&2	&2	&1	&1	&-1	&-1	&-1	&-1	&1	&1 \\	
		0	&0	&0	&0	&0	&0	&0	&1	&-1	&1	&-1	&0	&0	&0	&0	&-1	&1	&1	&-1	&0	&0	&0	&0	&0	&0	&0	&0 \\	
		0	&0	&0	&0	&0	&0	&0	&-1	&1	&1	&-1	&1	&-1	&1	&-1	&0	&0	&0	&0	&0	&0	&0	&0	&0	&0	&0	&0 \\	
		0	&0	&0	&0	&0	&0	&0	&0	&0	&0	&0	&-1	&1	&1	&-1	&1	&-1	&1	&-1	&0	&0	&0	&0	&0	&0	&0	&0 \\	
		0	&0	&0	&0	&0	&0	&0	&0	&0	&0	&0	&0	&0	&0	&0	&0	&0	&0	&0	&1	&-1	&-1	&1	&-1	&1	&-1	&1 \\	
	\end{array}
	\right] $},
\end{equation}
and its inverse is analytically given by
\begin{equation}
\scalebox{0.66}{$\displaystyle
{\bf M}^{-1} = 
\frac{1}{216}
\left [
	\begin{array}{ccccccccccccccccccccccccccc}
		8	&0	&0	&0	&-24	&0	&0	&0	&0	&0	&0	&0	&0	&0	&0	&0	&24	&-8	&0	&0	&0	&0	&0	&0	&0	&0	&0 \\	
		8	&12	&0	&0	&-12	&12	&0	&0	&0	&0	&-12	&0	&0	&12	&0	&0	&0	&4	&-12	&0	&0	&0	&0	&0	&0	&0	&0 \\	
		8	&-12	&0	&0	&-12	&12	&0	&0	&0	&0	&12	&0	&0	&-12	&0	&0	&0	&4	&-12	&0	&0	&0	&0	&0	&0	&0	&0 \\	
		8	&0	&12	&0	&-12	&-6	&18	&0	&0	&0	&0	&-12	&0	&0	&12	&0	&0	&4	&6	&-18	&0	&0	&0	&0	&0	&0	&0 \\	
		8	&0	&-12	&0	&-12	&-6	&18	&0	&0	&0	&0	&12	&0	&0	&-12	&0	&0	&4	&6	&-18	&0	&0	&0	&0	&0	&0	&0 \\	
		8	&0	&0	&12	&-12	&-6	&-18	&0	&0	&0	&0	&0	&-12	&0	&0	&12	&0	&4	&6	&18	&0	&0	&0	&0	&0	&0	&0 \\	
		8	&0	&0	&-12	&-12	&-6	&-18	&0	&0	&0	&0	&0	&12	&0	&0	&-12	&0	&4	&6	&18	&0	&0	&0	&0	&0	&0	&0 \\	
		8	&12	&12	&0	&0	&6	&18	&18	&0	&0	&-3	&-3	&0	&-6	&-6	&0	&-6	&-2	&3	&9	&-18	&0	&0	&27	&-27	&0	&0 \\	
		8	&-12	&-12	&0	&0	&6	&18	&18	&0	&0	&3	&3	&0	&6	&6	&0	&-6	&-2	&3	&9	&-18	&0	&0	&-27	&27	&0	&0 \\	
		8	&12	&-12	&0	&0	&6	&18	&-18	&0	&0	&-3	&3	&0	&-6	&6	&0	&-6	&-2	&3	&9	&18	&0	&0	&27	&27	&0	&0 \\	
		8	&-12	&12	&0	&0	&6	&18	&-18	&0	&0	&3	&-3	&0	&6	&-6	&0	&-6	&-2	&3	&9	&18	&0	&0	&-27	&-27	&0	&0 \\	
		8	&0	&12	&12	&0	&-12	&0	&0	&18	&0	&0	&-3	&-3	&0	&-6	&-6	&-6	&-2	&-6	&0	&0	&-18	&0	&0	&27	&-27	&0 \\	
		8	&0	&-12	&-12	&0	&-12	&0	&0	&18	&0	&0	&3	&3	&0	&6	&6	&-6	&-2	&-6	&0	&0	&-18	&0	&0	&-27	&27	&0 \\	
		8	&0	&12	&-12	&0	&-12	&0	&0	&-18	&0	&0	&-3	&3	&0	&-6	&6	&-6	&-2	&-6	&0	&0	&18	&0	&0	&27	&27	&0 \\	
		8	&0	&-12	&12	&0	&-12	&0	&0	&-18	&0	&0	&3	&-3	&0	&6	&-6	&-6	&-2	&-6	&0	&0	&18	&0	&0	&-27	&-27	&0 \\	
		8	&12	&0	&12	&0	&6	&-18	&0	&0	&18	&-3	&0	&-3	&-6	&0	&-6	&-6	&-2	&3	&-9	&0	&0	&-18	&-27	&0	&27	&0 \\	
		8	&-12	&0	&-12	&0	&6	&-18	&0	&0	&18	&3	&0	&3	&6	&0	&6	&-6	&-2	&3	&-9	&0	&0	&-18	&27	&0	&-27	&0 \\	
		8	&-12	&0	&12	&0	&6	&-18	&0	&0	&-18	&3	&0	&-3	&6	&0	&-6	&-6	&-2	&3	&-9	&0	&0	&18	&27	&0	&27	&0 \\	
		8	&12	&0	&-12	&0	&6	&-18	&0	&0	&-18	&-3	&0	&3	&-6	&0	&6	&-6	&-2	&3	&-9	&0	&0	&18	&-27	&0	&-27	&0 \\	
		8	&12	&12	&12	&12	&0	&0	&18	&18	&18	&6	&6	&6	&3	&3	&3	&6	&1	&0	&0	&9	&9	&9	&0	&0	&0	&27 \\	
		8	&-12	&-12	&-12	&12	&0	&0	&18	&18	&18	&-6	&-6	&-6	&-3	&-3	&-3	&6	&1	&0	&0	&9	&9	&9	&0	&0	&0	&-27 \\	
		8	&12	&12	&-12	&12	&0	&0	&18	&-18	&-18	&6	&6	&-6	&3	&3	&-3	&6	&1	&0	&0	&9	&-9	&-9&	0	&0	&0	&-27 \\	
		8	&-12	&-12	&12	&12	&0	&0	&18	&-18	&-18	&-6	&-6	&6	&-3	&-3	&3	&6	&1	&0	&0	&9	&-9	&-9	&0	&0	&0	&27 \\	
		8	&-12	&12	&12	&12	&0	&0	&-18	&18	&-18	&-6	&6	&6	&-3	&3	&3	&6	&1	&0	&0	&-9	&9	&-9	&0	&0	&0	&-27 \\	
		8	&12	&-12	&-12	&12	&0	&0	&-18	&18	&-18	&6	&-6	&-6	&3	&-3	&-3	&6	&1	&0	&0	&-9	&9	&-9	&0	&0	&0	&27 \\	
		8	&12	&-12	&12	&12	&0	&0	&-18	&-18	&18	&6	&-6	&6	&3	&-3	&3	&6	&1	&0	&0	&-9	&-9	&9	&0	&0	&0	&-27 \\	
		8	&-12	&12	&-12	&12	&0	&0	&-18	&-18	&18	&-6	&6	&-6	&-3	&3	&-3	&6	&1	&0	&0	&-9	&-9	&9	&0	&0	&0	&27 \\	
	\end{array}
	\right]  $},
\end{equation}
\end{widetext}
where $c=1$ is assumed.

\section{\label{sec:appB} Derivation of lattice-specific coefficients}

	The lattice-specific coefficients for the D3Q27 lattice are $\varphi_i^k$ in Eq.~(\ref{eq:varphi_ik}), $\Phi_i^k$ in Eq.~(\ref{eq:Phi}), and $B_i$ in Eq.~(\ref{eq:B_i}).
Derivations of these coefficients are provided in this appendix.

First, we focus on $\varphi_i^k$ in the equilibrium function Eq.~(\ref{eq:EES}).
The equilibrium function $f_i^{k(e)}$ can be chosen arbitrarily to satisfy the conservations of mass and momentum:
\begin{align}
	\rho &= \sum_i{\sum_k f_i^{k(e)} }, \label{eq:massEq}
	\\
	\rho {\bf u} &= \sum_i{\sum_k f_i^{k(e)} {\bf c}_i}. \label{eq:momentumEq}
\end{align}
The form of target equation here is
\begin{align}
	&f_i^{k(e)} \nonumber
	\\
	&= \rho_k \left(\varphi_i^k 
	+ w_i \left[\frac{3}{c^2} ({\bf c}_i \cdot {\bf u})
		+ \frac{9}{2c^4}({\bf c}_i \cdot {\bf u})^2
		- \frac{3}{2c^2}{\bf u}^2 \right] \right).
\end{align}
To satisfy Eqs.~(\ref{eq:massEq}) and (\ref{eq:momentumEq}), the required condition for $\varphi_i^k$ is
\begin{equation}
	\sum_i \varphi_i^k = 1.
\end{equation}
As in \citep{Reis2007}, we assume
\begin{equation}
	\varphi_0^k = \alpha_k,
\end{equation}
and
\begin{equation}
	\frac{\varphi_{2-7}^k}{\varphi_{8-19}^k} = \frac{\varphi_{8-19}^k}{\varphi_{20-27}^k} =  r,
\end{equation}
where $r$ is some constant. 
By choosing $r=4$ from the relations of weight factors $w_i$ (i.e., $w_{2-7}/w_{8-19}=w_{8-19}/w_{20-27}=4$), one can obtain the following form
\begin{equation}
	\varphi_i^k = 
	\begin{cases}
		~\alpha_k, & i=1 ,\\
		~2(1-\alpha_k)/19, & i = 2,3,\dots,7, \\
		~(1-\alpha_k)/38, & i = 8,9,\dots,19, \\
		~(1-\alpha_k)/152, & i = 20,21,\dots,27.
\end{cases}
\end{equation}

	Next, we derive coefficients of the additional term $\Phi_i^k$ in Eq.~(\ref{eq:EES}).
Let us consider the conservation of color-blinded variables $\Phi_i = \sum_k \Phi_i^k$ for simplicity.
From the conservation of the 0-th to third order moments in the equilibrium functions, $f_i^{(e)}=\sum_k f_i^{k(e)}$,
the relations to be satisfied are as follows~\citep{Nourgaliev2003,Leclaire2013}
\begin{align}
	&\sum_i \Phi_i = 0, \label{eq:Phi0th}
	\\
	&\sum_i \Phi_i {\bf c}_i = 0, 
	\\
	&\sum_i \Phi_i {\bf c}_i {\bf c}_i = \bar{\nu} [{\bf u}\otimes \nabla \rho + ({\bf u}\otimes \nabla \rho)^{\rm T} + {\bf u}\cdot (\nabla \rho){\bf I}], 
	\\
	&\sum_i \Phi_i {\bf c}_i {\bf c}_i {\bf c}_i = 0. \label{eq:Phi3rd}
\end{align}
To satisfy Eqs.~(\ref{eq:Phi0th})--(\ref{eq:Phi3rd}), we use a form similar to \citep{Leclaire2013}:
\begin{equation}
	\Phi_i = 
	\begin{cases}
		~A_i \bar{\nu}({\bf u}\cdot \nabla \rho_k)/c, & i=1 ,\\
		~A_i \bar{\nu}({\bf G} : {\bf c}_i \otimes {\bf c}_i)/c^3, & {\rm others}, 
\end{cases}
\end{equation}
with 
\begin{equation}
	{\bf G} = C \left[{\bf u}\otimes \nabla \rho 
	+ ({\bf u}\otimes \nabla \rho)^{{\rm T}} \right],
\end{equation}
where $A_i$ are the lattice-specific coefficients to be determined and $C$ is the arbitrary constants.
	We assume 
\begin{equation}
	\frac{A_{2-7}}{A_{8-19}} = \frac{A_{8-19}}{A_{20-27}} =  r,
\end{equation}
with $r=4$.
The resultant form is
\begin{equation}
	\Phi_i = 
	\begin{cases}
		~-3 \bar{\nu}({\bf u}\cdot \nabla \rho)/c, & i=1 ,\\
		~+16\bar{\nu}({\bf G} : {\bf c}_i \otimes {\bf c}_i)/c^3, & i = 2,3,\dots,7, \\
		~+4 \bar{\nu}({\bf G} : {\bf c}_i \otimes {\bf c}_i)/c^3, & i = 8,9,\dots,19, \\
		~+1 \bar{\nu}({\bf G} : {\bf c}_i \otimes {\bf c}_i)/c^3, & i = 20,21,\dots,27,
\end{cases}
\end{equation}
with
\begin{equation}
	{\bf G} = \frac{1}{48} \left[{\bf u}\otimes \nabla \rho
	+ ({\bf u}\otimes \nabla \rho)^{{\rm T}} \right].
\end{equation}

	Finally, we move on $B_i$ in Eq.~(\ref{eq:B_i}).
	Following Ref.~\citep{Liu2012}, the conditions for $B_i$ to be satisfied are
\begin{align}
	&\sum_i B_i = \frac{1}{3}c^2, \label{eq:Bi_0th}
	\\
	&\sum_i {B_i {\bf c}_i} = 0, 
	\\
	&\sum_i {B_i {\bf c}_i {\bf c}_i} = \frac{1}{3}c^4 {\bf I}. \label{eq:Bi_2nd}
\end{align}
	In addition to the relations~(\ref{eq:Bi_0th})--(\ref{eq:Bi_2nd}), we assume
\begin{equation}
	\frac{B_{2-7}}{B_{8-19}} = \frac{B_{8-19}}{B_{20-27}} =  r,
\end{equation}
with $r=4$.
	One can choose the following coefficients
\begin{equation}
	B_i = 
	\begin{cases}
		~-(10/27)c^2, & i=1 ,\\
		~+(2/27)c^2, & i = 2,3,\dots,7, \\
		~+(1/54)c^2, & i = 8,9,\dots,19, \\
		~+(1/216)c^2, & i = 20,21,\dots,27, \label{eq:B_i2}
\end{cases}
\end{equation} 
 	to satisfy Eqs.~(\ref{eq:Bi_0th})--(\ref{eq:Bi_2nd}).
	Note that the choice of coefficients in Eq.~(\ref{eq:B_i2}) is somewhat arbitrary as in \citep{Liu2012}.

\bibliography{refs}

\begin{thebibliography}{89}%
\makeatletter
\providecommand \@ifxundefined [1]{%
 \@ifx{#1\undefined}
}%
\providecommand \@ifnum [1]{%
 \ifnum #1\expandafter \@firstoftwo
 \else \expandafter \@secondoftwo
 \fi
}%
\providecommand \@ifx [1]{%
 \ifx #1\expandafter \@firstoftwo
 \else \expandafter \@secondoftwo
 \fi
}%
\providecommand \natexlab [1]{#1}%
\providecommand \enquote  [1]{``#1''}%
\providecommand \bibnamefont  [1]{#1}%
\providecommand \bibfnamefont [1]{#1}%
\providecommand \citenamefont [1]{#1}%
\providecommand \href@noop [0]{\@secondoftwo}%
\providecommand \href [0]{\begingroup \@sanitize@url \@href}%
\providecommand \@href[1]{\@@startlink{#1}\@@href}%
\providecommand \@@href[1]{\endgroup#1\@@endlink}%
\providecommand \@sanitize@url [0]{\catcode `\\12\catcode `\$12\catcode
  `\&12\catcode `\#12\catcode `\^12\catcode `\_12\catcode `\%12\relax}%
\providecommand \@@startlink[1]{}%
\providecommand \@@endlink[0]{}%
\providecommand \url  [0]{\begingroup\@sanitize@url \@url }%
\providecommand \@url [1]{\endgroup\@href {#1}{\urlprefix }}%
\providecommand \urlprefix  [0]{URL }%
\providecommand \Eprint [0]{\href }%
\providecommand \doibase [0]{http://dx.doi.org/}%
\providecommand \selectlanguage [0]{\@gobble}%
\providecommand \bibinfo  [0]{\@secondoftwo}%
\providecommand \bibfield  [0]{\@secondoftwo}%
\providecommand \translation [1]{[#1]}%
\providecommand \BibitemOpen [0]{}%
\providecommand \bibitemStop [0]{}%
\providecommand \bibitemNoStop [0]{.\EOS\space}%
\providecommand \EOS [0]{\spacefactor3000\relax}%
\providecommand \BibitemShut  [1]{\csname bibitem#1\endcsname}%
\let\auto@bib@innerbib\@empty
\bibitem [{\citenamefont {Plateau}(1873)}]{Plateau1873}%
  \BibitemOpen
  \bibfield  {author} {\bibinfo {author} {\bibfnamefont {J.}~\bibnamefont
  {Plateau}},\ }\href
  {http://www.susqu.edu/brakke/aux/downloads/plateau-fr.pdf} {\emph {\bibinfo
  {title} {Gauthier-Villars}}}\ (\bibinfo {address} {Paris},\ \bibinfo {year}
  {1873})\BibitemShut {NoStop}%
\bibitem [{\citenamefont {Rayleigh}(1878)}]{Rayleigh1878}%
  \BibitemOpen
  \bibfield  {author} {\bibinfo {author} {\bibfnamefont {L.}~\bibnamefont
  {Rayleigh}},\ }\href {\doibase 10.1112/plms/s1-10.1.4} {\bibfield  {journal}
  {\bibinfo  {journal} {Proc. London Math. Soc.}\ }\textbf {\bibinfo {volume}
  {s1-10}},\ \bibinfo {pages} {4} (\bibinfo {year} {1878})}\BibitemShut
  {NoStop}%
\bibitem [{\citenamefont {McCarthy}\ and\ \citenamefont
  {Molloy}(1974)}]{McCarthy1974}%
  \BibitemOpen
  \bibfield  {author} {\bibinfo {author} {\bibfnamefont {M.~J.}\ \bibnamefont
  {McCarthy}}\ and\ \bibinfo {author} {\bibfnamefont {N.~A.}\ \bibnamefont
  {Molloy}},\ }\href {\doibase 10.1016/0300-9467(74)80021-3} {\bibfield
  {journal} {\bibinfo  {journal} {Chem. Eng. J.}\ }\textbf {\bibinfo {volume}
  {7}},\ \bibinfo {pages} {1} (\bibinfo {year} {1974})}\BibitemShut {NoStop}%
\bibitem [{\citenamefont {Lin}\ and\ \citenamefont {Reitz}(1998)}]{Lin1998}%
  \BibitemOpen
  \bibfield  {author} {\bibinfo {author} {\bibfnamefont {S.~P.}\ \bibnamefont
  {Lin}}\ and\ \bibinfo {author} {\bibfnamefont {R.~D.}\ \bibnamefont
  {Reitz}},\ }\href {\doibase 10.1146/annurev.fluid.30.1.85} {\bibfield
  {journal} {\bibinfo  {journal} {Annu. Rev. Fluid Mech.}\ }\textbf {\bibinfo
  {volume} {30}},\ \bibinfo {pages} {85} (\bibinfo {year} {1998})}\BibitemShut
  {NoStop}%
\bibitem [{\citenamefont {Villermaux}(2007)}]{Villermaux2007}%
  \BibitemOpen
  \bibfield  {author} {\bibinfo {author} {\bibfnamefont {E.}~\bibnamefont
  {Villermaux}},\ }\href {\doibase 10.1146/annurev.fluid.39.050905.110214}
  {\bibfield  {journal} {\bibinfo  {journal} {Annu. Rev. Fluid Mech.}\ }\textbf
  {\bibinfo {volume} {39}},\ \bibinfo {pages} {419} (\bibinfo {year}
  {2007})}\BibitemShut {NoStop}%
\bibitem [{\citenamefont {Eggers}\ and\ \citenamefont
  {Villermaux}(2008)}]{Eggers2008}%
  \BibitemOpen
  \bibfield  {author} {\bibinfo {author} {\bibfnamefont {J.}~\bibnamefont
  {Eggers}}\ and\ \bibinfo {author} {\bibfnamefont {E.}~\bibnamefont
  {Villermaux}},\ }\href {\doibase 10.1088/0034-4885/71/3/036601} {\bibfield
  {journal} {\bibinfo  {journal} {Rep. Prog. Phys.}\ }\textbf {\bibinfo
  {volume} {71}},\ \bibinfo {pages} {036601} (\bibinfo {year}
  {2008})}\BibitemShut {NoStop}%
\bibitem [{\citenamefont {Ohnesorge}(1936)}]{Ohnesorge1936}%
  \BibitemOpen
  \bibfield  {author} {\bibinfo {author} {\bibfnamefont {W.~v.}\ \bibnamefont
  {Ohnesorge}},\ }\href {\doibase 10.1002/zamm.19360160611} {\bibfield
  {journal} {\bibinfo  {journal} {J. Appl. Math. Mech.}\ }\textbf {\bibinfo
  {volume} {16}},\ \bibinfo {pages} {355} (\bibinfo {year} {1936})}\BibitemShut
  {NoStop}%
\bibitem [{\citenamefont {Kolev}(2005)}]{Kolev2005}%
  \BibitemOpen
  \bibfield  {author} {\bibinfo {author} {\bibfnamefont {N.~I.}\ \bibnamefont
  {Kolev}},\ }\href {\doibase 10.1007/b138146} {\emph {\bibinfo {title}
  {{Multiphase Flow Dynamics 2}}}},\ \bibinfo {edition} {3rd}\ ed.\ (\bibinfo
  {publisher} {Springer-Verlag},\ \bibinfo {address} {Berlin/Heidelberg},\
  \bibinfo {year} {2005})\BibitemShut {NoStop}%
\bibitem [{\citenamefont {McKinley}\ and\ \citenamefont
  {Renardy}(2011)}]{McKinley2011}%
  \BibitemOpen
  \bibfield  {author} {\bibinfo {author} {\bibfnamefont {G.~H.}\ \bibnamefont
  {McKinley}}\ and\ \bibinfo {author} {\bibfnamefont {M.}~\bibnamefont
  {Renardy}},\ }\href {\doibase 10.1063/1.3663616} {\bibfield  {journal}
  {\bibinfo  {journal} {Phys. Fluids}\ }\textbf {\bibinfo {volume} {23}},\
  \bibinfo {pages} {127101} (\bibinfo {year} {2011})}\BibitemShut {NoStop}%
\bibitem [{\citenamefont {Merrington}\ and\ \citenamefont
  {Richardson}(1947)}]{Merrington1947}%
  \BibitemOpen
  \bibfield  {author} {\bibinfo {author} {\bibfnamefont {A.~C.}\ \bibnamefont
  {Merrington}}\ and\ \bibinfo {author} {\bibfnamefont {E.~G.}\ \bibnamefont
  {Richardson}},\ }\href {\doibase 10.1088/0959-5309/59/1/302} {\bibfield
  {journal} {\bibinfo  {journal} {Proc. Phys. Soc.}\ }\textbf {\bibinfo
  {volume} {59}},\ \bibinfo {pages} {1} (\bibinfo {year} {1947})}\BibitemShut
  {NoStop}%
\bibitem [{\citenamefont {Tanasawa}\ and\ \citenamefont
  {Toyoda}(1954)}]{Tanasawa1954}%
  \BibitemOpen
  \bibfield  {author} {\bibinfo {author} {\bibfnamefont {Y.}~\bibnamefont
  {Tanasawa}}\ and\ \bibinfo {author} {\bibfnamefont {S.}~\bibnamefont
  {Toyoda}},\ }\href {\doibase 10.1299/kikai1938.20.306} {\bibfield  {journal}
  {\bibinfo  {journal} {Trans. JSME}\ }\textbf {\bibinfo {volume} {20}},\
  \bibinfo {pages} {306} (\bibinfo {year} {1954})}\BibitemShut {NoStop}%
\bibitem [{\citenamefont {Grant}\ and\ \citenamefont
  {Middleman}(1966)}]{Grant1966}%
  \BibitemOpen
  \bibfield  {author} {\bibinfo {author} {\bibfnamefont {R.~P.}\ \bibnamefont
  {Grant}}\ and\ \bibinfo {author} {\bibfnamefont {S.}~\bibnamefont
  {Middleman}},\ }\href {\doibase 10.1002/aic.690120411} {\bibfield  {journal}
  {\bibinfo  {journal} {AIChE J.}\ }\textbf {\bibinfo {volume} {12}},\ \bibinfo
  {pages} {669} (\bibinfo {year} {1966})}\BibitemShut {NoStop}%
\bibitem [{\citenamefont {Meister}\ and\ \citenamefont
  {Scheele}(1969)}]{Meister1969}%
  \BibitemOpen
  \bibfield  {author} {\bibinfo {author} {\bibfnamefont {B.~J.}\ \bibnamefont
  {Meister}}\ and\ \bibinfo {author} {\bibfnamefont {G.~F.}\ \bibnamefont
  {Scheele}},\ }\href {\doibase 10.1002/aic.690150513} {\bibfield  {journal}
  {\bibinfo  {journal} {AIChE J.}\ }\textbf {\bibinfo {volume} {15}},\ \bibinfo
  {pages} {700} (\bibinfo {year} {1969})}\BibitemShut {NoStop}%
\bibitem [{\citenamefont {Takahashi}\ and\ \citenamefont
  {Kitamura}(1971)}]{Takahashi1971}%
  \BibitemOpen
  \bibfield  {author} {\bibinfo {author} {\bibfnamefont {T.}~\bibnamefont
  {Takahashi}}\ and\ \bibinfo {author} {\bibfnamefont {Y.}~\bibnamefont
  {Kitamura}},\ }\href {\doibase 10.1252/kakoronbunshu1953.35.637} {\bibfield
  {journal} {\bibinfo  {journal} {Chem. eng.}\ }\textbf {\bibinfo {volume}
  {35}},\ \bibinfo {pages} {637} (\bibinfo {year} {1971})}\BibitemShut
  {NoStop}%
\bibitem [{\citenamefont {Das}(1997)}]{Das1997}%
  \BibitemOpen
  \bibfield  {author} {\bibinfo {author} {\bibfnamefont {T.~K.}\ \bibnamefont
  {Das}},\ }\href {\doibase 10.1615/AtomizSpr.v7.i5.70} {\bibfield  {journal}
  {\bibinfo  {journal} {Atomization Sprays}\ }\textbf {\bibinfo {volume} {7}},\
  \bibinfo {pages} {549} (\bibinfo {year} {1997})}\BibitemShut {NoStop}%
\bibitem [{\citenamefont {Riestenberg}\ \emph {et~al.}(2004)\citenamefont
  {Riestenberg}, \citenamefont {Chiu}, \citenamefont {Gborigi}, \citenamefont
  {Liang}, \citenamefont {West},\ and\ \citenamefont
  {Tsouris}}]{Riestenberg2004}%
  \BibitemOpen
  \bibfield  {author} {\bibinfo {author} {\bibfnamefont {D.}~\bibnamefont
  {Riestenberg}}, \bibinfo {author} {\bibfnamefont {E.}~\bibnamefont {Chiu}},
  \bibinfo {author} {\bibfnamefont {M.}~\bibnamefont {Gborigi}}, \bibinfo
  {author} {\bibfnamefont {L.}~\bibnamefont {Liang}}, \bibinfo {author}
  {\bibfnamefont {O.~R.}\ \bibnamefont {West}}, \ and\ \bibinfo {author}
  {\bibfnamefont {C.}~\bibnamefont {Tsouris}},\ }\href {\doibase
  10.2138/am-2004-8-911} {\bibfield  {journal} {\bibinfo  {journal} {Am.
  Mineral.}\ }\textbf {\bibinfo {volume} {89}},\ \bibinfo {pages} {1240}
  (\bibinfo {year} {2004})}\BibitemShut {NoStop}%
\bibitem [{\citenamefont {Tsouris}\ \emph {et~al.}(2007)\citenamefont
  {Tsouris}, \citenamefont {McCallum}, \citenamefont {Aaron}, \citenamefont
  {Riestenberg}, \citenamefont {Gabitto}, \citenamefont {Chow},\ and\
  \citenamefont {Adams}}]{Tsouris2007}%
  \BibitemOpen
  \bibfield  {author} {\bibinfo {author} {\bibfnamefont {C.}~\bibnamefont
  {Tsouris}}, \bibinfo {author} {\bibfnamefont {S.}~\bibnamefont {McCallum}},
  \bibinfo {author} {\bibfnamefont {D.}~\bibnamefont {Aaron}}, \bibinfo
  {author} {\bibfnamefont {D.}~\bibnamefont {Riestenberg}}, \bibinfo {author}
  {\bibfnamefont {J.}~\bibnamefont {Gabitto}}, \bibinfo {author} {\bibfnamefont
  {A.}~\bibnamefont {Chow}}, \ and\ \bibinfo {author} {\bibfnamefont
  {E.}~\bibnamefont {Adams}},\ }\href {\doibase 10.1002/aic.11117} {\bibfield
  {journal} {\bibinfo  {journal} {AIChE J.}\ }\textbf {\bibinfo {volume}
  {53}},\ \bibinfo {pages} {1017} (\bibinfo {year} {2007})}\BibitemShut
  {NoStop}%
\bibitem [{\citenamefont {Kondo}\ \emph {et~al.}(1995)\citenamefont {Kondo},
  \citenamefont {Konishi}, \citenamefont {Isozaki}, \citenamefont {Imahori},
  \citenamefont {Furutani},\ and\ \citenamefont {Brear}}]{Kondo1995}%
  \BibitemOpen
  \bibfield  {author} {\bibinfo {author} {\bibfnamefont {S.}~\bibnamefont
  {Kondo}}, \bibinfo {author} {\bibfnamefont {K.}~\bibnamefont {Konishi}},
  \bibinfo {author} {\bibfnamefont {M.}~\bibnamefont {Isozaki}}, \bibinfo
  {author} {\bibfnamefont {S.}~\bibnamefont {Imahori}}, \bibinfo {author}
  {\bibfnamefont {A.}~\bibnamefont {Furutani}}, \ and\ \bibinfo {author}
  {\bibfnamefont {D.}~\bibnamefont {Brear}},\ }\href {\doibase
  10.1016/0029-5493(94)00870-5} {\bibfield  {journal} {\bibinfo  {journal}
  {Nucl. Eng. Des.}\ }\textbf {\bibinfo {volume} {155}},\ \bibinfo {pages} {73}
  (\bibinfo {year} {1995})}\BibitemShut {NoStop}%
\bibitem [{\citenamefont {Dinh}\ \emph {et~al.}(1999)\citenamefont {Dinh},
  \citenamefont {Bui}, \citenamefont {Nourgaliev}, \citenamefont {Green},\ and\
  \citenamefont {Sehgal}}]{Dinh1999}%
  \BibitemOpen
  \bibfield  {author} {\bibinfo {author} {\bibfnamefont {T.}~\bibnamefont
  {Dinh}}, \bibinfo {author} {\bibfnamefont {V.}~\bibnamefont {Bui}}, \bibinfo
  {author} {\bibfnamefont {R.}~\bibnamefont {Nourgaliev}}, \bibinfo {author}
  {\bibfnamefont {J.}~\bibnamefont {Green}}, \ and\ \bibinfo {author}
  {\bibfnamefont {B.}~\bibnamefont {Sehgal}},\ }\href {\doibase
  10.1016/S0029-5493(98)00275-1} {\bibfield  {journal} {\bibinfo  {journal}
  {Nucl. Eng. Des.}\ }\textbf {\bibinfo {volume} {189}},\ \bibinfo {pages}
  {299} (\bibinfo {year} {1999})}\BibitemShut {NoStop}%
\bibitem [{\citenamefont {Abe}\ \emph {et~al.}(2006)\citenamefont {Abe},
  \citenamefont {Matsuo}, \citenamefont {Arai}, \citenamefont {Nariai},
  \citenamefont {Chitose}, \citenamefont {Koyama},\ and\ \citenamefont
  {Itoh}}]{Abe2006}%
  \BibitemOpen
  \bibfield  {author} {\bibinfo {author} {\bibfnamefont {Y.}~\bibnamefont
  {Abe}}, \bibinfo {author} {\bibfnamefont {E.}~\bibnamefont {Matsuo}},
  \bibinfo {author} {\bibfnamefont {T.}~\bibnamefont {Arai}}, \bibinfo {author}
  {\bibfnamefont {H.}~\bibnamefont {Nariai}}, \bibinfo {author} {\bibfnamefont
  {K.}~\bibnamefont {Chitose}}, \bibinfo {author} {\bibfnamefont
  {K.}~\bibnamefont {Koyama}}, \ and\ \bibinfo {author} {\bibfnamefont
  {K.}~\bibnamefont {Itoh}},\ }\href {\doibase 10.1016/j.nucengdes.2006.04.008}
  {\bibfield  {journal} {\bibinfo  {journal} {Nucl. Eng. Des.}\ }\textbf
  {\bibinfo {volume} {236}},\ \bibinfo {pages} {1668} (\bibinfo {year}
  {2006})}\BibitemShut {NoStop}%
\bibitem [{\citenamefont {Chandrasekhar}(1961)}]{Chandrasekhar1961}%
  \BibitemOpen
  \bibfield  {author} {\bibinfo {author} {\bibfnamefont {S.}~\bibnamefont
  {Chandrasekhar}},\ }\href@noop {} {\emph {\bibinfo {title} {{Hydrodynamic and
  hydromagnetic stability}}}}\ (\bibinfo  {publisher} {Oxford University
  Press},\ \bibinfo {address} {New York},\ \bibinfo {year} {1961})\BibitemShut
  {NoStop}%
\bibitem [{\citenamefont {Saito}\ \emph {et~al.}(2017)\citenamefont {Saito},
  \citenamefont {Abe},\ and\ \citenamefont {Koyama}}]{Saito2017}%
  \BibitemOpen
  \bibfield  {author} {\bibinfo {author} {\bibfnamefont {S.}~\bibnamefont
  {Saito}}, \bibinfo {author} {\bibfnamefont {Y.}~\bibnamefont {Abe}}, \ and\
  \bibinfo {author} {\bibfnamefont {K.}~\bibnamefont {Koyama}},\ }\href
  {\doibase 10.1016/j.nucengdes.2017.02.011} {\bibfield  {journal} {\bibinfo
  {journal} {Nucl. Eng. Des.}\ }\textbf {\bibinfo {volume} {315}},\ \bibinfo
  {pages} {128} (\bibinfo {year} {2017})}\BibitemShut {NoStop}%
\bibitem [{\citenamefont {Richards}\ \emph {et~al.}(1993)\citenamefont
  {Richards}, \citenamefont {Beris},\ and\ \citenamefont
  {Lenhoff}}]{Richards1993}%
  \BibitemOpen
  \bibfield  {author} {\bibinfo {author} {\bibfnamefont {J.~R.}\ \bibnamefont
  {Richards}}, \bibinfo {author} {\bibfnamefont {A.~N.}\ \bibnamefont {Beris}},
  \ and\ \bibinfo {author} {\bibfnamefont {A.~M.}\ \bibnamefont {Lenhoff}},\
  }\href {\doibase 10.1063/1.858847} {\bibfield  {journal} {\bibinfo  {journal}
  {Phys Fluids A: Fluid Dyn.}\ }\textbf {\bibinfo {volume} {5}},\ \bibinfo
  {pages} {1703} (\bibinfo {year} {1993})}\BibitemShut {NoStop}%
\bibitem [{\citenamefont {Hirt}\ and\ \citenamefont
  {Nichols}(1981)}]{Hirt1981}%
  \BibitemOpen
  \bibfield  {author} {\bibinfo {author} {\bibfnamefont {C.}~\bibnamefont
  {Hirt}}\ and\ \bibinfo {author} {\bibfnamefont {B.}~\bibnamefont {Nichols}},\
  }\href {\doibase 10.1016/0021-9991(81)90145-5} {\bibfield  {journal}
  {\bibinfo  {journal} {J. Comput. Phys.}\ }\textbf {\bibinfo {volume} {39}},\
  \bibinfo {pages} {201} (\bibinfo {year} {1981})}\BibitemShut {NoStop}%
\bibitem [{\citenamefont {Thakre}\ \emph {et~al.}(2015)\citenamefont {Thakre},
  \citenamefont {Manickam},\ and\ \citenamefont {Ma}}]{Thakre2015}%
  \BibitemOpen
  \bibfield  {author} {\bibinfo {author} {\bibfnamefont {S.}~\bibnamefont
  {Thakre}}, \bibinfo {author} {\bibfnamefont {L.}~\bibnamefont {Manickam}}, \
  and\ \bibinfo {author} {\bibfnamefont {W.}~\bibnamefont {Ma}},\ }\href
  {\doibase 10.1016/j.anucene.2015.02.038} {\bibfield  {journal} {\bibinfo
  {journal} {Ann. Nucl. Energy}\ }\textbf {\bibinfo {volume} {80}},\ \bibinfo
  {pages} {467} (\bibinfo {year} {2015})}\BibitemShut {NoStop}%
\bibitem [{\citenamefont {B{\"{u}}rger}\ \emph {et~al.}(1995)\citenamefont
  {B{\"{u}}rger}, \citenamefont {Cho}, \citenamefont {Berg},\ and\
  \citenamefont {Schatz}}]{Burger1995}%
  \BibitemOpen
  \bibfield  {author} {\bibinfo {author} {\bibfnamefont {M.}~\bibnamefont
  {B{\"{u}}rger}}, \bibinfo {author} {\bibfnamefont {S.}~\bibnamefont {Cho}},
  \bibinfo {author} {\bibfnamefont {E.}~\bibnamefont {Berg}}, \ and\ \bibinfo
  {author} {\bibfnamefont {A.}~\bibnamefont {Schatz}},\ }\href {\doibase
  10.1016/0029-5493(94)00875-Y} {\bibfield  {journal} {\bibinfo  {journal}
  {Nucl. Eng. Des.}\ }\textbf {\bibinfo {volume} {155}},\ \bibinfo {pages}
  {215} (\bibinfo {year} {1995})}\BibitemShut {NoStop}%
\bibitem [{\citenamefont {Saito}\ \emph {et~al.}(2015)\citenamefont {Saito},
  \citenamefont {Abe},\ and\ \citenamefont {Koyama}}]{Saito2015}%
  \BibitemOpen
  \bibfield  {author} {\bibinfo {author} {\bibfnamefont {S.}~\bibnamefont
  {Saito}}, \bibinfo {author} {\bibfnamefont {Y.}~\bibnamefont {Abe}}, \ and\
  \bibinfo {author} {\bibfnamefont {K.}~\bibnamefont {Koyama}},\ }\href
  {\doibase 10.1299/transjsme.15-00395} {\bibfield  {journal} {\bibinfo
  {journal} {Trans. JSME}\ }\textbf {\bibinfo {volume} {81}},\ \bibinfo {pages}
  {15} (\bibinfo {year} {2015})}\BibitemShut {NoStop}%
\bibitem [{\citenamefont {Homma}\ \emph {et~al.}(2006)\citenamefont {Homma},
  \citenamefont {Koga}, \citenamefont {Matsumoto}, \citenamefont {Song},\ and\
  \citenamefont {Tryggvason}}]{Homma2006}%
  \BibitemOpen
  \bibfield  {author} {\bibinfo {author} {\bibfnamefont {S.}~\bibnamefont
  {Homma}}, \bibinfo {author} {\bibfnamefont {J.}~\bibnamefont {Koga}},
  \bibinfo {author} {\bibfnamefont {S.}~\bibnamefont {Matsumoto}}, \bibinfo
  {author} {\bibfnamefont {M.}~\bibnamefont {Song}}, \ and\ \bibinfo {author}
  {\bibfnamefont {G.}~\bibnamefont {Tryggvason}},\ }\href {\doibase
  10.1016/j.ces.2006.01.029} {\bibfield  {journal} {\bibinfo  {journal} {Chem.
  Eng. Sci.}\ }\textbf {\bibinfo {volume} {61}},\ \bibinfo {pages} {3986}
  (\bibinfo {year} {2006})}\BibitemShut {NoStop}%
\bibitem [{\citenamefont {Unverdi}\ and\ \citenamefont
  {Tryggvason}(1992)}]{Unverdi1992}%
  \BibitemOpen
  \bibfield  {author} {\bibinfo {author} {\bibfnamefont {S.~O.}\ \bibnamefont
  {Unverdi}}\ and\ \bibinfo {author} {\bibfnamefont {G.}~\bibnamefont
  {Tryggvason}},\ }\href {\doibase 10.1016/0021-9991(92)90307-K} {\bibfield
  {journal} {\bibinfo  {journal} {J. Comput. Phys.}\ }\textbf {\bibinfo
  {volume} {100}},\ \bibinfo {pages} {25} (\bibinfo {year} {1992})}\BibitemShut
  {NoStop}%
\bibitem [{\citenamefont {Tryggvason}\ \emph {et~al.}(2001)\citenamefont
  {Tryggvason}, \citenamefont {Bunner}, \citenamefont {Esmaeeli}, \citenamefont
  {Juric}, \citenamefont {Al-Rawahi}, \citenamefont {Tauber}, \citenamefont
  {Han}, \citenamefont {Nas},\ and\ \citenamefont {Jan}}]{Tryggvason2001}%
  \BibitemOpen
  \bibfield  {author} {\bibinfo {author} {\bibfnamefont {G.}~\bibnamefont
  {Tryggvason}}, \bibinfo {author} {\bibfnamefont {B.}~\bibnamefont {Bunner}},
  \bibinfo {author} {\bibfnamefont {A.}~\bibnamefont {Esmaeeli}}, \bibinfo
  {author} {\bibfnamefont {D.}~\bibnamefont {Juric}}, \bibinfo {author}
  {\bibfnamefont {N.}~\bibnamefont {Al-Rawahi}}, \bibinfo {author}
  {\bibfnamefont {W.}~\bibnamefont {Tauber}}, \bibinfo {author} {\bibfnamefont
  {J.}~\bibnamefont {Han}}, \bibinfo {author} {\bibfnamefont {S.}~\bibnamefont
  {Nas}}, \ and\ \bibinfo {author} {\bibfnamefont {Y.-J.}\ \bibnamefont
  {Jan}},\ }\href {\doibase 10.1006/jcph.2001.6726} {\bibfield  {journal}
  {\bibinfo  {journal} {J. Comput. Phys.}\ }\textbf {\bibinfo {volume} {169}},\
  \bibinfo {pages} {708} (\bibinfo {year} {2001})}\BibitemShut {NoStop}%
\bibitem [{\citenamefont {McCracken}\ and\ \citenamefont
  {Abraham}(2005{\natexlab{a}})}]{McCracken2005a}%
  \BibitemOpen
  \bibfield  {author} {\bibinfo {author} {\bibfnamefont {M.~E.}\ \bibnamefont
  {McCracken}}\ and\ \bibinfo {author} {\bibfnamefont {J.}~\bibnamefont
  {Abraham}},\ }\href {\doibase 10.1142/S0129183105008291} {\bibfield
  {journal} {\bibinfo  {journal} {Int. J. Mod. Phys. C}\ }\textbf {\bibinfo
  {volume} {16}},\ \bibinfo {pages} {1671} (\bibinfo {year}
  {2005}{\natexlab{a}})}\BibitemShut {NoStop}%
\bibitem [{\citenamefont {Matsuo}\ \emph {et~al.}(2015)\citenamefont {Matsuo},
  \citenamefont {Abe}, \citenamefont {Iwasawa}, \citenamefont {Ebihara},\ and\
  \citenamefont {Koyama}}]{Matsuo2015}%
  \BibitemOpen
  \bibfield  {author} {\bibinfo {author} {\bibfnamefont {E.}~\bibnamefont
  {Matsuo}}, \bibinfo {author} {\bibfnamefont {Y.}~\bibnamefont {Abe}},
  \bibinfo {author} {\bibfnamefont {Y.}~\bibnamefont {Iwasawa}}, \bibinfo
  {author} {\bibfnamefont {K.}~\bibnamefont {Ebihara}}, \ and\ \bibinfo
  {author} {\bibfnamefont {K.}~\bibnamefont {Koyama}},\ }\href {\doibase
  10.1299/transjsme.14-00409} {\bibfield  {journal} {\bibinfo  {journal}
  {Trans. JSME}\ }\textbf {\bibinfo {volume} {81}},\ \bibinfo {pages} {1}
  (\bibinfo {year} {2015})}\BibitemShut {NoStop}%
\bibitem [{\citenamefont {Saito}\ \emph {et~al.}(2016)\citenamefont {Saito},
  \citenamefont {Abe}, \citenamefont {Kaneko}, \citenamefont {Kanagawa},
  \citenamefont {Iwasawa},\ and\ \citenamefont {Koyama}}]{Saito2016}%
  \BibitemOpen
  \bibfield  {author} {\bibinfo {author} {\bibfnamefont {S.}~\bibnamefont
  {Saito}}, \bibinfo {author} {\bibfnamefont {Y.}~\bibnamefont {Abe}}, \bibinfo
  {author} {\bibfnamefont {A.}~\bibnamefont {Kaneko}}, \bibinfo {author}
  {\bibfnamefont {T.}~\bibnamefont {Kanagawa}}, \bibinfo {author}
  {\bibfnamefont {Y.}~\bibnamefont {Iwasawa}}, \ and\ \bibinfo {author}
  {\bibfnamefont {K.}~\bibnamefont {Koyama}},\ }\href {\doibase
  10.3811/jjmf.29.433} {\bibfield  {journal} {\bibinfo  {journal} {Jpn. J.
  Multiphase Flow}\ }\textbf {\bibinfo {volume} {29}},\ \bibinfo {pages} {433}
  (\bibinfo {year} {2016})}\BibitemShut {NoStop}%
\bibitem [{\citenamefont {Aidun}\ and\ \citenamefont
  {Clausen}(2010)}]{Aidun2010}%
  \BibitemOpen
  \bibfield  {author} {\bibinfo {author} {\bibfnamefont {C.~K.}\ \bibnamefont
  {Aidun}}\ and\ \bibinfo {author} {\bibfnamefont {J.~R.}\ \bibnamefont
  {Clausen}},\ }\href {\doibase 10.1146/annurev-fluid-121108-145519} {\bibfield
   {journal} {\bibinfo  {journal} {Annu. Rev. Fluid Mech.}\ }\textbf {\bibinfo
  {volume} {42}},\ \bibinfo {pages} {439} (\bibinfo {year} {2010})}\BibitemShut
  {NoStop}%
\bibitem [{\citenamefont {Gunstensen}\ \emph {et~al.}(1991)\citenamefont
  {Gunstensen}, \citenamefont {Rothman}, \citenamefont {Zaleski},\ and\
  \citenamefont {Zanetti}}]{Gunstensen1991}%
  \BibitemOpen
  \bibfield  {author} {\bibinfo {author} {\bibfnamefont {A.~K.}\ \bibnamefont
  {Gunstensen}}, \bibinfo {author} {\bibfnamefont {D.~H.}\ \bibnamefont
  {Rothman}}, \bibinfo {author} {\bibfnamefont {S.}~\bibnamefont {Zaleski}}, \
  and\ \bibinfo {author} {\bibfnamefont {G.}~\bibnamefont {Zanetti}},\ }\href
  {\doibase 10.1103/PhysRevA.43.4320} {\bibfield  {journal} {\bibinfo
  {journal} {Phys. Rev. A}\ }\textbf {\bibinfo {volume} {43}},\ \bibinfo
  {pages} {4320} (\bibinfo {year} {1991})}\BibitemShut {NoStop}%
\bibitem [{\citenamefont {Grunau}\ \emph {et~al.}(1993)\citenamefont {Grunau},
  \citenamefont {Chen},\ and\ \citenamefont {Eggert}}]{Grunau1993}%
  \BibitemOpen
  \bibfield  {author} {\bibinfo {author} {\bibfnamefont {D.}~\bibnamefont
  {Grunau}}, \bibinfo {author} {\bibfnamefont {S.}~\bibnamefont {Chen}}, \ and\
  \bibinfo {author} {\bibfnamefont {K.}~\bibnamefont {Eggert}},\ }\href
  {\doibase 10.1063/1.858769} {\bibfield  {journal} {\bibinfo  {journal} {Phys.
  Fluids A: Fluid Dyn.}\ }\textbf {\bibinfo {volume} {5}},\ \bibinfo {pages}
  {2557} (\bibinfo {year} {1993})}\BibitemShut {NoStop}%
\bibitem [{\citenamefont {Shan}\ and\ \citenamefont {Chen}(1993)}]{Shan1993}%
  \BibitemOpen
  \bibfield  {author} {\bibinfo {author} {\bibfnamefont {X.}~\bibnamefont
  {Shan}}\ and\ \bibinfo {author} {\bibfnamefont {H.}~\bibnamefont {Chen}},\
  }\href {\doibase 10.1103/PhysRevE.47.1815} {\bibfield  {journal} {\bibinfo
  {journal} {Phys. Rev. E}\ }\textbf {\bibinfo {volume} {47}},\ \bibinfo
  {pages} {1815} (\bibinfo {year} {1993})}\BibitemShut {NoStop}%
\bibitem [{\citenamefont {Shan}\ and\ \citenamefont {Chen}(1994)}]{Shan1994}%
  \BibitemOpen
  \bibfield  {author} {\bibinfo {author} {\bibfnamefont {X.}~\bibnamefont
  {Shan}}\ and\ \bibinfo {author} {\bibfnamefont {H.}~\bibnamefont {Chen}},\
  }\href {\doibase 10.1103/PhysRevE.49.2941} {\bibfield  {journal} {\bibinfo
  {journal} {Phys. Rev. E}\ }\textbf {\bibinfo {volume} {49}},\ \bibinfo
  {pages} {2941} (\bibinfo {year} {1994})}\BibitemShut {NoStop}%
\bibitem [{\citenamefont {Swift}\ \emph {et~al.}(1995)\citenamefont {Swift},
  \citenamefont {Osborn},\ and\ \citenamefont {Yeomans}}]{Swift1995}%
  \BibitemOpen
  \bibfield  {author} {\bibinfo {author} {\bibfnamefont {M.~R.}\ \bibnamefont
  {Swift}}, \bibinfo {author} {\bibfnamefont {W.~R.}\ \bibnamefont {Osborn}}, \
  and\ \bibinfo {author} {\bibfnamefont {J.~M.}\ \bibnamefont {Yeomans}},\
  }\href {\doibase 10.1103/PhysRevLett.75.830} {\bibfield  {journal} {\bibinfo
  {journal} {Phys. Rev. Lett.}\ }\textbf {\bibinfo {volume} {75}},\ \bibinfo
  {pages} {830} (\bibinfo {year} {1995})}\BibitemShut {NoStop}%
\bibitem [{\citenamefont {Swift}\ \emph {et~al.}(1996)\citenamefont {Swift},
  \citenamefont {Orlandini}, \citenamefont {Osborn},\ and\ \citenamefont
  {Yeomans}}]{Swift1996}%
  \BibitemOpen
  \bibfield  {author} {\bibinfo {author} {\bibfnamefont {M.~R.}\ \bibnamefont
  {Swift}}, \bibinfo {author} {\bibfnamefont {E.}~\bibnamefont {Orlandini}},
  \bibinfo {author} {\bibfnamefont {W.~R.}\ \bibnamefont {Osborn}}, \ and\
  \bibinfo {author} {\bibfnamefont {J.~M.}\ \bibnamefont {Yeomans}},\ }\href
  {\doibase 10.1103/PhysRevE.54.5041} {\bibfield  {journal} {\bibinfo
  {journal} {Phys. Rev. E}\ }\textbf {\bibinfo {volume} {54}},\ \bibinfo
  {pages} {5041} (\bibinfo {year} {1996})}\BibitemShut {NoStop}%
\bibitem [{\citenamefont {He}\ \emph {et~al.}(1999{\natexlab{a}})\citenamefont
  {He}, \citenamefont {Chen},\ and\ \citenamefont {Zhang}}]{He1999}%
  \BibitemOpen
  \bibfield  {author} {\bibinfo {author} {\bibfnamefont {X.}~\bibnamefont
  {He}}, \bibinfo {author} {\bibfnamefont {S.}~\bibnamefont {Chen}}, \ and\
  \bibinfo {author} {\bibfnamefont {R.}~\bibnamefont {Zhang}},\ }\href
  {\doibase 10.1006/jcph.1999.6257} {\bibfield  {journal} {\bibinfo  {journal}
  {J. Comput. Phys.}\ }\textbf {\bibinfo {volume} {152}},\ \bibinfo {pages}
  {642} (\bibinfo {year} {1999}{\natexlab{a}})}\BibitemShut {NoStop}%
\bibitem [{\citenamefont {Li}\ \emph {et~al.}(2016)\citenamefont {Li},
  \citenamefont {Luo}, \citenamefont {Kang}, \citenamefont {He}, \citenamefont
  {Chen},\ and\ \citenamefont {Liu}}]{Li2016}%
  \BibitemOpen
  \bibfield  {author} {\bibinfo {author} {\bibfnamefont {Q.}~\bibnamefont
  {Li}}, \bibinfo {author} {\bibfnamefont {K.}~\bibnamefont {Luo}}, \bibinfo
  {author} {\bibfnamefont {Q.}~\bibnamefont {Kang}}, \bibinfo {author}
  {\bibfnamefont {Y.}~\bibnamefont {He}}, \bibinfo {author} {\bibfnamefont
  {Q.}~\bibnamefont {Chen}}, \ and\ \bibinfo {author} {\bibfnamefont
  {Q.}~\bibnamefont {Liu}},\ }\href {\doibase 10.1016/j.pecs.2015.10.001}
  {\bibfield  {journal} {\bibinfo  {journal} {Prog. Energy Combust. Sci.}\
  }\textbf {\bibinfo {volume} {52}},\ \bibinfo {pages} {62} (\bibinfo {year}
  {2016})}\BibitemShut {NoStop}%
\bibitem [{\citenamefont {Chen}\ and\ \citenamefont {Doolen}(1998)}]{Chen1998}%
  \BibitemOpen
  \bibfield  {author} {\bibinfo {author} {\bibfnamefont {S.}~\bibnamefont
  {Chen}}\ and\ \bibinfo {author} {\bibfnamefont {G.~D.}\ \bibnamefont
  {Doolen}},\ }\href {\doibase 10.1146/annurev.fluid.30.1.329} {\bibfield
  {journal} {\bibinfo  {journal} {Annu. Rev. Fluid Mech.}\ }\textbf {\bibinfo
  {volume} {30}},\ \bibinfo {pages} {329} (\bibinfo {year} {1998})}\BibitemShut
  {NoStop}%
\bibitem [{\citenamefont {Nourgaliev}\ \emph {et~al.}(2003)\citenamefont
  {Nourgaliev}, \citenamefont {Dinh}, \citenamefont {Theofanous},\ and\
  \citenamefont {Joseph}}]{Nourgaliev2003}%
  \BibitemOpen
  \bibfield  {author} {\bibinfo {author} {\bibfnamefont {R.}~\bibnamefont
  {Nourgaliev}}, \bibinfo {author} {\bibfnamefont {T.}~\bibnamefont {Dinh}},
  \bibinfo {author} {\bibfnamefont {T.}~\bibnamefont {Theofanous}}, \ and\
  \bibinfo {author} {\bibfnamefont {D.}~\bibnamefont {Joseph}},\ }\href
  {\doibase 10.1016/S0301-9322(02)00108-8} {\bibfield  {journal} {\bibinfo
  {journal} {Int. J. Multiphase Flow}\ }\textbf {\bibinfo {volume} {29}},\
  \bibinfo {pages} {117} (\bibinfo {year} {2003})}\BibitemShut {NoStop}%
\bibitem [{\citenamefont {Chen}\ \emph {et~al.}(2014)\citenamefont {Chen},
  \citenamefont {Kang}, \citenamefont {Mu}, \citenamefont {He},\ and\
  \citenamefont {Tao}}]{Chen2014}%
  \BibitemOpen
  \bibfield  {author} {\bibinfo {author} {\bibfnamefont {L.}~\bibnamefont
  {Chen}}, \bibinfo {author} {\bibfnamefont {Q.}~\bibnamefont {Kang}}, \bibinfo
  {author} {\bibfnamefont {Y.}~\bibnamefont {Mu}}, \bibinfo {author}
  {\bibfnamefont {Y.-L.}\ \bibnamefont {He}}, \ and\ \bibinfo {author}
  {\bibfnamefont {W.-Q.}\ \bibnamefont {Tao}},\ }\href {\doibase
  10.1016/j.ijheatmasstransfer.2014.04.032} {\bibfield  {journal} {\bibinfo
  {journal} {Int. J. Heat Mass Transf.}\ }\textbf {\bibinfo {volume} {76}},\
  \bibinfo {pages} {210} (\bibinfo {year} {2014})}\BibitemShut {NoStop}%
\bibitem [{\citenamefont {Liu}\ \emph {et~al.}(2016)\citenamefont {Liu},
  \citenamefont {Kang}, \citenamefont {Leonardi}, \citenamefont {Schmieschek},
  \citenamefont {Narv{\'{a}}ez}, \citenamefont {Jones}, \citenamefont
  {Williams}, \citenamefont {Valocchi},\ and\ \citenamefont
  {Harting}}]{Liu2015}%
  \BibitemOpen
  \bibfield  {author} {\bibinfo {author} {\bibfnamefont {H.}~\bibnamefont
  {Liu}}, \bibinfo {author} {\bibfnamefont {Q.}~\bibnamefont {Kang}}, \bibinfo
  {author} {\bibfnamefont {C.~R.}\ \bibnamefont {Leonardi}}, \bibinfo {author}
  {\bibfnamefont {S.}~\bibnamefont {Schmieschek}}, \bibinfo {author}
  {\bibfnamefont {A.}~\bibnamefont {Narv{\'{a}}ez}}, \bibinfo {author}
  {\bibfnamefont {B.~D.}\ \bibnamefont {Jones}}, \bibinfo {author}
  {\bibfnamefont {J.~R.}\ \bibnamefont {Williams}}, \bibinfo {author}
  {\bibfnamefont {A.~J.}\ \bibnamefont {Valocchi}}, \ and\ \bibinfo {author}
  {\bibfnamefont {J.}~\bibnamefont {Harting}},\ }\href {\doibase
  10.1007/s10596-015-9542-3} {\bibfield  {journal} {\bibinfo  {journal}
  {Computat. Geosci.}\ }\textbf {\bibinfo {volume} {20}},\ \bibinfo {pages}
  {777} (\bibinfo {year} {2016})}\BibitemShut {NoStop}%
\bibitem [{\citenamefont {Ba}\ \emph {et~al.}(2016)\citenamefont {Ba},
  \citenamefont {Liu}, \citenamefont {Li}, \citenamefont {Kang},\ and\
  \citenamefont {Sun}}]{Ba2016}%
  \BibitemOpen
  \bibfield  {author} {\bibinfo {author} {\bibfnamefont {Y.}~\bibnamefont
  {Ba}}, \bibinfo {author} {\bibfnamefont {H.}~\bibnamefont {Liu}}, \bibinfo
  {author} {\bibfnamefont {Q.}~\bibnamefont {Li}}, \bibinfo {author}
  {\bibfnamefont {Q.}~\bibnamefont {Kang}}, \ and\ \bibinfo {author}
  {\bibfnamefont {J.}~\bibnamefont {Sun}},\ }\href {\doibase
  10.1103/PhysRevE.94.023310} {\bibfield  {journal} {\bibinfo  {journal} {Phys.
  Rev. E}\ }\textbf {\bibinfo {volume} {94}},\ \bibinfo {pages} {023310}
  (\bibinfo {year} {2016})}\BibitemShut {NoStop}%
\bibitem [{\citenamefont {Rothman}\ and\ \citenamefont
  {Keller}(1988)}]{Rothman1988}%
  \BibitemOpen
  \bibfield  {author} {\bibinfo {author} {\bibfnamefont {D.~H.}\ \bibnamefont
  {Rothman}}\ and\ \bibinfo {author} {\bibfnamefont {J.~M.}\ \bibnamefont
  {Keller}},\ }\href {\doibase 10.1007/BF01019743} {\bibfield  {journal}
  {\bibinfo  {journal} {J. Stat. Phys.}\ }\textbf {\bibinfo {volume} {52}},\
  \bibinfo {pages} {1119} (\bibinfo {year} {1988})}\BibitemShut {NoStop}%
\bibitem [{\citenamefont {Latva-Kokko}\ and\ \citenamefont
  {Rothman}(2005)}]{Latva-Kokko2005}%
  \BibitemOpen
  \bibfield  {author} {\bibinfo {author} {\bibfnamefont {M.}~\bibnamefont
  {Latva-Kokko}}\ and\ \bibinfo {author} {\bibfnamefont {D.~H.}\ \bibnamefont
  {Rothman}},\ }\href {\doibase 10.1103/PhysRevE.71.056702} {\bibfield
  {journal} {\bibinfo  {journal} {Phys. Rev. E}\ }\textbf {\bibinfo {volume}
  {71}},\ \bibinfo {pages} {056702} (\bibinfo {year} {2005})}\BibitemShut
  {NoStop}%
\bibitem [{\citenamefont {Reis}\ and\ \citenamefont
  {Phillips}(2007)}]{Reis2007}%
  \BibitemOpen
  \bibfield  {author} {\bibinfo {author} {\bibfnamefont {T.}~\bibnamefont
  {Reis}}\ and\ \bibinfo {author} {\bibfnamefont {T.~N.}\ \bibnamefont
  {Phillips}},\ }\href {\doibase 10.1088/1751-8113/40/14/018} {\bibfield
  {journal} {\bibinfo  {journal} {J. Phys. A: Math. Theor.}\ }\textbf {\bibinfo
  {volume} {40}},\ \bibinfo {pages} {4033} (\bibinfo {year}
  {2007})}\BibitemShut {NoStop}%
\bibitem [{\citenamefont {Leclaire}\ \emph {et~al.}(2012)\citenamefont
  {Leclaire}, \citenamefont {Reggio},\ and\ \citenamefont
  {Tr{\'{e}}panier}}]{Leclaire2012}%
  \BibitemOpen
  \bibfield  {author} {\bibinfo {author} {\bibfnamefont {S.}~\bibnamefont
  {Leclaire}}, \bibinfo {author} {\bibfnamefont {M.}~\bibnamefont {Reggio}}, \
  and\ \bibinfo {author} {\bibfnamefont {J.-Y.}\ \bibnamefont
  {Tr{\'{e}}panier}},\ }\href {\doibase 10.1016/j.apm.2011.08.027} {\bibfield
  {journal} {\bibinfo  {journal} {Appl. Math. Modell.}\ }\textbf {\bibinfo
  {volume} {36}},\ \bibinfo {pages} {2237} (\bibinfo {year}
  {2012})}\BibitemShut {NoStop}%
\bibitem [{\citenamefont {Leclaire}\ \emph {et~al.}(2011)\citenamefont
  {Leclaire}, \citenamefont {Reggio},\ and\ \citenamefont
  {Tr{\'{e}}panier}}]{Leclaire2011}%
  \BibitemOpen
  \bibfield  {author} {\bibinfo {author} {\bibfnamefont {S.}~\bibnamefont
  {Leclaire}}, \bibinfo {author} {\bibfnamefont {M.}~\bibnamefont {Reggio}}, \
  and\ \bibinfo {author} {\bibfnamefont {J.-Y.}\ \bibnamefont
  {Tr{\'{e}}panier}},\ }\href {\doibase 10.1016/j.compfluid.2011.04.001}
  {\bibfield  {journal} {\bibinfo  {journal} {Comput. Fluids}\ }\textbf
  {\bibinfo {volume} {48}},\ \bibinfo {pages} {98} (\bibinfo {year}
  {2011})}\BibitemShut {NoStop}%
\bibitem [{\citenamefont {Liu}\ \emph {et~al.}(2012)\citenamefont {Liu},
  \citenamefont {Valocchi},\ and\ \citenamefont {Kang}}]{Liu2012}%
  \BibitemOpen
  \bibfield  {author} {\bibinfo {author} {\bibfnamefont {H.}~\bibnamefont
  {Liu}}, \bibinfo {author} {\bibfnamefont {A.~J.}\ \bibnamefont {Valocchi}}, \
  and\ \bibinfo {author} {\bibfnamefont {Q.}~\bibnamefont {Kang}},\ }\href
  {\doibase 10.1103/PhysRevE.85.046309} {\bibfield  {journal} {\bibinfo
  {journal} {Phys. Rev. E}\ }\textbf {\bibinfo {volume} {85}},\ \bibinfo
  {pages} {046309} (\bibinfo {year} {2012})}\BibitemShut {NoStop}%
\bibitem [{\citenamefont {Leclaire}\ \emph {et~al.}(2017)\citenamefont
  {Leclaire}, \citenamefont {Parmigiani}, \citenamefont {Malaspinas},
  \citenamefont {Chopard},\ and\ \citenamefont {Latt}}]{Leclaire2017}%
  \BibitemOpen
  \bibfield  {author} {\bibinfo {author} {\bibfnamefont {S.}~\bibnamefont
  {Leclaire}}, \bibinfo {author} {\bibfnamefont {A.}~\bibnamefont
  {Parmigiani}}, \bibinfo {author} {\bibfnamefont {O.}~\bibnamefont
  {Malaspinas}}, \bibinfo {author} {\bibfnamefont {B.}~\bibnamefont {Chopard}},
  \ and\ \bibinfo {author} {\bibfnamefont {J.}~\bibnamefont {Latt}},\ }\href
  {\doibase 10.1103/PhysRevE.95.033306} {\bibfield  {journal} {\bibinfo
  {journal} {Phys. Rev. E}\ }\textbf {\bibinfo {volume} {95}},\ \bibinfo
  {pages} {033306} (\bibinfo {year} {2017})}\BibitemShut {NoStop}%
\bibitem [{\citenamefont {Holdych}\ \emph {et~al.}(1998)\citenamefont
  {Holdych}, \citenamefont {Rovas}, \citenamefont {Georgiadis},\ and\
  \citenamefont {Buckius}}]{Holdych1998}%
  \BibitemOpen
  \bibfield  {author} {\bibinfo {author} {\bibfnamefont {D.~J.}\ \bibnamefont
  {Holdych}}, \bibinfo {author} {\bibfnamefont {D.}~\bibnamefont {Rovas}},
  \bibinfo {author} {\bibfnamefont {J.~G.}\ \bibnamefont {Georgiadis}}, \ and\
  \bibinfo {author} {\bibfnamefont {R.~O.}\ \bibnamefont {Buckius}},\ }\href
  {\doibase 10.1142/S0129183198001266} {\bibfield  {journal} {\bibinfo
  {journal} {Int. J.Mod. Phys. C}\ }\textbf {\bibinfo {volume} {09}},\ \bibinfo
  {pages} {1393} (\bibinfo {year} {1998})}\BibitemShut {NoStop}%
\bibitem [{\citenamefont {Leclaire}\ \emph
  {et~al.}(2013{\natexlab{a}})\citenamefont {Leclaire}, \citenamefont
  {Pellerin}, \citenamefont {Reggio},\ and\ \citenamefont
  {Tr{\'{e}}panier}}]{Leclaire2013}%
  \BibitemOpen
  \bibfield  {author} {\bibinfo {author} {\bibfnamefont {S.}~\bibnamefont
  {Leclaire}}, \bibinfo {author} {\bibfnamefont {N.}~\bibnamefont {Pellerin}},
  \bibinfo {author} {\bibfnamefont {M.}~\bibnamefont {Reggio}}, \ and\ \bibinfo
  {author} {\bibfnamefont {J.-Y.}\ \bibnamefont {Tr{\'{e}}panier}},\ }\href
  {\doibase 10.1016/j.ijmultiphaseflow.2013.07.001} {\bibfield  {journal}
  {\bibinfo  {journal} {Int. J. Multiphase Flow}\ }\textbf {\bibinfo {volume}
  {57}},\ \bibinfo {pages} {159} (\bibinfo {year}
  {2013}{\natexlab{a}})}\BibitemShut {NoStop}%
\bibitem [{\citenamefont {Luo}\ \emph {et~al.}(2011)\citenamefont {Luo},
  \citenamefont {Liao}, \citenamefont {Chen}, \citenamefont {Peng},\ and\
  \citenamefont {Zhang}}]{Luo2011}%
  \BibitemOpen
  \bibfield  {author} {\bibinfo {author} {\bibfnamefont {L.-S.}\ \bibnamefont
  {Luo}}, \bibinfo {author} {\bibfnamefont {W.}~\bibnamefont {Liao}}, \bibinfo
  {author} {\bibfnamefont {X.}~\bibnamefont {Chen}}, \bibinfo {author}
  {\bibfnamefont {Y.}~\bibnamefont {Peng}}, \ and\ \bibinfo {author}
  {\bibfnamefont {W.}~\bibnamefont {Zhang}},\ }\href {\doibase
  10.1103/PhysRevE.83.056710} {\bibfield  {journal} {\bibinfo  {journal} {Phys.
  Rev. E}\ }\textbf {\bibinfo {volume} {83}},\ \bibinfo {pages} {056710}
  (\bibinfo {year} {2011})}\BibitemShut {NoStop}%
\bibitem [{\citenamefont {d'Humi{\`{e}}res}(1994)}]{dHumieres1994}%
  \BibitemOpen
  \bibfield  {author} {\bibinfo {author} {\bibfnamefont {D.}~\bibnamefont
  {d'Humi{\`{e}}res}},\ }in\ \href {\doibase 10.2514/5.9781600866319.0450.0458}
  {\emph {\bibinfo {booktitle} {Rarefied Gas Dynamics: Theory and
  Simulations}}}\ (\bibinfo  {publisher} {American Institute of Aeronautics and
  Astronautics},\ \bibinfo {address} {Washington DC},\ \bibinfo {year} {1994})\
  pp.\ \bibinfo {pages} {450--458}\BibitemShut {NoStop}%
\bibitem [{\citenamefont {Lallemand}\ and\ \citenamefont
  {Luo}(2000)}]{Lallemand2000}%
  \BibitemOpen
  \bibfield  {author} {\bibinfo {author} {\bibfnamefont {P.}~\bibnamefont
  {Lallemand}}\ and\ \bibinfo {author} {\bibfnamefont {L.-S.}\ \bibnamefont
  {Luo}},\ }\href {\doibase 10.1103/PhysRevE.61.6546} {\bibfield  {journal}
  {\bibinfo  {journal} {Phys. Rev. E}\ }\textbf {\bibinfo {volume} {61}},\
  \bibinfo {pages} {6546} (\bibinfo {year} {2000})}\BibitemShut {NoStop}%
\bibitem [{\citenamefont {d'Humieres}\ \emph {et~al.}(2002)\citenamefont
  {d'Humieres}, \citenamefont {Ginzburg}, \citenamefont {Krafczyk},
  \citenamefont {Lallemand},\ and\ \citenamefont {Luo}}]{dHumieres2002}%
  \BibitemOpen
  \bibfield  {author} {\bibinfo {author} {\bibfnamefont {D.}~\bibnamefont
  {d'Humieres}}, \bibinfo {author} {\bibfnamefont {I.}~\bibnamefont
  {Ginzburg}}, \bibinfo {author} {\bibfnamefont {M.}~\bibnamefont {Krafczyk}},
  \bibinfo {author} {\bibfnamefont {P.}~\bibnamefont {Lallemand}}, \ and\
  \bibinfo {author} {\bibfnamefont {L.-S.}\ \bibnamefont {Luo}},\ }\href
  {\doibase 10.1098/rsta.2001.0955} {\bibfield  {journal} {\bibinfo  {journal}
  {Philos. Trans. R. Soc. London, Ser. A: Math., Phys. Eng. Sci.}\ }\textbf
  {\bibinfo {volume} {360}},\ \bibinfo {pages} {437} (\bibinfo {year}
  {2002})}\BibitemShut {NoStop}%
\bibitem [{\citenamefont {McCracken}\ and\ \citenamefont
  {Abraham}(2005{\natexlab{b}})}]{McCracken2005b}%
  \BibitemOpen
  \bibfield  {author} {\bibinfo {author} {\bibfnamefont {M.~E.}\ \bibnamefont
  {McCracken}}\ and\ \bibinfo {author} {\bibfnamefont {J.}~\bibnamefont
  {Abraham}},\ }\href {\doibase 10.1103/PhysRevE.71.036701} {\bibfield
  {journal} {\bibinfo  {journal} {Phys. Rev. E}\ }\textbf {\bibinfo {volume}
  {71}},\ \bibinfo {pages} {036701} (\bibinfo {year}
  {2005}{\natexlab{b}})}\BibitemShut {NoStop}%
\bibitem [{\citenamefont {Ebihara}\ and\ \citenamefont
  {Watanabe}(2003)}]{Ebihara2003}%
  \BibitemOpen
  \bibfield  {author} {\bibinfo {author} {\bibfnamefont {K.}~\bibnamefont
  {Ebihara}}\ and\ \bibinfo {author} {\bibfnamefont {T.}~\bibnamefont
  {Watanabe}},\ }\href {\doibase 10.1142/S0217979203017175} {\bibfield
  {journal} {\bibinfo  {journal} {Int. J. Mod. Phys. B}\ }\textbf {\bibinfo
  {volume} {17}},\ \bibinfo {pages} {113} (\bibinfo {year} {2003})}\BibitemShut
  {NoStop}%
\bibitem [{\citenamefont {Geier}\ \emph {et~al.}(2015)\citenamefont {Geier},
  \citenamefont {Sch{\"{o}}nherr}, \citenamefont {Pasquali},\ and\
  \citenamefont {Krafczyk}}]{Geier2015}%
  \BibitemOpen
  \bibfield  {author} {\bibinfo {author} {\bibfnamefont {M.}~\bibnamefont
  {Geier}}, \bibinfo {author} {\bibfnamefont {M.}~\bibnamefont
  {Sch{\"{o}}nherr}}, \bibinfo {author} {\bibfnamefont {A.}~\bibnamefont
  {Pasquali}}, \ and\ \bibinfo {author} {\bibfnamefont {M.}~\bibnamefont
  {Krafczyk}},\ }\href {\doibase 10.1016/j.camwa.2015.05.001} {\bibfield
  {journal} {\bibinfo  {journal} {Comput. Math. Appl.}\ }\textbf {\bibinfo
  {volume} {70}},\ \bibinfo {pages} {507} (\bibinfo {year} {2015})}\BibitemShut
  {NoStop}%
\bibitem [{\citenamefont {Suga}\ \emph {et~al.}(2015)\citenamefont {Suga},
  \citenamefont {Kuwata}, \citenamefont {Takashima},\ and\ \citenamefont
  {Chikasue}}]{Suga2015}%
  \BibitemOpen
  \bibfield  {author} {\bibinfo {author} {\bibfnamefont {K.}~\bibnamefont
  {Suga}}, \bibinfo {author} {\bibfnamefont {Y.}~\bibnamefont {Kuwata}},
  \bibinfo {author} {\bibfnamefont {K.}~\bibnamefont {Takashima}}, \ and\
  \bibinfo {author} {\bibfnamefont {R.}~\bibnamefont {Chikasue}},\ }\href
  {\doibase 10.1016/j.camwa.2015.01.010} {\bibfield  {journal} {\bibinfo
  {journal} {Comput. Math. Appl.}\ }\textbf {\bibinfo {volume} {69}},\ \bibinfo
  {pages} {518} (\bibinfo {year} {2015})}\BibitemShut {NoStop}%
\bibitem [{\citenamefont {He}\ and\ \citenamefont {Luo}(1997)}]{He1997}%
  \BibitemOpen
  \bibfield  {author} {\bibinfo {author} {\bibfnamefont {X.}~\bibnamefont
  {He}}\ and\ \bibinfo {author} {\bibfnamefont {L.-S.}\ \bibnamefont {Luo}},\
  }\href {\doibase 10.1103/PhysRevE.56.6811} {\bibfield  {journal} {\bibinfo
  {journal} {Phys. Rev. E}\ }\textbf {\bibinfo {volume} {56}},\ \bibinfo
  {pages} {6811} (\bibinfo {year} {1997})}\BibitemShut {NoStop}%
\bibitem [{\citenamefont {T{\"{o}}lke}\ \emph {et~al.}(2002)\citenamefont
  {T{\"{o}}lke}, \citenamefont {Krafczyk}, \citenamefont {Schulz},\ and\
  \citenamefont {Rank}}]{Tolke2002}%
  \BibitemOpen
  \bibfield  {author} {\bibinfo {author} {\bibfnamefont {J.}~\bibnamefont
  {T{\"{o}}lke}}, \bibinfo {author} {\bibfnamefont {M.}~\bibnamefont
  {Krafczyk}}, \bibinfo {author} {\bibfnamefont {M.}~\bibnamefont {Schulz}}, \
  and\ \bibinfo {author} {\bibfnamefont {E.}~\bibnamefont {Rank}},\ }\href
  {\doibase 10.1098/rsta.2001.0944} {\bibfield  {journal} {\bibinfo  {journal}
  {Philos. Trans. R. Soc. London, Ser. A: Math., Phys. Eng. Sci.}\ }\textbf
  {\bibinfo {volume} {360}},\ \bibinfo {pages} {535} (\bibinfo {year}
  {2002})}\BibitemShut {NoStop}%
\bibitem [{\citenamefont {Guo}\ \emph {et~al.}(2002)\citenamefont {Guo},
  \citenamefont {Zheng},\ and\ \citenamefont {Shi}}]{Guo2002}%
  \BibitemOpen
  \bibfield  {author} {\bibinfo {author} {\bibfnamefont {Z.}~\bibnamefont
  {Guo}}, \bibinfo {author} {\bibfnamefont {C.}~\bibnamefont {Zheng}}, \ and\
  \bibinfo {author} {\bibfnamefont {B.}~\bibnamefont {Shi}},\ }\href {\doibase
  10.1103/PhysRevE.65.046308} {\bibfield  {journal} {\bibinfo  {journal} {Phys.
  Rev. E}\ }\textbf {\bibinfo {volume} {65}},\ \bibinfo {pages} {046308}
  (\bibinfo {year} {2002})}\BibitemShut {NoStop}%
\bibitem [{\citenamefont {Dubois}\ and\ \citenamefont
  {Lallemand}(2011)}]{Dubois2011}%
  \BibitemOpen
  \bibfield  {author} {\bibinfo {author} {\bibfnamefont {F.}~\bibnamefont
  {Dubois}}\ and\ \bibinfo {author} {\bibfnamefont {P.}~\bibnamefont
  {Lallemand}},\ }\href {\doibase 10.1016/j.camwa.2011.01.011} {\bibfield
  {journal} {\bibinfo  {journal} {Comput. Math. Appl.}\ }\textbf {\bibinfo
  {volume} {61}},\ \bibinfo {pages} {3404} (\bibinfo {year}
  {2011})}\BibitemShut {NoStop}%
\bibitem [{\citenamefont {Premnath}\ and\ \citenamefont
  {Banerjee}(2011)}]{Premnath2011}%
  \BibitemOpen
  \bibfield  {author} {\bibinfo {author} {\bibfnamefont {K.~N.}\ \bibnamefont
  {Premnath}}\ and\ \bibinfo {author} {\bibfnamefont {S.}~\bibnamefont
  {Banerjee}},\ }\href {\doibase 10.1007/s10955-011-0208-9} {\bibfield
  {journal} {\bibinfo  {journal} {J. Stat. Phys.}\ }\textbf {\bibinfo {volume}
  {143}},\ \bibinfo {pages} {747} (\bibinfo {year} {2011})}\BibitemShut
  {NoStop}%
\bibitem [{\citenamefont {Leclaire}\ \emph {et~al.}(2014)\citenamefont
  {Leclaire}, \citenamefont {Pellerin}, \citenamefont {Reggio},\ and\
  \citenamefont {Tr{\'{e}}panier}}]{Leclaire2014}%
  \BibitemOpen
  \bibfield  {author} {\bibinfo {author} {\bibfnamefont {S.}~\bibnamefont
  {Leclaire}}, \bibinfo {author} {\bibfnamefont {N.}~\bibnamefont {Pellerin}},
  \bibinfo {author} {\bibfnamefont {M.}~\bibnamefont {Reggio}}, \ and\ \bibinfo
  {author} {\bibfnamefont {J.-Y.}\ \bibnamefont {Tr{\'{e}}panier}},\ }\href
  {\doibase 10.1088/1751-8113/47/10/105501} {\bibfield  {journal} {\bibinfo
  {journal} {J. Phys. A: Math. Theor.}\ }\textbf {\bibinfo {volume} {47}},\
  \bibinfo {pages} {105501} (\bibinfo {year} {2014})}\BibitemShut {NoStop}%
\bibitem [{\citenamefont {Leclaire}\ \emph {et~al.}(2015)\citenamefont
  {Leclaire}, \citenamefont {Pellerin}, \citenamefont {Reggio},\ and\
  \citenamefont {Tr{\'{e}}panier}}]{Leclaire2015}%
  \BibitemOpen
  \bibfield  {author} {\bibinfo {author} {\bibfnamefont {S.}~\bibnamefont
  {Leclaire}}, \bibinfo {author} {\bibfnamefont {N.}~\bibnamefont {Pellerin}},
  \bibinfo {author} {\bibfnamefont {M.}~\bibnamefont {Reggio}}, \ and\ \bibinfo
  {author} {\bibfnamefont {J.-Y.}\ \bibnamefont {Tr{\'{e}}panier}},\ }\href
  {\doibase 10.1002/fld.4002} {\bibfield  {journal} {\bibinfo  {journal} {Int.
  J. Num. Methods Fluids}\ }\textbf {\bibinfo {volume} {77}},\ \bibinfo {pages}
  {732} (\bibinfo {year} {2015})}\BibitemShut {NoStop}%
\bibitem [{\citenamefont {Dellar}(2014)}]{Dellar2014}%
  \BibitemOpen
  \bibfield  {author} {\bibinfo {author} {\bibfnamefont {P.~J.}\ \bibnamefont
  {Dellar}},\ }\href {\doibase 10.1016/j.jcp.2013.11.021} {\bibfield  {journal}
  {\bibinfo  {journal} {J. Comput. Phys.}\ }\textbf {\bibinfo {volume} {259}},\
  \bibinfo {pages} {270} (\bibinfo {year} {2014})}\BibitemShut {NoStop}%
\bibitem [{\citenamefont {Yu}\ and\ \citenamefont {Fan}(2010)}]{Yu2010}%
  \BibitemOpen
  \bibfield  {author} {\bibinfo {author} {\bibfnamefont {Z.}~\bibnamefont
  {Yu}}\ and\ \bibinfo {author} {\bibfnamefont {L.-S.}\ \bibnamefont {Fan}},\
  }\href {\doibase 10.1103/PhysRevE.82.046708} {\bibfield  {journal} {\bibinfo
  {journal} {Phys. Rev. E}\ }\textbf {\bibinfo {volume} {82}},\ \bibinfo
  {pages} {046708} (\bibinfo {year} {2010})}\BibitemShut {NoStop}%
\bibitem [{\citenamefont {Brackbill}\ \emph {et~al.}(1992)\citenamefont
  {Brackbill}, \citenamefont {Kothe},\ and\ \citenamefont
  {Zemach}}]{Brackbill1992}%
  \BibitemOpen
  \bibfield  {author} {\bibinfo {author} {\bibfnamefont {J.~U.}\ \bibnamefont
  {Brackbill}}, \bibinfo {author} {\bibfnamefont {D.~B.}\ \bibnamefont
  {Kothe}}, \ and\ \bibinfo {author} {\bibfnamefont {C.}~\bibnamefont
  {Zemach}},\ }\href {\doibase 10.1016/0021-9991(92)90240-Y} {\bibfield
  {journal} {\bibinfo  {journal} {J. Comput. Phys.}\ }\textbf {\bibinfo
  {volume} {100}},\ \bibinfo {pages} {335} (\bibinfo {year}
  {1992})}\BibitemShut {NoStop}%
\bibitem [{\citenamefont {Halliday}\ \emph {et~al.}(2007)\citenamefont
  {Halliday}, \citenamefont {Hollis},\ and\ \citenamefont
  {Care}}]{Halliday2007}%
  \BibitemOpen
  \bibfield  {author} {\bibinfo {author} {\bibfnamefont {I.}~\bibnamefont
  {Halliday}}, \bibinfo {author} {\bibfnamefont {A.~P.}\ \bibnamefont
  {Hollis}}, \ and\ \bibinfo {author} {\bibfnamefont {C.~M.}\ \bibnamefont
  {Care}},\ }\href {\doibase 10.1103/PhysRevE.76.026708} {\bibfield  {journal}
  {\bibinfo  {journal} {Phys. Rev. E}\ }\textbf {\bibinfo {volume} {76}},\
  \bibinfo {pages} {026708} (\bibinfo {year} {2007})}\BibitemShut {NoStop}%
\bibitem [{\citenamefont {Kr{\"{u}}ger}\ \emph {et~al.}(2017)\citenamefont
  {Kr{\"{u}}ger}, \citenamefont {Kusumaatmaja}, \citenamefont {Kuzmin},
  \citenamefont {Shardt}, \citenamefont {Silva},\ and\ \citenamefont
  {Viggen}}]{Kruger2017}%
  \BibitemOpen
  \bibfield  {author} {\bibinfo {author} {\bibfnamefont {T.}~\bibnamefont
  {Kr{\"{u}}ger}}, \bibinfo {author} {\bibfnamefont {H.}~\bibnamefont
  {Kusumaatmaja}}, \bibinfo {author} {\bibfnamefont {A.}~\bibnamefont
  {Kuzmin}}, \bibinfo {author} {\bibfnamefont {O.}~\bibnamefont {Shardt}},
  \bibinfo {author} {\bibfnamefont {G.}~\bibnamefont {Silva}}, \ and\ \bibinfo
  {author} {\bibfnamefont {E.~M.}\ \bibnamefont {Viggen}},\ }\href {\doibase
  10.1007/978-3-319-44649-3} {\emph {\bibinfo {title} {{The Lattice Boltzmann
  Method}}}},\ Graduate Texts in Physics\ (\bibinfo  {publisher} {Springer
  International Publishing},\ \bibinfo {year} {2017})\BibitemShut {NoStop}%
\bibitem [{\citenamefont {Guo}\ \emph {et~al.}(2011)\citenamefont {Guo},
  \citenamefont {Zheng},\ and\ \citenamefont {Shi}}]{Guo2011}%
  \BibitemOpen
  \bibfield  {author} {\bibinfo {author} {\bibfnamefont {Z.}~\bibnamefont
  {Guo}}, \bibinfo {author} {\bibfnamefont {C.}~\bibnamefont {Zheng}}, \ and\
  \bibinfo {author} {\bibfnamefont {B.}~\bibnamefont {Shi}},\ }\href {\doibase
  10.1103/PhysRevE.83.036707} {\bibfield  {journal} {\bibinfo  {journal} {Phys.
  Rev. E}\ }\textbf {\bibinfo {volume} {83}},\ \bibinfo {pages} {036707}
  (\bibinfo {year} {2011})}\BibitemShut {NoStop}%
\bibitem [{\citenamefont {Liang}\ \emph {et~al.}(2014)\citenamefont {Liang},
  \citenamefont {Shi}, \citenamefont {Guo},\ and\ \citenamefont
  {Chai}}]{Liang2014}%
  \BibitemOpen
  \bibfield  {author} {\bibinfo {author} {\bibfnamefont {H.}~\bibnamefont
  {Liang}}, \bibinfo {author} {\bibfnamefont {B.~C.}\ \bibnamefont {Shi}},
  \bibinfo {author} {\bibfnamefont {Z.~L.}\ \bibnamefont {Guo}}, \ and\
  \bibinfo {author} {\bibfnamefont {Z.~H.}\ \bibnamefont {Chai}},\ }\href
  {\doibase 10.1103/PhysRevE.89.053320} {\bibfield  {journal} {\bibinfo
  {journal} {Phys. Rev. E}\ }\textbf {\bibinfo {volume} {89}},\ \bibinfo
  {pages} {053320} (\bibinfo {year} {2014})}\BibitemShut {NoStop}%
\bibitem [{\citenamefont {Lou}\ \emph {et~al.}(2012)\citenamefont {Lou},
  \citenamefont {Guo},\ and\ \citenamefont {Shi}}]{Lou2012}%
  \BibitemOpen
  \bibfield  {author} {\bibinfo {author} {\bibfnamefont {Q.}~\bibnamefont
  {Lou}}, \bibinfo {author} {\bibfnamefont {Z.~L.}\ \bibnamefont {Guo}}, \ and\
  \bibinfo {author} {\bibfnamefont {B.~C.}\ \bibnamefont {Shi}},\ }\href
  {\doibase 10.1209/0295-5075/99/64005} {\bibfield  {journal} {\bibinfo
  {journal} {Europhys. Lett.}\ }\textbf {\bibinfo {volume} {99}},\ \bibinfo
  {pages} {64005} (\bibinfo {year} {2012})}\BibitemShut {NoStop}%
\bibitem [{\citenamefont {Leclaire}\ \emph
  {et~al.}(2013{\natexlab{b}})\citenamefont {Leclaire}, \citenamefont
  {Reggio},\ and\ \citenamefont {Tr{\'{e}}panier}}]{Leclaire2013b}%
  \BibitemOpen
  \bibfield  {author} {\bibinfo {author} {\bibfnamefont {S.}~\bibnamefont
  {Leclaire}}, \bibinfo {author} {\bibfnamefont {M.}~\bibnamefont {Reggio}}, \
  and\ \bibinfo {author} {\bibfnamefont {J.-Y.}\ \bibnamefont
  {Tr{\'{e}}panier}},\ }\href {\doibase 10.1016/j.jcp.2013.03.039} {\bibfield
  {journal} {\bibinfo  {journal} {J. Comput. Phys.}\ }\textbf {\bibinfo
  {volume} {246}},\ \bibinfo {pages} {318} (\bibinfo {year}
  {2013}{\natexlab{b}})}\BibitemShut {NoStop}%
\bibitem [{\citenamefont {Premnath}\ and\ \citenamefont
  {Abraham}(2007)}]{Premnath2007}%
  \BibitemOpen
  \bibfield  {author} {\bibinfo {author} {\bibfnamefont {K.~N.}\ \bibnamefont
  {Premnath}}\ and\ \bibinfo {author} {\bibfnamefont {J.}~\bibnamefont
  {Abraham}},\ }\href {\doibase 10.1016/j.jcp.2006.10.023} {\bibfield
  {journal} {\bibinfo  {journal} {J. Comput. Phys.}\ }\textbf {\bibinfo
  {volume} {224}},\ \bibinfo {pages} {539} (\bibinfo {year}
  {2007})}\BibitemShut {NoStop}%
\bibitem [{\citenamefont {Miller}\ and\ \citenamefont
  {Scriven}(1968)}]{Miller1968}%
  \BibitemOpen
  \bibfield  {author} {\bibinfo {author} {\bibfnamefont {C.~A.}\ \bibnamefont
  {Miller}}\ and\ \bibinfo {author} {\bibfnamefont {L.~E.}\ \bibnamefont
  {Scriven}},\ }\href {\doibase 10.1017/S0022112068000832} {\bibfield
  {journal} {\bibinfo  {journal} {J. Fluid Mech.}\ }\textbf {\bibinfo {volume}
  {32}},\ \bibinfo {pages} {417} (\bibinfo {year} {1968})}\BibitemShut
  {NoStop}%
\bibitem [{\citenamefont {Lamb}(1945)}]{Lamb1945}%
  \BibitemOpen
  \bibfield  {author} {\bibinfo {author} {\bibfnamefont {H.}~\bibnamefont
  {Lamb}},\ }\href@noop {} {\emph {\bibinfo {title} {{Hydrodynamics}}}}\
  (\bibinfo {address} {New York, USA},\ \bibinfo {year} {1945})\BibitemShut
  {NoStop}%
\bibitem [{\citenamefont {He}\ \emph {et~al.}(1999{\natexlab{b}})\citenamefont
  {He}, \citenamefont {Zhang}, \citenamefont {Chen},\ and\ \citenamefont
  {Doolen}}]{He1999b}%
  \BibitemOpen
  \bibfield  {author} {\bibinfo {author} {\bibfnamefont {X.}~\bibnamefont
  {He}}, \bibinfo {author} {\bibfnamefont {R.}~\bibnamefont {Zhang}}, \bibinfo
  {author} {\bibfnamefont {S.}~\bibnamefont {Chen}}, \ and\ \bibinfo {author}
  {\bibfnamefont {G.~D.}\ \bibnamefont {Doolen}},\ }\href {\doibase
  10.1063/1.869984} {\bibfield  {journal} {\bibinfo  {journal} {Phys. Fluids}\
  }\textbf {\bibinfo {volume} {11}},\ \bibinfo {pages} {1143} (\bibinfo {year}
  {1999}{\natexlab{b}})}\BibitemShut {NoStop}%
\bibitem [{\citenamefont {Wang}\ \emph {et~al.}(2016)\citenamefont {Wang},
  \citenamefont {Liu},\ and\ \citenamefont {Zhang}}]{Wang2016}%
  \BibitemOpen
  \bibfield  {author} {\bibinfo {author} {\bibfnamefont {N.}~\bibnamefont
  {Wang}}, \bibinfo {author} {\bibfnamefont {H.}~\bibnamefont {Liu}}, \ and\
  \bibinfo {author} {\bibfnamefont {C.}~\bibnamefont {Zhang}},\ }\href
  {\doibase 10.1016/j.jocs.2016.04.012} {\bibfield  {journal} {\bibinfo
  {journal} {J. Comput. Sci.}\ }\textbf {\bibinfo {volume} {17}},\ \bibinfo
  {pages} {340} (\bibinfo {year} {2016})}\BibitemShut {NoStop}%
\bibitem [{\citenamefont {Lee}\ and\ \citenamefont {Kim}(2013)}]{Lee2013}%
  \BibitemOpen
  \bibfield  {author} {\bibinfo {author} {\bibfnamefont {H.~G.}\ \bibnamefont
  {Lee}}\ and\ \bibinfo {author} {\bibfnamefont {J.}~\bibnamefont {Kim}},\
  }\href {\doibase 10.1016/j.camwa.2013.08.021} {\bibfield  {journal} {\bibinfo
   {journal} {Comput. Math. Appl.}\ }\textbf {\bibinfo {volume} {66}},\
  \bibinfo {pages} {1466} (\bibinfo {year} {2013})}\BibitemShut {NoStop}%
\bibitem [{\citenamefont {Succi}(2001)}]{Succi2001}%
  \BibitemOpen
  \bibfield  {author} {\bibinfo {author} {\bibfnamefont {S.}~\bibnamefont
  {Succi}},\ }\href@noop {} {\emph {\bibinfo {title} {{The Lattice Boltzmann
  Equation for Fluid Dynamics and Beyond}}}}\ (\bibinfo  {publisher} {Oxford
  University Press},\ \bibinfo {year} {2001})\BibitemShut {NoStop}%
\bibitem [{\citenamefont {Lou}\ \emph {et~al.}(2013)\citenamefont {Lou},
  \citenamefont {Guo},\ and\ \citenamefont {Shi}}]{Lou2013}%
  \BibitemOpen
  \bibfield  {author} {\bibinfo {author} {\bibfnamefont {Q.}~\bibnamefont
  {Lou}}, \bibinfo {author} {\bibfnamefont {Z.}~\bibnamefont {Guo}}, \ and\
  \bibinfo {author} {\bibfnamefont {B.}~\bibnamefont {Shi}},\ }\href {\doibase
  10.1103/PhysRevE.87.063301} {\bibfield  {journal} {\bibinfo  {journal} {Phys.
  Rev. E}\ }\textbf {\bibinfo {volume} {87}},\ \bibinfo {pages} {063301}
  (\bibinfo {year} {2013})}\BibitemShut {NoStop}%
\bibitem [{\citenamefont {Orlanski}(1976)}]{Orlanski1976}%
  \BibitemOpen
  \bibfield  {author} {\bibinfo {author} {\bibfnamefont {I.}~\bibnamefont
  {Orlanski}},\ }\href {\doibase 10.1016/0021-9991(76)90023-1} {\bibfield
  {journal} {\bibinfo  {journal} {J. Comput. Phys.}\ }\textbf {\bibinfo
  {volume} {21}},\ \bibinfo {pages} {251} (\bibinfo {year} {1976})}\BibitemShut
  {NoStop}%
\end{thebibliography}%

\end{document}